\documentclass[aps,pra,reprint,amsmath,amssymb,longbibliography]{revtex4-2}
\usepackage[T1]{fontenc}
\usepackage[utf8]{inputenc}
\ifdefined\DeclareUnicodeCharacter
\DeclareUnicodeCharacter{2009}{\,}
\fi
\usepackage{xcolor,physics,bm,empheq,url,quantikz,enumitem,booktabs,array}
\usepackage{algorithm,algpseudocode,subfigure,float,graphicx,microtype}
\usepackage{tikz}
\usetikzlibrary{positioning,calc,decorations.pathreplacing,arrows.meta,fit}
\usepackage{hyperref}
\hypersetup{colorlinks=true,linkcolor=blue,citecolor=blue,urlcolor=blue,pdftitle={Syndrome extraction for quantum LDPC codes in silicon: Distributed entanglement and mobile ancillas}}

\definecolor{Lcol}{RGB}{124,110,221}
\definecolor{Rcol}{RGB}{29,158,117}
\definecolor{Ocol}{RGB}{216,90,48}
\begin{document}
\setcounter{dbltopnumber}{2}
\setcounter{topnumber}{3}
\renewcommand{\dbltopfraction}{0.85}
\renewcommand{\topfraction}{0.85}
\renewcommand{\textfraction}{0.1}
\renewcommand{\floatpagefraction}{0.75}
\renewcommand{\dblfloatpagefraction}{0.75}

\title{Syndrome extraction for quantum LDPC codes in silicon: Distributed entanglement and mobile ancillas}

\author{William Chow}
\affiliation{Institute for Quantum Computing and Department of Physics, University of Waterloo,
200 University Avenue West, Waterloo, Ontario N2L 3G1, Canada}
\author{Jonathan Baugh}
\affiliation{Institute for Quantum Computing and Department of Chemistry, University of Waterloo,
200 University Avenue West, Waterloo, Ontario N2L 3G1, Canada}
\begin{abstract}
Quantum low-density parity-check (qLDPC) codes can reduce the number of
data qubits required per logical qubit, but implementing their checks on
a planar device can require interactions between distant sites. We develop
a singlet fabric for bivariate-bicycle codes on a silicon node network.
Spin singlets are generated on shuttling lanes and distributed between
nodes, where local operations consume them to measure non-local
stabilizers. We construct two syndrome-extraction circuits, based on
distributed GHZ states and ancilla teleportation, and evaluate them using
hardware-constrained schedules and circuit-level memory simulations.
We compare these circuits with CAbLECAR, which transports a mobile ancilla
between the data qubits of each check. Teleportation can outperform
CAbLECAR on its original layout when transport includes spin flips as
well as dephasing. However, placing the mobile-ancilla protocol on our
node lattice changes the comparison. For the $[[144,12,12]]$ Gross code,
the mean scheduled route decreases from $350$ to $153~\mu\mathrm{m}$,
and the logical error rate improves by up to approximately a factor of
$50$ in the sampled bit-flip regime. Direct transport then achieves a
lower logical error rate than either singlet circuit at every sampled
Gross-code point. When transport bit flips are frequent, however, the GHZ
circuit overtakes direct transport once two-qubit gate and readout errors
fall below about $4$--$5\times10^{-4}$, roughly half the value assumed here. Fixing the ancilla's data-visit order also removes a
small low-noise penalty, revealing a trade-off between shorter routes and the
correlated errors produced by ancilla faults. Source heralding improves both singlet circuits, whereas phase
purification helps the GHZ circuit only above a noise-dependent crossover
and worsens teleportation throughout the sampled unsaturated range.
These results show that the physical layout and interaction order can
be as consequential as the choice of extraction circuit, and that cleaner
entangled resources do not necessarily produce a more reliable logical
memory.
\end{abstract}
\maketitle
\section{Introduction}

Following Kitaev's introduction of topological quantum codes in 1997 \cite{Kitaev_2003}, the
surface code has emerged as one of the leading architectures for fault-tolerant quantum error
correction (QEC) \cite{Dennis_2002, Fowler_2012, PhysRevA.86.032324}. This stems
primarily from its comparatively high fault-tolerance threshold and its ability to be
implemented on a two-dimensional lattice using only local, nearest-neighbour interactions,
making it particularly compatible with the connectivity and fidelity constraints of many
experimental quantum-computing platforms. This has motivated extensive experimental
implementations, including superconducting-qubit demonstrations by Google and IBM and
trapped-ion experiments by Quantinuum \cite{Google_SurfaceCode_2025, IBM_SurfaceCode_2024,
Quantinuum_SurfaceCode_2024}.

More recently, advances in quantum low-density parity-check (qLDPC) codes have demonstrated the
possibility of achieving fault-tolerance thresholds comparable to those of the surface code
while substantially reducing the physical-qubit overhead required to encode and protect logical
information \cite{Bravyi_2024}. For example, Bravyi \textit{et al.} demonstrated a qLDPC
construction with a circuit-level threshold of approximately $0.7\%$ and estimated that
preserving 12 logical qubits at a physical error rate of $10^{-3}$ would require 288 physical
qubits, compared with nearly 3,000 physical qubits using the surface code. The potential
reduction in overhead becomes particularly significant for large-scale fault-tolerant
algorithms. Recent estimates indicate that factoring a 2048-bit RSA integer using a
surface-code-based architecture would require on the order of $10^6$ physical qubits
\cite{gidney2025factor2048bitrsa}. In contrast, recent qLDPC-based fault-tolerant architectures
have produced resource estimates ranging from approximately $10^5$ physical qubits
\cite{webster2026pinnaclearchitecturereducingcost} to as few as $10^4$ reconfigurable atomic
qubits for cryptographically relevant implementations of Shor's algorithm
\cite{cain2026shorsalgorithmpossible10000}, although these estimates rely on different hardware
architectures, error models, and runtime assumptions.

Despite their favourable encoding properties, qLDPC codes remain challenging to implement on
existing quantum hardware because their stabilizer measurements generally require non-local
interactions that must be realized without introducing prohibitive time or error overheads
\cite{Bravyi_2024, Xu_2024}. Superconducting-qubit platforms support fast gate operations, but
their predominantly fixed and local connectivity makes the long-range interactions required by
many qLDPC codes difficult to realize directly \cite{Kjaergaard_2020, Bravyi_2024}. Trapped-ion
and neutral-atom platforms, by contrast, provide more flexible connectivity through collective
interactions and physical qubit rearrangement, respectively, although these approaches can
introduce additional gate or transport times into the syndrome-extraction cycle
\cite{Bruzewicz_2019, Bluvstein_2024, Xu_2024, Pecorari_2025}.

Semiconductor spin qubits offer a potentially attractive compromise. Fast electrically
controlled local operations can be combined with coherent electron shuttling to dynamically
connect spatially separated regions of the processor \cite{Noiri_2022, DeSmet_2025,
Matsumoto_2026}. Recent experiments have further demonstrated that spin shuttling can be
incorporated directly into multi-qubit parity-check circuits, illustrating its potential as a
connectivity resource for quantum error correction \cite{Undseth_2026}. In addition,
semiconductor spin qubits have substantial compatibility with existing complementary
metal--oxide--semiconductor (CMOS) fabrication techniques, providing a potential route to
leverage the mature infrastructure, process control, and large-scale manufacturing capabilities
of the semiconductor industry \cite{VanDorpe_2024, Steinacker_2025, DumoulinStuyck_2026}. Together, these properties make semiconductor spin qubits a promising platform for realizing
the reduced-overhead advantages of qLDPC codes while mitigating some of the connectivity and
scalability constraints encountered in other architectures.

In this work, we develop a syndrome-extraction architecture for bivariate-bicycle codes on a
shuttling-based silicon spin-qubit lattice, combining the reduced encoding overhead of qLDPC
codes with distributed entanglement generated through coherent electron transport. We build on
the network architecture of Ref.~\cite{Buonacorsi_2019} by introducing the \textit{singlet fabric} 
in which singlets are generated locally and distributed by shuttling one member of each pair
between spatially separated nodes.
Rather than transporting a syndrome ancilla sequentially between the data qubits participating
in a stabilizer, we distribute singlet halves to establish the non-local entanglement required
by the long-range legs of each check. Because the entanglement is established before any
data qubit is touched, the preparation of each singlet can be verified locally, and a
distributed pair can additionally be purified, before it is consumed. The outcomes of these
checks are passed to the decoder as heralded information.

We construct resource schedules for the $[[72,12,6]]$ and $[[144,12,12]]$
bivariate-bicycle codes, determine the corresponding per-node hardware requirements and
syndrome-round durations, and evaluate their circuit-level logical error rates as a function
of the shuttling error rate. We compare this
distributed-entanglement approach with direct ancilla transport as implemented by CAbLECAR
\cite{chadwick2026cablecarefficientlyschedulingqldpc}, a recent scheduler designed for
qLDPC codes on shuttling-based spin-qubit hardware.

A central question in such a comparison is what is actually being compared. The two
architectures differ not only in how a check is measured but also in where the data qubits
sit, and therefore in how far any electron has to travel. We separate these effects by also
running the CAbLECAR scheduler on our own node lattice, and then by fixing the order in
which its ancillas visit their data qubits. These controls change the conclusion one would
draw from the direct comparison. On the original CAbLECAR layout, the teleport gadget wins
when transport frequently flips spins. On a common layout, direct transport is better at
every sampled Gross-code point, and the lattice we designed for the singlet fabric turns out
to be a good layout for mobile-ancilla schemes as well. We also find that, for the mobile
ancilla, the order of its interactions with the data controls a trade-off between transport
faults and correlated (hook) errors that is separate from route length. Finally, lowering the
two-qubit gate and readout errors together shows when distributed entanglement does pay off on a
common layout: with frequent transport bit flips, the GHZ gadget overtakes direct transport once
these errors fall below about $4$--$5\times10^{-4}$.

\section{Background}

\subsection{QEC and Bivariate-Bicycle Codes}

A stabilizer code on $n$ physical qubits is the joint $+1$ eigenspace of an
Abelian group of Pauli operators, generated by mutually commuting
\emph{checks}; a code encoding $k$ logical qubits with distance $d$ is written
$[[n,k,d]]$ \cite{Gottesman_1997, Terhal_2015}. For a
Calderbank--Shor--Steane (CSS) code the checks split into $X$-type and $Z$-type
generators with parity-check matrices satisfying $H_XH_Z^T=0 \pmod 2$
\cite{CalderbankShor_1996, Steane_1996}. Measuring the checks gives a binary
syndrome, from which a classical decoder infers a correction up to a
stabilizer.

\begin{figure*}[t]
\centering
\begin{tabular}{@{}c@{\hspace{1.0cm}}c@{}}
\begin{quantikz}[row sep={0.5cm,between origins}, column sep=0.16cm]
\lstick{$\ket{0}$} & \gate{H} & \ctrl{1} & \ctrl{2} & \ctrl{3} & \ctrl{4} & \gate{H} & \meter{} \\
\lstick[4]{$\ket{\psi_L}$} & \qw & \targ{} & \qw & \qw & \qw & \qw & \qw \\
 & \qw & \qw & \targ{} & \qw & \qw & \qw & \qw \\
 & \qw & \qw & \qw & \targ{} & \qw & \qw & \qw \\
 & \qw & \qw & \qw & \qw & \targ{} & \qw & \qw
\end{quantikz}
&
\begin{quantikz}[row sep={0.5cm,between origins}, column sep=0.16cm]
\lstick{$\ket{0}$} & \gate{H} & \ctrl{1} & \ctrl{2} & \ctrl{3} & \ctrl{4} & \gate{H} & \meter{} \\
\lstick[4]{$\ket{\psi_L}$} & \qw & \control{} & \qw & \qw & \qw & \qw & \qw \\
 & \qw & \qw & \control{} & \qw & \qw & \qw & \qw \\
 & \qw & \qw & \qw & \control{} & \qw & \qw & \qw \\
 & \qw & \qw & \qw & \qw & \control{} & \qw & \qw
\end{quantikz}
\\[2pt]
{\footnotesize (a) $X$-type check $X_1X_2X_3X_4$} & {\footnotesize (b) $Z$-type check $Z_1Z_2Z_3Z_4$}
\end{tabular}
\caption{Measuring a weight-4 stabilizer with a single ancilla. The ancilla is prepared in $\ket{+}=H\ket{0}$, coupled once to each data qubit of the check, and read out in the $X$ basis; the outcome is the eigenvalue $\pm1$ of the check and the encoded state $\ket{\psi_L}$ is left in the corresponding eigenspace. (a) For an $X$-type check the couplings are CNOTs from the ancilla onto the data. (b) For a $Z$-type check the couplings are controlled-$Z$ gates.}
\label{fig:stabilizer}
\end{figure*}

Each check is measured by coupling its data qubits to an ancilla that is then read out
(Fig.~\ref{fig:stabilizer}); these circuits are the building blocks of the gadgets of
Sec.~\ref{sec:gadgets}. Because the measurement circuits are themselves noisy, the checks are
measured repeatedly, and the decoder uses the space--time syndrome history to distinguish
persistent data errors from transient measurement faults \cite{Dennis_2002, Fowler_2012}.
The rotated surface code encodes one logical qubit in $d^2$ data qubits using weight-$4$ and
weight-$2$ checks between nearest neighbours \cite{Horsman_2012, Fowler_2012}, so its encoding
rate $k/n=1/d^2$ vanishes as the distance grows. Quantum low-density parity-check (qLDPC) codes
keep the checks sparse but drop geometric locality, and can reach a non-vanishing rate while the
distance grows \cite{TillichZemor_2014, PanteleevKalachev_2022, Breuckmann_2021}.

The qLDPC codes considered in this work are the bivariate-bicycle (BB) codes introduced by
Bravyi \textit{et al.} \cite{Bravyi_2024}. Let $S_l$ and $S_m$ denote cyclic shift matrices of
dimensions $l$ and $m$, respectively, and define
\begin{align}
x &= S_l \otimes I_m ,\hspace{20pt}  
y = I_l \otimes S_m ,
\end{align}
so that $x$ and $y$ are commuting $lm \times lm$ binary matrices satisfying
\begin{align}
x^l = y^m = I_{lm}.
\end{align}
A BB code is specified by two binary matrices
\begin{align}
A &= A_1 + A_2 + A_3, \hspace{20pt}  
B = B_1 + B_2 + B_3,
\end{align}
where each $A_i$ and $B_i$ is a power of either $x$ or $y$. For the class of codes considered
here, these may be written in the form
\begin{align}
A &= x^{a_1} + y^{a_2} + y^{a_3}, \hspace{20pt}  
B = y^{b_1} + x^{b_2} + x^{b_3}.
\end{align}

The resulting CSS code contains $n = 2lm$ data qubits, partitioned into two blocks, denoted $L$
and $R$, each containing $lm$ qubits. Its parity-check matrices are
\begin{align}
H_X &= [A \mid B], \hspace{20pt}
H_Z = [B^T \mid A^T].
\end{align}
Since $x$ and $y$ commute, $AB=BA$, and therefore
\begin{align}
H_X H_Z^T = AB + BA = 0
\end{align}
over $\mathbb{F}_2$, ensuring that all $X$- and $Z$-type stabilizer generators commute. Because
$A$ and $B$ each contain three monomials, every stabilizer generator has weight six, acting on
three qubits in the $L$ block and three qubits in the $R$ block. Correspondingly, each data
qubit participates in three $X$-type and three $Z$-type stabilizer checks \cite{Bravyi_2024}.

Following Ref.~\cite{Berthusen_2025}, we choose embeddings for the BB codes, summarized in 
Table~\ref{tab:embeddings}, such that under periodic boundary conditions each stabilizer check can be represented
using four nearest-neighbour interactions and two non-local interactions. In this work,
however, we model the layout as it would appear on a finite physical chip and therefore impose
open boundary conditions. Consequently, interactions that would wrap across the periodic
boundary must instead be treated explicitly as non-local connections in the hardware
architecture.   
\begin{table}[htbp]
\centering
\begin{tabular}{lcc}
\toprule
 & $[[72,12,6]]$ & $[[144,12,12]]$ (Gross) \\
\midrule
$(l,m)$ & $(6,6)$ & $(12,6)$ \\
$A$ & $x^{3}+y+y^{2}$ & $x^{3}+y+y^{2}$ \\
$B$ & $y^{3}+x+x^{2}$ & $y^{3}+x+x^{2}$ \\
$T_h$ & $(3,1)$ & $(9,1)$ \\
$T_v$ & $(4,3)$ & $(10,3)$ \\
\bottomrule
\end{tabular}
\caption{Toric embeddings used for the two bivariate-bicycle codes studied. Each embedding was found by exhaustive search subject to every check having exactly four legs at lattice distance one and two non-local legs, and admitting a linear three-colouring (six subcycles per round)~\cite{Berthusen_2025}. The translations are written as exponent pairs, $T_h=(u,v)$ denoting the monomial $x^uy^v$.}
\label{tab:embeddings}
\end{table}

\subsection{Silicon Spin Qubits, Shuttling and Native Singlets}

Semiconductor spin qubits encode information in the spins of electrons confined in gate-defined
quantum dots in Si/SiGe or Si-MOS heterostructures \cite{Burkard_2023}. Single-qubit gates are
driven by electron- or electric-dipole spin resonance, two-qubit gates by electrically controlled
exchange between neighbouring dots, and spins are read out by Pauli spin blockade (PSB), which
converts a two-spin configuration into a charge configuration detected by a nearby sensor
\cite{Noiri_2022, Veldhorst_2018}; a single spin is read by first mapping its state onto such a
pair with a two-qubit gate~\cite{Buonacorsi_2019}. Single-qubit fidelities above $99.9\%$ \cite{Yoneda_2018},
two-qubit fidelities at the $99.5$--$99.9\%$ level \cite{Noiri_2022, Xue_2022, Mills_2022,
Tanttu_2024}, PSB parity readout above $99\%$ on microsecond timescales
\cite{Takeda_2024_Readout}, and dynamically decoupled coherence times of tens of milliseconds
in isotopically enriched silicon \cite{Veldhorst_2014, Steinacker_2025} have been reported, and
devices from industrial 300-mm process lines now approach research-grade performance
\cite{Zwerver_2022, Koch_2025, Steinacker_2025}.

Two properties of the platform are central here. The first is coherent conveyor-mode shuttling:
phase-shifted voltages on a periodic array of clavier gates create a travelling potential minimum
that carries an electron along a channel \cite{Seidler_2022}. Because the gates are interleaved
into a few sets driven by common sinusoids, the number of control lines per lane does not grow
with its length, which makes a conveyor lane on every lattice edge plausible. Coherent transport
of spins over micrometre distances, preserving coherence and entanglement, has been demonstrated,
with phase accumulation and dephasing as important error channels \cite{Struck_2024,
DeSmet_2025}. We therefore consider transport noise that is dephasing only, and dephasing
accompanied by spin flips, which can arise from the spatially varying $g$-factor and spin--orbit
field and from spin--valley mixing at valley-splitting minima \cite{Langrock_2023, Losert_2023}.

The second is the two-electron spin singlet
\begin{align}
\ket{\Psi^-}=\frac{1}{\sqrt{2}}\left(\ket{\uparrow\downarrow}-\ket{\downarrow\uparrow}\right),
\end{align}
the ground state of two electrons in the lowest orbital of one dot. Its preparation can be
verified locally by Pauli spin blockade: after loading into $(0,2)$, the pair is separated into $(1,1)$ and
returned towards $(0,2)$, which a singlet can re-enter but a triplet cannot, and a charge sensor
distinguishes the two outcomes. The verified pair can then be separated adiabatically while preserving
its correlations \cite{Petta_2005, Veldhorst_2018, Struck_2024}. This check is necessarily local: once the halves have been separated, the pair can only be
checked non-locally, using a second pair and conditional gates (Sec.~\ref{sec:singlets}).
Singlets are thus a native distributed-entanglement resource: generated locally, distributed by
shuttling one half, and consumed in stabilizer-measurement circuits \cite{Buonacorsi_2019}.

\subsection{Non-local Checks: Mobile Ancillas versus Distributed Entanglement}

Two strategies exist for measuring checks whose data qubits are far apart. The first moves a
syndrome ancilla between them. CAbLECAR \cite{chadwick2026cablecarefficientlyschedulingqldpc}
develops this for a tileable shuttling processor inspired by SpinBus \cite{Kunne_2024}: each
ancilla is routed by a collision-aware path planner, interacts with its data qubits in turn
through CNOT gates, and is measured at the end of its route. Because shuttling noise is expected
to be strongly dephasing-biased, Hadamard gates around the shuttling intervals convert transport
phase errors into ancilla bit flips that only corrupt the measured syndrome, increasing the
tolerable shuttle error by about a factor of $5$--$10$ in that noise model. The approach needs a
single ancilla and readout per check, but the ancilla must stay coherent along its whole route,
its transport errors are not heralded, and spin flips during transport, which the tailoring does
not convert into harmless syndrome errors, are not part of that noise model; we add them in
Sec.~\ref{sec:results_setup}.

The second strategy, from distributed quantum computing, uses pre-shared entanglement. A Bell
pair with local operations and classical communication suffices to teleport a state
\cite{Bennett_1993} or a gate \cite{Gottesman_1999}, and thus to build remote parity
measurements \cite{Eisert_2000, Jiang_2007}. The entangled resource can be checked before it
touches the data, and noisy pairs can be purified \cite{Bennett_1996, Deutsch_1996, Dur_2007}.
Nickerson \textit{et al.} used purified Bell pairs fused into GHZ states to measure surface-code
checks in a networked architecture, with fault tolerance at network-link error rates of
$10\%$ and above \cite{Nickerson_2013, Nickerson_2014}. The cost is the extra hardware and time
for generating, storing, purifying and measuring the resource, and residual resource errors still
reach the data.

The present work combines elements of these two approaches. Spin singlets
provide a native entangled-pair resource in semiconductor quantum dots, while
conveyor-mode electron shuttling distributes one member of each pair across the
same two-dimensional lattice that could otherwise be traversed by a mobile
syndrome ancilla. The distributed singlets are subsequently fused into small
GHZ resources, or used to teleport the state of a locally generated ancilla,
to realize the non-local interactions required by the BB-code stabilizers.
In the GHZ gadget, a stabilizer is measured using a distributed GHZ ancilla,
with each GHZ qubit coupling locally to one participating data qubit.
Singlet distribution and fusion establish the required entanglement before
these data interactions, avoiding the need for a single syndrome ancilla to
visit every data qubit of the check in sequence. In the teleport gadget, an
ancilla visits the nearby data qubits itself and is teleported through a
distributed singlet to the distant ones. Both constructions separate
resource preparation from data coupling, which permits verification or
purification where these are included in the protocol. They do not remove
transport errors: the quantum state that finally couples to a data qubit
still carries any error the distributed singlet acquired during transport,
regardless of which physical electron holds that state. Residual resource
errors can therefore still affect syndrome extraction, and the simulations
include them.

\section{Architecture}

This section follows the steps required to measure the stabilizer on the
proposed device. We first describe where the data qubits are stored and
which operations each node can perform. We then explain how singlets are
prepared, checked and transported between nodes. These pairs are used in
two different measurement circuits: one first assembles a distributed GHZ
ancilla, while the other teleports the state of a moving ancilla between
distant data sites. Finally, we describe how the operations are scheduled
on shared hardware and how that schedule is translated into a noisy
quantum circuit. This order separates what each circuit does from when
its operations can actually take place.

\subsection{Lattice and Node Model}
\label{sec:lattice}

We realize the BB-code embeddings on a finite $2l\times 2m$ square lattice with open boundary conditions. 
Each lattice site corresponds to a physical node and each nearest-neighbour lattice edge to a conveyor-mode
electron-shuttling lane of length $L_{\mathrm{lane}}=5~\mu\mathrm{m}$. 
Sites with even coordinate sum are assigned as check centres, with $X$-type checks placed at
$(\mathrm{even},\mathrm{even})$ coordinates and $Z$-type checks at
$(\mathrm{odd},\mathrm{odd})$ coordinates. 
Sites with odd coordinate sum are data sites and contain one stationary data qubit.

For a bulk check centre, four of the six data qubits participating in the corresponding BB stabilizer
occupy the four neighbouring lattice sites, while the remaining two are displaced by the non-local
translations $T_h$ and $T_v$ defined by the embedding. At an open boundary, an interaction that would
cross a periodic boundary in the toric embedding is instead treated as an additional routed non-local
connection. Here a ``leg'' means one required connection between a check and a data
qubit in its support. A connection that is short when opposite edges of
the code drawing are identified may become a long route on the actual
chip, whose edges are not connected. Throughout the scheduling, we use
these physical routes rather than treating a periodic boundary connection
as a shortcut.

With this construction, the $[[72,12,6]]$ and $[[144,12,12]]$ codes occupy
144  and 288 nodes respectively. The corresponding open square lattices contain
264 and 540 nearest-neighbour shuttling lanes, respectively, as listed in
Table~\ref{tab:resources}.

Each lattice node is implemented as a small array of gate-defined quantum dots with four functional
components, illustrated in Fig.~\ref{fig:node}, and, when link purification is enabled, a fifth
(Fig.~\ref{fig:node_purifier}):

\begin{figure*}[t]
\centering
\begin{tikzpicture}[every node/.style={font=\footnotesize}]
\begin{scope}
  \def\s{0.78}
  \foreach \x in {0,...,9}{\foreach \y in {0,...,5}{
    \ifnum\x<9 \draw[gray!70, line width=2pt] (\x*\s,\y*\s) -- ({(\x+1)*\s},\y*\s);
      \fill[brown!80!black] ({(\x+0.5)*\s},\y*\s) circle (0.07); \fi
    \ifnum\y<5 \draw[gray!70, line width=2pt] (\x*\s,\y*\s) -- (\x*\s,{(\y+1)*\s});
      \fill[brown!80!black] (\x*\s,{(\y+0.5)*\s}) circle (0.07); \fi }}
  \foreach \x in {0,...,9}{\foreach \y in {0,...,5}{
    \pgfmathtruncatemacro{\ps}{mod(\x+\y,2)}
    \pgfmathtruncatemacro{\xpar}{mod(\x,2)}
    \ifnum\ps=0
      \ifnum\xpar=0 \fill[blue!60] (\x*\s,\y*\s) circle (0.15);
      \else \fill[red!70] (\x*\s,\y*\s) circle (0.15); \fi
    \else
      \ifnum\xpar=1 \fill[green!55!black] (\x*\s,\y*\s) circle (0.15);
      \else \fill[violet!80] (\x*\s,\y*\s) circle (0.15); \fi
    \fi }}
  \begin{scope}[shift={(0.3,-0.75)}]
    \node[anchor=west] at (0,0)   {\tikz\fill[blue!60] (0,0) circle (0.1); $X$ centre};
    \node[anchor=west] at (1.9,0) {\tikz\fill[red!70] (0,0) circle (0.1); $Z$ centre};
    \node[anchor=west] at (3.8,0) {\tikz\fill[green!55!black] (0,0) circle (0.1); $L$ data};
    \node[anchor=west] at (0,-0.45)   {\tikz\fill[violet!80] (0,0) circle (0.1); $R$ data};
    \node[anchor=west] at (1.9,-0.45) {\tikz\fill[brown!80!black] (0,0) circle (0.08); generator};
    \node[anchor=west] at (3.8,-0.45) {\tikz\draw[gray!70,line width=2pt] (0,0)--(0.35,0); lane};
  \end{scope}
  \node[anchor=north, text width=7.2cm, align=justify] at (4.5*\s,-1.45) {(a) Singlet fabric. Every lattice site is a node and every edge a 5\,$\mu$m conveyor lane carrying a singlet generator.};
\end{scope}

\begin{scope}[shift={(10.9cm,1.95cm)}]
  \def\d{0.72}
  \draw[rounded corners=6pt, line width=0.8pt] (-2.3,-2.3) rectangle (2.3,2.3);
  \foreach \a/\l in {90/N, 270/S, 0/E, 180/W}{
    \draw[gray!70, line width=2pt] (\a:2.3) -- (\a:3.0);
    \node at (\a:3.25) {\l};
    \draw[fill=white, line width=0.8pt] (\a:1.9) circle (0.16);
  }
  \node[anchor=west] at (0.2,1.9) {\scriptsize lane port};
  \foreach \i in {0,...,2}{\foreach \j in {0,1}{
    \draw[red!40, line width=0.6pt] ({(\i-1.5)*\d+0.2},{(\j-0.5)*\d}) -- ({(\i-0.5)*\d-0.2},{(\j-0.5)*\d});}}
  \foreach \i in {0,...,3}{
    \draw[red!40, line width=0.6pt] ({(\i-1.5)*\d},{-0.5*\d+0.2}) -- ({(\i-1.5)*\d},{0.5*\d-0.2});}
  \draw[red!70!black, line width=2.4pt] ({-1.5*\d+0.2},{-0.5*\d}) -- ({-0.5*\d-0.2},{-0.5*\d});
  \draw[red!70!black, line width=2.4pt] ({1.5*\d},{-0.5*\d+0.2}) -- ({1.5*\d},{0.5*\d-0.2});
  \foreach \s in {1,-1}{
    \draw[red!40, line width=0.6pt] ({-1.5*\d*\s},{\s*(0.5*\d+0.2)}) -- ({-1.5*\d*\s},{\s*0.85});
    \draw[cyan!60!black, line width=0.9pt, rounded corners=3pt] ({-1.5*\d*\s-0.22*\s},{\s*0.78}) rectangle ({-0.38*\s},{\s*1.18});
    \draw[fill=cyan!30, line width=0.8pt] ({-1.5*\d*\s},{\s*0.98}) circle (0.13);
    \draw[fill=cyan!30, line width=0.8pt] ({-0.6*\s},{\s*0.98}) circle (0.13);}
  \foreach \i in {0,...,3}{\foreach \j in {0,1}{
    \draw[fill=yellow!35, line width=0.8pt] ({(\i-1.5)*\d},{(\j-0.5)*\d}) circle (0.2);}}
  \draw[fill=green!55!black, line width=0.8pt] ({-1.5*\d},{-0.5*\d}) circle (0.2);
  \draw[fill=cyan!30, line width=0.8pt] (-1.19,1.28) rectangle (-0.49,1.64); \node at (-0.84,1.46) {\scriptsize SET};
  \draw[fill=cyan!30, line width=0.8pt] (1.40,-1.16) rectangle (2.10,-0.80); \node at (1.75,-0.98) {\scriptsize SET};
  \node[anchor=east, green!40!black] at ({-1.5*\d-0.25},{-0.5*\d}) {\scriptsize data};
  \draw[dashed, gray, rounded corners=4pt, line width=0.5pt] ({-1.5*\d-0.38},{-0.5*\d-0.38}) rectangle ({1.5*\d+0.38},{0.5*\d+0.38});
  \node[anchor=north east, gray!40!black] at ({1.5*\d+0.38},{0.5*\d+0.78}) {\scriptsize 8-dot pool};
  \draw[red!40, line width=0.6pt] (-2.1,-1.55) -- (-1.8,-1.55); \node[anchor=west] at (-1.78,-1.55) {\scriptsize possible exchange};
  \draw[red!70!black, line width=2.4pt] (-2.1,-1.9) -- (-1.8,-1.9); \node[anchor=west] at (-1.78,-1.9) {\scriptsize active (${\le}2$)};
  \node[anchor=north, text width=6.4cm, align=justify] at (0,-3.6) {(b) A bulk data-site node (resource model). Lane ports (white) hand electrons between the conveyor and the static node. Eight pool dots (yellow) hold the resident data electron (green) and, as needed, singlet halves, parked relay electrons and gadget ancillas. Any two neighbouring dots can be exchange coupled (thin red); at most two couplings are pulsed at a time (thick red, one shown at the data electron). Each charge sensor (SET) reads a double dot (cyan) by Pauli spin blockade; to read a pool electron, the double dot is prepared in a singlet and one of its spins is exchange coupled to the adjacent pool dot (Sec.~\ref{sec:lattice}). A check centre is the same with a six-dot pool and no data electron.};
\end{scope}
\end{tikzpicture}
\caption{Singlet-fabric layout and node model. (a)~A representative region of the open array, not one complete code block. Every lattice site is a node and every edge a $5~\mu$m conveyor lane (grey). Check-centre nodes host the measurement of one stabilizer each: blue for $X$-type and red for $Z$-type checks. Data nodes hold one data qubit each: green for the left ($L$) block and violet for the right ($R$) block of the BB code, the two halves of $H_X=[A\,|\,B]$. Brown markers are the dual singlet-generator banks at the lane midpoints. (b)~The node as modelled by the scheduler: a budget of dots, exchange operations and readouts rather than a gate-level device layout. Electrons are moved within the node to sit beside their exchange or readout partner, at the intra-node move cost of Sec.~\ref{sec:lattice}. The pool sizes ($8$ at a data site, $6$ at a check centre) are a sufficient inventory for the validated schedules; the peak occupancy is given in Table~\ref{tab:resources}. Verification and purification provide information about resource reliability but do not remove all residual errors.}
\label{fig:node}
\end{figure*}
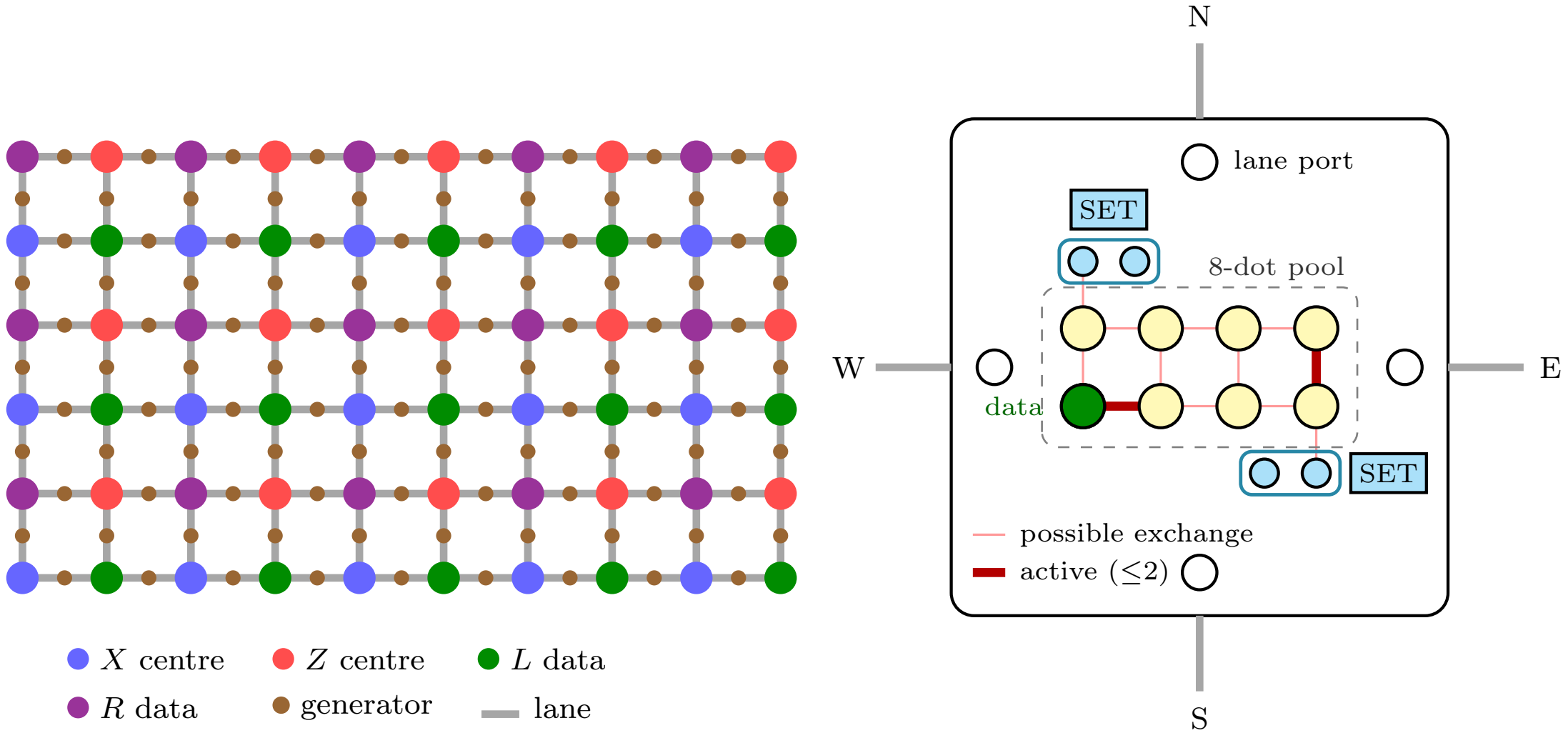

\begin{itemize}
    \item \textbf{Lane ports.}
    Each incident shuttling lane terminates at a dedicated port dot that provides the interface between
    the travelling conveyor potential and the static quantum dots within the node. A bulk node therefore
    has four lane ports, while boundary nodes have fewer incident physical lanes.

    \item \textbf{Static dot pool.}
    A set of local quantum dots stores electrons that must remain within the node during syndrome
    extraction. Depending on the node type and gadget, these may include the resident data electron,
    delivered singlet halves, including raw halves waiting to be purified, electrons temporarily parked
    while changing direction, and stabilizer-measurement ancillas.

    \item \textbf{Exchange-coupling zones.}
    Electrically tunable tunnel barriers between neighbouring dots allow any two electrons that have been
    moved next to each other, including an ancilla and the resident data electron, to be exchange coupled.
    The scheduler treats this as a capacity: at most two exchange operations run concurrently in a node.

    \item \textbf{Spin readout.}
    Each node contains two Pauli-spin-blockade (PSB) readout channels. Each channel is a double dot
    next to a charge sensor, outside the static pool, one of whose dots can be exchange coupled to an
    adjacent pool dot. A singlet--triplet measurement of two spins can be made directly in the double
    dot, but a single spin cannot be read by spin blockade against a reference spin: with a reference
    $\ket{\uparrow}$, an unknown $\ket{\uparrow}$ gives $\ket{T_+}$, whereas $\ket{\downarrow}$ gives an
    equal superposition of $\ket{S}$ and $\ket{T_0}$. Following Ref.~\cite{Buonacorsi_2019}, the
    electron to be read is therefore moved into the adjacent pool dot, the double dot is prepared in
    $\ket{S}$, and a CZ between the electron and one spin of the pair leaves the pair in $\ket{S}$ if
    the electron is up and rotates it to $\ket{T_0}$ if it is down; PSB readout of the pair then gives
    the spin. Every readout in our gadgets is of a single spin and uses this sequence, so each costs one
    CZ in addition to the PSB measurement (Sec.~\ref{sec:results_setup}). The two channels are both in
    use at the peak of every schedule except that of the unpurified GHZ gadget
    (Table~\ref{tab:resources}). These channels are used for purification measurements, fusion
    measurements, and final ancilla readout. The source test, a two-spin measurement, is done at the
    generator bank on the lane, not in the node.

    \item \textbf{Purification module.}
    When link purification is enabled, each node contains one purification
    module: a pair of work dots with a dedicated exchange coupling, into which
    the retained and sacrificial halves of a purified link are loaded. The
    module performs this endpoint's half of the phase purification of
    Fig.~\ref{fig:purification}, a CNOT compiled from a CZ and single-qubit
    layers, after which the sacrificial half is read out using one of the
    node's Pauli-spin-blockade channels. The module can serve only one link at a time, so
    links terminating at the same node are purified sequentially.
\end{itemize}

In addition, each physical lane contains a dual singlet-generator bank positioned on its midpoint.
Each generator prepares a two-electron singlet in a doubly occupied dot and verifies the
preparation locally, before the pair is distributed (Sec.~\ref{sec:singlets}). The pair is
then separated and its two halves are launched toward opposite endpoints of the lane. The
associated source-verification result is retained as classical side information and accompanies the distributed pair through the
syndrome-extraction protocol.

Longer-range transport is implemented by concatenating nearest-neighbour conveyor segments.
When an electron changes direction at an intermediate node, it is unloaded from the incoming lane into
a static pool dot and subsequently relaunched onto the outgoing lane. We refer to this procedure as
\emph{park-and-relaunch}. Transport within a node, including transfer between a lane port and the static
dot pool, is modelled as a fixed dwell time of $0.1~\mu\mathrm{s}$ together with the same dephasing cost
assigned to $1~\mu\mathrm{m}$ of conveyor transport.

The number of storage dots and simultaneously available exchange and readout resources is chosen from
the scheduling analysis of Sec.~\ref{sec:scheduling}. For the schedules considered here, a bulk data site is provisioned
with an eight-dot static pool and a bulk check centre with a six-dot pool, in addition to their lane
ports. Each node contains two exchange-coupling zones and two readout channels. These capacities support the schedules obtained here without increasing
the reported round duration; the peak simultaneous use of each resource is listed in
Table~\ref{tab:resources}. A data site needs the larger pool because it holds the data
electron as well as singlet halves waiting for their deadline or for purification and
electrons parked between lanes. They describe a sufficient inventory
for those schedules, rather than a proof of the minimum hardware required
by any possible schedule.

\subsection{Singlet Distribution: Heralding and Purification}
\label{sec:singlets}

Every distributed resource originates from a dual singlet-generator bank on a shuttling lane.
Each of its two generators loads a second electron into a singly occupied dot, $(0,1)\to(0,2)$, and
then verifies the pair in two steps before it is distributed. A charge sensor first confirms the
$(0,2)$ charge state. A failed load leaves the dot in $(0,1)$: there is no pair, so this is not an
error on a pair but a missing resource. The pair is then separated into $(1,1)$ and returned towards
$(0,2)$ in the spin-blockade regime: a singlet returns, while a triplet remains blocked in $(1,1)$, and
the sensor reads which occurred. This spin-sensitive step detects pairs in the wrong spin state, such
as a thermally loaded or unrelaxed triplet. It cannot see errors that occur after it, in the final
separation, transport or storage.

The two outcomes are therefore treated differently. A \emph{load failure} means no pair exists; the
other generator of the bank supplies the link, and if both fail to load, which happens with
probability of order $10^{-6}$ per link at the parameters below, the load is repeated. A \emph{spin
rejection} means a pair exists but is likely faulty. We describe the spin verification by the
probability $p_{\mathrm{src}}$ that a loaded pair is faulty, the fraction $f$ of faulty pairs that the
blockade step cannot detect, and the efficiency $\eta$ with which it flags the rest.
Appendix~\ref{app:source} gives the resulting probabilities that accepted and rejected pairs are
faulty. The main results use the most favourable case $f=0$, with $p_{\mathrm{src}}=10^{-3}$ and
$\eta=0.999$, for which an accepted pair is faulty with probability $\approx10^{-6}$. The opposite
limit $f=1$, in which the check detects nothing, is equivalent to switching heralding off, which we
also simulate; the two bound the benefit of heralding (Sec.~\ref{sec:results_heralding}).

If either generator produces an accepted pair, an accepted pair is selected
for distribution. If both pairs are rejected by the spin check, one of them is
nevertheless selected in order to preserve a fixed syndrome-extraction
schedule, and a \emph{double-reject flag} is attached to the link. The decoder
therefore distinguishes the overwhelmingly likely accepted branch from the
rare double-reject branch and assigns different conditional error probabilities
to the two cases. In this way, redundant parallel singlet generation suppresses
the probability of being forced to use a rejected pair quadratically
without requiring variable-latency retries. The flag warns that a link
is unreliable; it does not mean the check is skipped. The verification adds a separation, a return
and a readout to each generation. We add $0.5~\mu$s, one PSB measurement time of Table~\ref{tab:params},
to every generation, treating the separation and return pulses as fast, so a generator bank is
occupied for $2.0~\mu$s per pair. Because generation is pipelined ahead of consumption, the extra
time is absorbed by the teleport gadget, whose round is unchanged, but the GHZ gadget draws more
pairs per round and becomes generation-limited (Table~\ref{tab:rounds}).

After preparation, the two members of the singlet are transported toward
their respective endpoints. A local link connects nodes separated by a single
lattice edge and requires no intermediate relay. For a routed link of Manhattan
length $h$, the source is chosen near the midpoint of a shortest path so that
the transport distance is approximately balanced between the two singlet
halves. At corners, an electron is parked in an intermediate node and
relaunched onto the outgoing conveyor lane.

We consider two models for transport noise. In the first, transport causes only dephasing,
that is, random phase ($Z$) errors on the moving spin. If a single transport segment
has phase-flip probability $p_s$, then for $h$ independent equivalent segments
the accumulated phase-flip probability is
\begin{align}
    p_Z(h)
    =
    \frac{1-(1-2p_s)^h}{2}
    \simeq hp_s ,
\end{align}
The expression counts the probability of an odd number of phase flips:
two $Z$ errors on the same spin cancel. The approximation on the right
applies when $hp_s\ll1$, so that multiple flips are unlikely. Additional local gate,
preparation, and measurement errors are included independently in the
circuit-level noise model. In the second model, transport also causes spin flips ($X$
errors), with a probability proportional to the dephasing probability. We use two ratios,
$p_X(h)=p_Z(h)/10$ and $p_X(h)=p_Z(h)/3$, to see how sensitive each scheme is to this
less well characterized channel. In Sec.~\ref{sec:results_setup} these are specified per
$10~\mu\mathrm{m}$ of transport as $p_\mathrm{sh}$ (phase flip) and $p_x$ (bit flip), from which
the per-lane probability $p_s$ follows. 

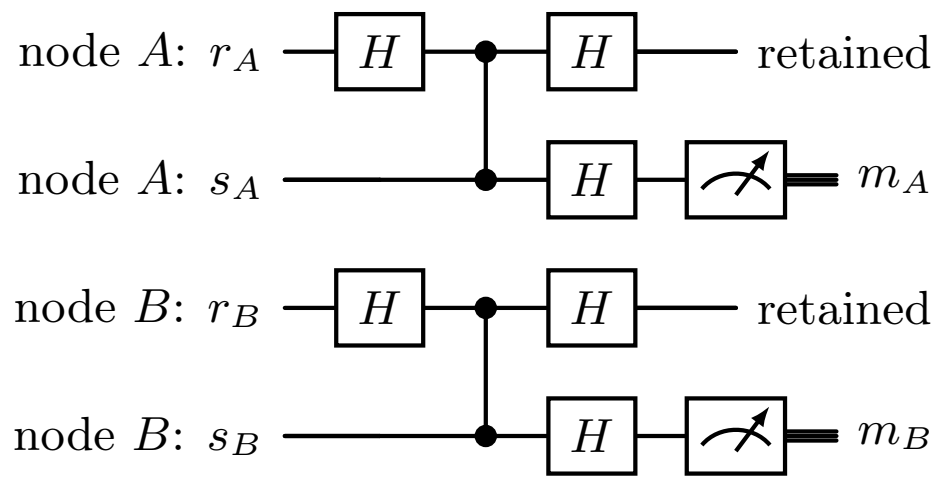
\begin{figure*}[t]
\centering
\begin{quantikz}[row sep=0.35cm,column sep=0.34cm]
\lstick{node $A$: $r_A$} & \gate{H} & \ctrl{1} & \gate{H} & \qw \rstick{retained} \\
\lstick{node $A$: $s_A$} & \qw & \control{} & \gate{H} & \meter{} & \cw \rstick{$m_A$} \\
\lstick{node $B$: $r_B$} & \gate{H} & \ctrl{1} & \gate{H} & \qw \rstick{retained} \\
\lstick{node $B$: $s_B$} & \qw & \control{} & \gate{H} & \meter{} & \cw \rstick{$m_B$}
\end{quantikz}
\caption{Local operations for phase purification of retained pair
$(r_A,r_B)$ using sacrificial pair $(s_A,s_B)$, in compiled form. The
retained halves are rotated to the $X$ basis, each endpoint applies a CZ
between its retained and sacrificial halves, the retained halves are rotated
back and the sacrificial halves are measured in the $X$ basis. This is the
circuit obtained from Hadamards on all four halves followed by a CNOT compiled
as $H(s)$--CZ--$H(s)$, after the two adjacent Hadamards on each sacrificial
wire cancel. In the simulations the uncompiled form is used, with both
Hadamards kept; this adds two single-qubit error locations, about
$2p_1=2\times10^{-4}$, to each sacrificial half and slightly overstates the
cost of purification. Outcome parity $m_A\oplus m_B$ supplies the
accept/reject flag under the chosen Bell-frame convention. Both branches
retain an output in the fixed-latency policy.}
\label{fig:purification}
\end{figure*}

For non-local routed links, two raw singlets are generated at different times between
the same pair of endpoints and subjected to a phase-error parity check (Fig.~\ref{fig:purification}). At
each endpoint, a bilateral CNOT is implemented in the node's purification
module (Sec.~\ref{sec:lattice}), after which the sacrificial pair is measured. Comparison of
the two endpoint outcomes reveals whether the two distributed pairs have the
same phase-error parity; we call the result the \emph{purification flag}. On the
accepted branch the targeted phase error is suppressed from first to second order
(Appendix~\ref{app:source}).

Unlike conventional recurrence purification~\cite{Bennett_1996, Deutsch_1996}, our fixed-latency
protocol does not necessarily discard a pair following an unsuccessful parity
test. If the test rejects, the retained pair may still be used, but the reject
outcome is supplied to the decoder as classical side information. The decoder
therefore conditions the prior error probability of each distributed link on
both its source and purification flags. In the full circuit model, the
conditional probabilities of the accepted and rejected branches are computed
including preparation, transport, two-qubit-gate, and measurement errors.
These probabilities tell the decoder how much to trust each delivered
resource. They do not remove its errors from the quantum circuit, and an
accepted purification outcome is not a guarantee of a fault-free pair.

This approach trades additional singlet generation, local two-qubit gates,
and readout operations for information about the reliability of the
non-local entanglement resource. Whether this trade is worthwhile is not
obvious in advance, because the purification circuit itself adds faults. We
therefore treat purification as an option and test it in the full memory
simulation (Sec.~\ref{sec:results_purification}). There we find that it helps
the GHZ gadget at high shuttle noise but makes the teleport gadget worse at
every sampled point.

\subsection{Fusion}
\label{sec:fusion}

We assemble distributed ancilla resources using a singlet-mediated
measurement of $Z_aZ_b$, where the two retained spins $a$ and $b$ are
called the \emph{arms}. A distributed singlet
\[
    \ket{\Psi^-}_{x_ax_b}=\frac{\ket{01}-\ket{10}}{\sqrt{2}}
\]
(with $\ket{0}\equiv\ket{\uparrow}$ and $\ket{1}\equiv\ket{\downarrow}$)
provides the \emph{bridge}, with $x_a$ adjacent to $a$ and $x_b$
adjacent to $b$. We apply $\mathrm{CZ}(a,x_a)$ and
$\mathrm{CZ}(b,x_b)$, then measure both bridge spins in $X$
(Fig.~\ref{fig:fusion}). The arms are retained, while the bridge
is consumed.

Writing the bridge outcomes as $m_a,m_b\in\{0,1\}$, with measured
eigenvalues $s_a=(-1)^{m_a}$ and $s_b=(-1)^{m_b}$, the arm parity is
\begin{equation}
    s_{ab}=-s_as_b=(-1)^{m_a\oplus m_b\oplus1}.
\end{equation}
The minus sign follows from the singlet correlation
$X_{x_a}X_{x_b}=-1$: conjugation by the two CZ gates maps this
operator to $Z_aZ_bX_{x_a}X_{x_b}$. Thus the bridge readouts
determine the joint arm parity without measuring either arm
individually. The $X$ readouts are implemented by Hadamard
rotations followed by the modelled Pauli-spin-blockade readout.

The measurement also produces a known Pauli byproduct. Together
with the measured parity sector, this determines the corrections
needed to express the retained arms in a common reference frame.
We track these corrections classically as a \emph{Pauli frame}~\cite{Knill_2005}.
Successive fusions join initially separate entangled components
into the GHZ resource used below.

Frame tracking must be included in the fault analysis. A faulty
bridge readout can change the inferred arm correction even when
the fault does not directly propagate to an arm through the
fusion CZ. An incorrectly inferred arm $X$ subsequently propagates
through an arm--data CZ as
\begin{equation}
    X_a \longrightarrow X_aZ_d.
\end{equation}
Bridge-record errors can therefore produce data errors, rather
than just changing the reported stabilizer outcome.

\begin{figure*}[t]
\centering
\begin{tikzpicture}[every node/.style={font=\footnotesize}]
\node[anchor=north west] at (0,0) {\begin{quantikz}[row sep=0.35cm, column sep=0.45cm]
\lstick{$a$}   & \ctrl{1}   & \qw      & \qw      & \qw \rstick{kept} \\
\lstick{$x_a$} & \control{} & \gate{H} & \meter{} & \cw \rstick{$m_a$} \\
\lstick{$x_b$} & \control{} & \gate{H} & \meter{} & \cw \rstick{$m_b$} \\
\lstick{$b$}   & \ctrl{-1}  & \qw      & \qw      & \qw \rstick{kept}
\end{quantikz}};
\draw[decorate, decoration={brace, amplitude=4pt, mirror}] (-0.55,-1.05) -- (-0.55,-1.85) node[midway, xshift=-14pt, rotate=90, anchor=south] {\scriptsize $\ket{\Psi^-}$};
\node[anchor=north, text width=5.6cm, align=justify] at (2.6,-3.2) {(a) Fusion of arms $a$ and $b$ through a bridge singlet $(x_a,x_b)$. Each CZ is local; the bridge halves are read in the $X$ basis at their own sites. The parity eigenvalue is $(-1)^{m_a\oplus m_b\oplus1}$. The arms remain available in the measured parity sector, with known Pauli byproducts.};

\begin{scope}[shift={(7.6cm,-0.4cm)}]
  \def\w{0.95}
  \foreach \i/\n in {0/a_4,1/a_1,2/a_0,3/a_2,4/a_3,5/a_5}{
    \node[draw, circle, fill=green!25, minimum size=0.55cm, inner sep=0] (a\i) at (\i*\w,0) {\scriptsize $\n$};}
  \draw[line width=2.2pt, teal!70!black] (a1) -- (a2); \node[teal!60!black] at (1.5*\w,0.42) {\scriptsize pairA};
  \draw[line width=2.2pt, teal!70!black] (a3) -- (a4); \node[teal!60!black] at (3.5*\w,0.42) {\scriptsize pairB};
  \node[green!40!black] at (0,0.42) {\scriptsize $\ket{+}$}; \node[green!40!black] at (5*\w,0.42) {\scriptsize $\ket{+}$};
  \colorlet{cNL1}{orange!30}\colorlet{cF}{blue!25}\colorlet{cNL2}{violet!30}
  \colorlet{lNL1}{orange}\colorlet{lF}{blue!60}\colorlet{lNL2}{violet}
  \foreach \l/\r/\nm/\fc/\lc in {0/1/fuseNL1/cNL1/lNL1, 2/3/fuse/cF/lF, 4/5/fuseNL2/cNL2/lNL2}{
    \node[draw, circle, fill=\fc, minimum size=0.4cm, inner sep=0] (xl\nm) at (\l*\w,-1.1) {};
    \node[draw, circle, fill=\fc, minimum size=0.4cm, inner sep=0] (xr\nm) at (\r*\w,-1.1) {};
    \draw[line width=2.2pt, \lc] (xl\nm) -- (xr\nm);
    \draw[thick] (a\l) -- (xl\nm); \draw[thick] (a\r) -- (xr\nm);
    \node[\lc] at ({(\l+0.5)*\w},-1.5) {\scriptsize \nm};}
  \node at (0,-1.1) {\scriptsize $x_2$}; \node at (\w,-1.1) {\scriptsize $x_3$};
  \node at (2*\w,-1.1) {\scriptsize $x_0$}; \node at (3*\w,-1.1) {\scriptsize $x_1$};
  \node at (4*\w,-1.1) {\scriptsize $x_4$}; \node at (5*\w,-1.1) {\scriptsize $x_5$};
  \node[anchor=east] at (-0.55,0) {\scriptsize arms};
  \node[anchor=east] at (-0.55,-1.1) {\scriptsize bridges};
  \node[anchor=north, text width=6.2cm, align=justify] at (2.5*\w,-2.05) {(b) Chain construction of $\ket{\mathrm{GHZ}_6}$. Thick bonds are singlets; thin lines are the CZs of three fusions. pairA and pairB supply four arms, $a_4,a_5$ are fresh $\ket{+}$ states, and the three bridges are consumed. In the bulk, fuseNL1 and fuseNL2 are the routed bridges; purification is a separate policy choice. Boundary routes require additional accounting.};
\end{scope}

\end{tikzpicture}
\caption{Fusion and the chain GHZ gadget. The propagation of errors through the fusion is discussed in Sec.~\protect\ref{sec:fusion}.}
\label{fig:fusion}
\end{figure*}

\subsection{Syndrome-extraction gadgets}
\label{sec:gadgets}

We use fusion and teleportation to construct two measurements
of the same weight-six BB stabilizers. We describe the
$Z$-type measurement using ancilla--data CZ gates; the
$X$-type measurement includes the corresponding data-basis
changes. Data qubits remain at their assigned sites in both
constructions.

\subsubsection{Chain GHZ gadget}
\label{sec:chain}

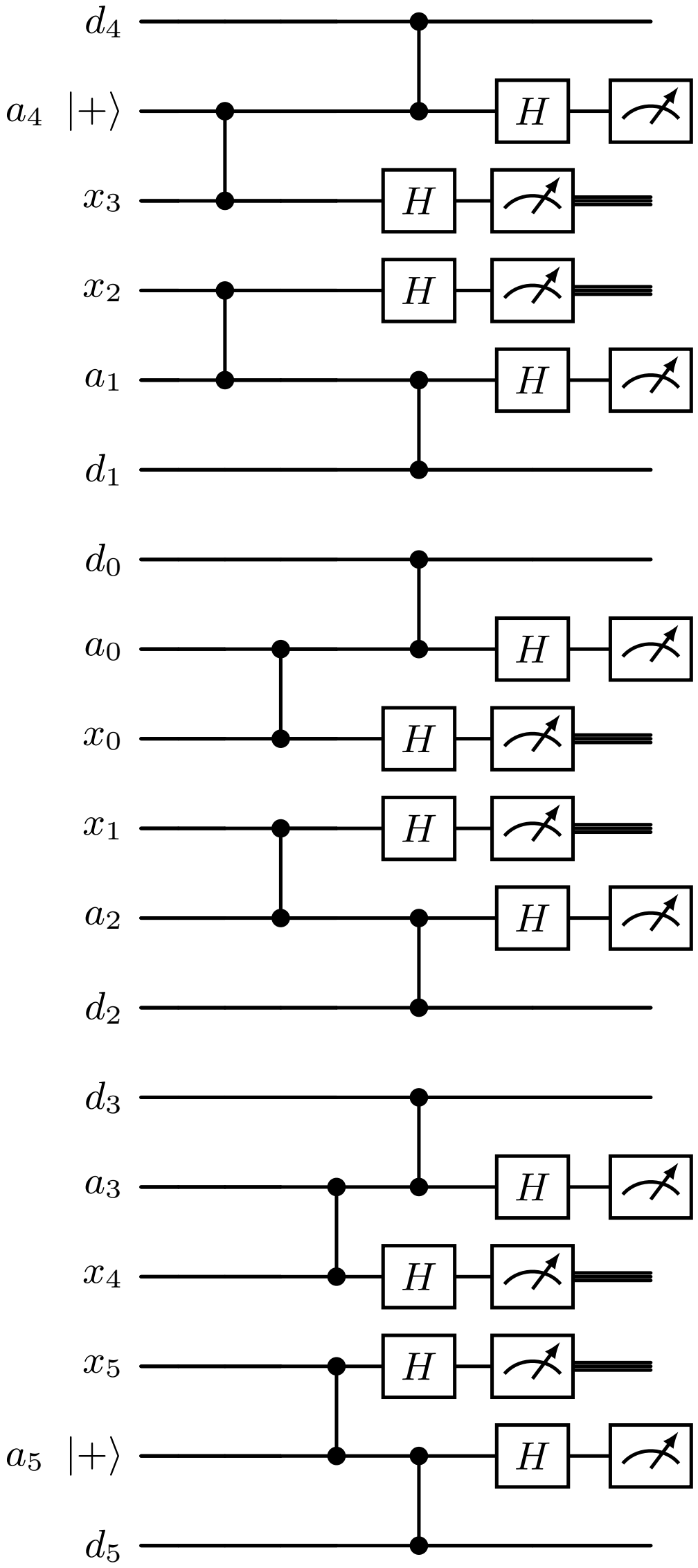
\begin{figure*}[t]
\centering
\resizebox*{!}{0.64\textheight}{%
\begin{quantikz}[row sep={0.72cm,between origins}, column sep=0.3cm]
\lstick{$d_4$}                & \qw & \qw      & \qw      & \qw      & \ctrl{1}   & \qw      & \qw \\
\lstick{$a_4\ \ket{+}$}       & \qw & \ctrl{1} & \qw      & \qw      & \control{} & \gate{H} & \meter{} \\
\lstick{$x_3$}                & \qw & \control{} & \qw    & \qw      & \gate{H}   & \meter{} & \cw \\
\lstick{$x_2$}                & \qw & \ctrl{1} & \qw      & \qw      & \gate{H}   & \meter{} & \cw \\
\lstick{$a_1$}                & \qw & \control{} & \qw    & \qw      & \ctrl{1}   & \gate{H} & \meter{} \\
\lstick{$d_1$}                & \qw & \qw      & \qw      & \qw      & \control{} & \qw      & \qw \\
\lstick{$d_0$}                & \qw & \qw      & \qw      & \qw      & \ctrl{1}   & \qw      & \qw \\
\lstick{$a_0$}                & \qw & \qw      & \ctrl{1} & \qw      & \control{} & \gate{H} & \meter{} \\
\lstick{$x_0$}                & \qw & \qw      & \control{} & \qw    & \gate{H}   & \meter{} & \cw \\
\lstick{$x_1$}                & \qw & \qw      & \ctrl{1} & \qw      & \gate{H}   & \meter{} & \cw \\
\lstick{$a_2$}                & \qw & \qw      & \control{} & \qw    & \ctrl{1}   & \gate{H} & \meter{} \\
\lstick{$d_2$}                & \qw & \qw      & \qw      & \qw      & \control{} & \qw      & \qw \\
\lstick{$d_3$}                & \qw & \qw      & \qw      & \qw      & \ctrl{1}   & \qw      & \qw \\
\lstick{$a_3$}                & \qw & \qw      & \qw      & \ctrl{1} & \control{} & \gate{H} & \meter{} \\
\lstick{$x_4$}                & \qw & \qw      & \qw      & \control{} & \gate{H} & \meter{} & \cw \\
\lstick{$x_5$}                & \qw & \qw      & \qw      & \ctrl{1} & \gate{H}   & \meter{} & \cw \\
\lstick{$a_5\ \ket{+}$}       & \qw & \qw      & \qw      & \control{} & \ctrl{1} & \gate{H} & \meter{} \\
\lstick{$d_5$}                & \qw & \qw      & \qw      & \qw      & \control{} & \qw      & \qw
\end{quantikz}}
\caption{The chain GHZ gadget for one weight-6 $Z$-check, drawn as a circuit. Wires are ordered along the chain $a_4$--$a_1$--$a_0$--$a_2$--$a_3$--$a_5$, with each bridge singlet placed between the two arms it joins and each data qubit next to its arm: pairA $=(a_0,a_1)$ and pairB $=(a_2,a_3)$ are singlets, $a_4,a_5$ are prepared locally in $\ket{+}$, and the bridges are fuseNL1 $=(x_2,x_3)$, fuse $=(x_0,x_1)$ and fuseNL2 $=(x_4,x_5)$. The drawing shows the quantum interaction pattern: fusion CZs, data CZs, and $X$-basis readouts. Measurement-dependent Pauli-frame updates and physical timing are not drawn; their dependencies are part of the scheduled circuit. The bridge outcomes determine the Pauli frame; after its consistent propagation through the circuit, the arm readouts yield the eigenvalue of $Z_{d_0}\cdots Z_{d_5}$. For an $X$-check the data wires carry an $H$ before and after the data-CZ column. Every two-qubit gate acts between two electrons at the same node (Sec.~\ref{sec:chain}).}
\label{fig:chaincircuit}
\end{figure*}

Two singlets, $(a_0,a_1)$ and $(a_2,a_3)$, provide four ancilla
arms, while $a_4$ and $a_5$ are prepared in $\ket{+}$.
Three bridge-mediated fusions join these components along
the chain
\[
    a_4-a_1-a_0-a_2-a_3-a_5.
\]
The bridge outcomes specify the Pauli frame of the resulting
six-arm GHZ resource. Each arm $a_i$ then applies a CZ to its
data qubit $d_i$ and is measured in $X$. The parity of the
six arm outcomes, interpreted with the propagated frame,
gives the stabilizer result
(Fig.~\ref{fig:chaincircuit}).

In the bulk assignment, $a_1,a_2,a_4,a_5$ are located at
their data sites. Arms $a_0$ and $a_3$ begin at the check
centre and move one lane to their data sites after fusion.
The two outer bridges supply the long-range connections.
Their residual transport errors remain part of the
complete gadget noise model.

Before verification or purification, a bulk check consumes
five delivered singlets, twelve CZ gates and twelve readouts:
six CZs and six readouts for fusion, followed by six CZs
and six readouts for data coupling and arm measurement.
Purifying the two routed bridges adds two delivered raw
singlets, four CZs and four readouts. Delivered-singlet
counts exclude the parallel source candidates used in
source verification. Boundary checks are accounted for
using their actual connections and routes.

The central bridge $(x_0,x_1)$ measures the joint parity
$Z_{a_0}Z_{a_2}$, joining the two components of the GHZ
resource. A readout error on either bridge spin reverses
the inferred parity. In our Pauli-frame convention, this
gives an incorrect correction on the three arms
$a_2,a_3,a_5$, which form the component on one side of
the central fusion. The resulting frame error is
$X_{a_2}X_{a_3}X_{a_5}$. When these arms subsequently
couple to their data qubits through CZ gates, each arm
$X$ propagates a $Z$ onto its associated data qubit,
producing the correlated error
\begin{equation}
    Z_{d_2}Z_{d_3}Z_{d_5}.
\end{equation}
This error does not change the measured outcome of
the stabilizer being extracted, so that outcome alone
cannot reveal the fault. Other checks or later syndrome
rounds may detect its consequences, but an additional
measurement can identify the inconsistency directly.

Following the principle of verifying a cat state before
coupling it to data \cite{Shor_1996, DiVincenzo_1996},
we add a local parity measurement wherever $a_0$ and
$a_3$ occupy the same check centre. A resident verifier
$v$ is prepared in $\ket{+}$, coupled to both arms by
$\mathrm{CZ}(a_0,v)$ and $\mathrm{CZ}(a_3,v)$, and
measured in the $X$ basis. Its outcome measures
$Z_{a_0}Z_{a_3}$ and provides a consistency check on
the central fusion.

The initial singlet correlation between $a_2$ and
$a_3$, together with the central-fusion parity, fixes
the expected relation between the verifier outcome
and the two bridge outcomes. Writing all measurement
outcomes as bits $m\in\{0,1\}$, this relation is
\begin{equation}
    D_v=m_v\oplus m_{x_0}\oplus m_{x_1}=0
\end{equation}
in the absence of faults. We therefore include $D_v$
as a detector: $D_v=1$ signals that the three outcomes
are inconsistent with their expected relation
(Fig.~\ref{fig:verification} in
Appendix~\ref{app:figures}). It identifies an
inconsistency, rather than uniquely identifying
the fault that caused it.

An isolated central-bridge readout error changes one
bridge outcome and therefore triggers this detector.
An $X$ error on $a_3$ before fusion also triggers it.
However, an $X$ error on $a_0$ before fusion flips both
$m_v$ and $m_{x_0}$, leaving $D_v$ unchanged. The
verification therefore detects some, but not all,
faults in GHZ preparation. We target the central
bridge because an incorrect central parity can
produce a three-qubit data error, whereas an
incorrect outer-bridge parity produces a
single-qubit data error in the chosen frame convention.

Verification requires one resident dot, two CZ gates
and one readout. The verifier is reset and reused,
so no additional singlet delivery is required.
These operations fit within the validated resource
calendar without extending the syndrome round:
the purified GHZ gadget takes $14.1~\mu\mathrm{s}$
for the Gross code with or without verification
under the parameters of Table~\ref{tab:params}. The verification is
available for $110$ of the $144$ Gross-code checks,
where the two required arms share a check centre;
the remaining checks lack this co-location.
Its effect on logical performance is evaluated
in Sec.~\ref{sec:results_verification}.

\begin{table*}[t]
  \centering
  \caption{Hardware resources and per-round workload of the singlet-fabric
  gadgets. ``Tp'' is the teleported-ancilla gadget and ``GHZ'' the verified
  chain GHZ gadget; ``pur.'' denotes phase purification of every link of two
  or more hops. The GHZ gadget without purification was scheduled only for
  the Gross code. Per-check counts are for a bulk check; boundary checks
  need extra teleportations or routed links. GHZ counts include the
  verification circuit (two CZs and one readout); each purified link adds one
  raw singlet, and one CZ and one readout at each of its two ends. Every readout
  is a single-spin readout that also uses one CZ with the double dot of a
  readout channel (Sec.~\ref{sec:lattice}); these readout CZs are not included
  in the CZ row. A dash marks a configuration that was not scheduled.}
  \label{tab:resources}
  \footnotesize
  \setlength{\tabcolsep}{4pt}
  \begin{tabular}{lcccccccc}
    \toprule
    & \multicolumn{4}{c}{$[[72,12,6]]$} & \multicolumn{4}{c}{$[[144,12,12]]$ (Gross)} \\
    \cmidrule(lr){2-5}\cmidrule(lr){6-9}
    & Tp & Tp, pur. & GHZ & GHZ, pur. & Tp & Tp, pur. & GHZ & GHZ, pur. \\
    \midrule
    \multicolumn{9}{l}{\emph{Device}} \\
    Nodes (data + check) & \multicolumn{4}{c}{$144$ ($72+72$)} & \multicolumn{4}{c}{$288$ ($144+144$)} \\
    Lanes ($5~\mu$m), one generator bank each & \multicolumn{4}{c}{$264$} & \multicolumn{4}{c}{$540$} \\
    Static pool dots, data site / check centre & \multicolumn{8}{c}{$8$ / $6$, plus a verifier dot (GHZ) and a double dot per readout channel} \\
    Exchange zones and readout channels per node & \multicolumn{8}{c}{$2$ and $2$} \\
    Purification modules per node & $0$ & $1$ & -- & $1$ & $0$ & $1$ & $0$ & $1$ \\
    \midrule
    \multicolumn{9}{l}{\emph{Per bulk check}} \\
    Singlets delivered & $2$ & $4$ & -- & $7$ & $2$ & $4$ & $5$ & $7$ \\
    Two-qubit gates (CZ) & $8$ & $12$ & -- & $18$ & $8$ & $12$ & $14$ & $18$ \\
    Readouts & $5$ & $9$ & -- & $17$ & $5$ & $9$ & $13$ & $17$ \\
    \midrule
    \multicolumn{9}{l}{\emph{Per syndrome round}} \\
    Singlets per round & $168$ & $336$ & -- & $528$ & $324$ & $648$ & $720$ & $1044$ \\
    Local lane transits & $456$ & $456$ & -- & $122$ & $936$ & $936$ & $254$ & $254$ \\
    II, achieved / lower bound ($\mu$s) & $1.1/0.98$ & $3.0/1.35$ & -- & $2.33/1.35$ & $1.1/0.98$ & $3.0/1.35$ & $1.5/1.35$ & $2.35/2.0$ \\
    Round duration $6\,\mathrm{II}$ ($\mu$s) & $6.6$ & $18.0$ & -- & $13.95$ & $6.6$ & $18.0$ & $9.0$ & $14.1$ \\
    \midrule
    \multicolumn{9}{l}{\emph{Peak simultaneous use at any node or lane}} \\
    Pool dots occupied & $5$ & $6$ & -- & $7$ & $6$ & $6$ & $8$ & $8$ \\
    Readout channels & $2$ & $2$ & -- & $2$ & $2$ & $2$ & $1$ & $2$ \\
    Purifiers, generator banks, lanes & \multicolumn{8}{c}{at most $1$ each} \\
    \bottomrule
  \end{tabular}
\end{table*}

\subsubsection{Teleported-ancilla gadget}
\label{sec:teleport}

The teleport gadget accumulates the stabilizer parity
in a sequential ancilla state, reducing the measurement
overhead of the GHZ construction. For a bulk check,
an ancilla prepared in $\ket{+}$ visits the four local
data sites and applies a CZ at each. The two non-local
sites are visited last, using a distributed singlet
to teleport the ancilla state to each destination
(Fig.~\ref{fig:teleport}).

Each teleportation Bell-measures the current ancilla
with the near singlet half. The far half becomes the
next ancilla, with a Pauli frame determined by the
two readout bits and the singlet convention. It
interacts with the destination data qubit before the
next teleportation or final $X$ readout. The final
syndrome interpretation includes both teleportation
records.

A bulk check consumes two delivered singlets, eight
CZ gates and five readouts: six data CZs, one CZ and
two readouts per Bell measurement, and one final
ancilla readout. Source verification and optional
purification are additional. Local walking requires
no singlets or intermediate readouts.

The error propagation differs from that of the GHZ
frame. After accounting for the teleportation frame,
a residual phase error on a delivered pair can be
represented as $Z$ on the received ancilla. It
commutes with subsequent data CZs and flips the
final $X$ readout. A residual ancilla $X$, in contrast,
propagates $Z$ errors to subsequently visited data
qubits. Placing the non-local interactions last
limits this support to two data interactions for
an error introduced at the first teleportation
of a bulk check. This bound does not apply to
earlier ancilla faults or boundary checks requiring
additional teleportations.

Teleportation therefore retains order-dependent
hook errors. Its potential benefit comes from
distributing and checking the long-range resource
before transferring the accumulated ancilla state
onto it, while requiring fewer operations than
the GHZ construction. Residual resource faults
and their herald records are included in the
simulation.

\begin{figure*}[t]
\centering
\begin{tikzpicture}
\node[anchor=north west] at (0,0) {\begin{quantikz}[row sep=0.4cm, column sep=0.32cm]
\lstick{ancilla $a$} & \qw & \ctrl{1} & \qw & \gate{H} & \meter{} & \cw \rstick{$m_x$} \\
\lstick{near half $e_1$} & \gate{H} & \control{} & \gate{H} & \qw & \meter{} & \cw \rstick{$m_z$} \\
\lstick{far half $e_2$} & \qw & \qw & \qw & \qw & \qw & \qw \rstick{new ancilla; known Pauli frame}
\end{quantikz}};
\end{tikzpicture}

\vspace{4pt}
\parbox{0.94\textwidth}{\small (a) One teleportation step. The pair $(e_1,e_2)$ is a singlet. A CNOT from $a$ to $e_1$ is implemented as $H(e_1)$--CZ--$H(e_1)$, followed by $H(a)$ and two computational-basis readouts. The receiver's Pauli frame depends on both bits and includes the fixed offset associated with the singlet resource.}

\vspace{10pt}
\begin{tikzpicture}[every node/.style={font=\footnotesize}]
  \def\s{0.82}
  \foreach \x in {0,...,12}{\foreach \y in {0,...,5}{
    \ifnum\x<12 \draw[gray!55, line width=1.6pt] (\x*\s,\y*\s) -- ({(\x+1)*\s},\y*\s);
      \fill[brown!70!black] ({(\x+0.5)*\s},\y*\s) circle (0.055); \fi
    \ifnum\y<5 \draw[gray!55, line width=1.6pt] (\x*\s,\y*\s) -- (\x*\s,{(\y+1)*\s});
      \fill[brown!70!black] (\x*\s,{(\y+0.5)*\s}) circle (0.055); \fi }}
  \foreach \x in {0,...,12}{\foreach \y in {0,...,5}{
    \pgfmathtruncatemacro{\ps}{mod(\x+\y,2)}
    \pgfmathtruncatemacro{\xpar}{mod(\x,2)}
    \ifnum\ps=0
      \ifnum\xpar=0 \fill[blue!35] (\x*\s,\y*\s) circle (0.13);
      \else \fill[red!45] (\x*\s,\y*\s) circle (0.13); \fi
    \else
      \ifnum\xpar=1 \fill[green!40!black!45] (\x*\s,\y*\s) circle (0.13);
      \else \fill[violet!45] (\x*\s,\y*\s) circle (0.13); \fi
    \fi }}
  \coordinate (C)  at (2*\s,2*\s);
  \coordinate (D0) at (1*\s,2*\s); \coordinate (D1) at (3*\s,2*\s);
  \coordinate (D2) at (2*\s,3*\s); \coordinate (D3) at (2*\s,1*\s);
  \coordinate (D4) at (9*\s,4*\s); \coordinate (D5) at (11*\s,1*\s);
  \draw[orange, line width=2.4pt, opacity=0.9]
    (D1) -- (4*\s,2*\s) -- (4*\s,3*\s) -- (5*\s,3*\s) -- (6*\s,3*\s) -- (6*\s,4*\s) -- (7*\s,4*\s) -- (8*\s,4*\s) -- (D4);
  \node[draw, fill=orange!25, inner sep=1.2pt] at (5.5*\s,3*\s) {\tiny gen};
  \draw[violet, line width=2.4pt, opacity=0.9]
    (D4) -- (10*\s,4*\s) -- (10*\s,3*\s) -- (10*\s,2*\s) -- (11*\s,2*\s) -- (D5);
  \node[draw, fill=violet!20, inner sep=1.2pt] at (10*\s,2.5*\s) {\tiny gen};
  \draw[-{Stealth[length=5pt]}, teal!65!black, line width=1.6pt] (C) -- (D2);
  \draw[-{Stealth[length=5pt]}, teal!65!black, line width=1.6pt] (D2) -- (2*\s,2*\s) -- (D0);
  \draw[-{Stealth[length=5pt]}, teal!65!black, line width=1.6pt] (D0) -- (2*\s,2*\s) -- (D3);
  \draw[-{Stealth[length=5pt]}, teal!65!black, line width=1.6pt] (D3) -- (2*\s,2*\s) -- (D1);
  \node[draw, circle, fill=blue!60, minimum size=0.42cm, inner sep=0] at (C) {\scriptsize C};
  \foreach \p/\l in {D0/d_0, D1/d_1, D2/d_2, D3/d_3, D4/d_4, D5/d_5}{
    \node[draw, circle, fill=green!35, minimum size=0.42cm, inner sep=0] at (\p) {\scriptsize $\l$};}
  \node[draw, fill=white, inner sep=1.2pt, anchor=north] at ($(D1)+(0,-0.25)$) {\tiny BSM 1};
  \node[draw, fill=white, inner sep=1.2pt, anchor=south] at ($(D4)+(0,0.25)$) {\tiny BSM 2};
  \node[anchor=north, inner sep=1pt] at ($(D5)+(0,-0.25)$) {\scriptsize $X$ readout};
  \node[teal!65!black, anchor=east] at ($(D2)+(-0.2,0.15)$) {\scriptsize 1};
  \node[teal!65!black, anchor=south] at ($(D0)+(0,0.22)$) {\scriptsize 2};
  \node[teal!65!black, anchor=east] at ($(D3)+(-0.2,-0.1)$) {\scriptsize 3};
  \node[teal!65!black, anchor=north] at ($(D1)+(0.05,-0.6)$) {\scriptsize 4};
  \begin{scope}[shift={(0,-0.95)}]
    \draw[-{Stealth[length=4pt]}, teal!65!black, line width=1.6pt] (0,0) -- (0.6,0);
    \node[anchor=west] at (0.7,0) {local ancilla walk};
    \draw[orange, line width=2.4pt] (5.2,0) -- (5.8,0);
    \node[anchor=west] at (5.9,0) {singlet 1: $d_1\to d_4$};
    \draw[violet, line width=2.4pt] (0,-0.55) -- (0.6,-0.55);
    \node[anchor=west] at (0.7,-0.55) {singlet 2: $d_4\to d_5$};
    \node[draw, fill=white, inner sep=1.2pt] at (5.5,-0.55) {\tiny gen};
    \node[anchor=west] at (5.9,-0.55) {source bank};
  \end{scope}
\end{tikzpicture}

\vspace{2pt}
\parbox{0.9\textwidth}{\footnotesize (b) The gadget for one weight-6 check drawn on the lattice. The ancilla is prepared at the centre $C$ and walks the four local legs $d_2\!\to\!d_0\!\to\!d_3\!\to\!d_1$ (teal, two lanes per step through the centre), applying a CZ at each. Meanwhile two singlets have been generated on lanes near the middle of their routes and their halves conveyed outward (orange, violet), parking at intermediate nodes. At $d_1$ the ancilla is Bell-measured with the near half of singlet 1 and its state reappears on the far half, already parked at $d_4$; that electron applies the CZ to $d_4$, is Bell-measured with the near half of singlet 2, and the state reappears at $d_5$, where the final CZ and the single $X$ readout take place. For this bulk example: 2 delivered singlets, 8 CZs, and 5 readouts, before source verification and purification.}
\caption{The teleported-ancilla gadget. Panel (b) is a schematic route example rather than a coordinate-level drawing of one selected BB embedding.}
\label{fig:teleport}
\end{figure*}

\subsection{Scheduling}
\label{sec:scheduling}

Scheduling decides when each singlet is generated, which lanes its halves use, and when it is
consumed. Different checks share lanes, generators and node resources, so these choices must be
coordinated, and the resulting timetable also supplies every duration used by the noise model.
We solve the problem in two stages; the details are given in Appendix~\ref{app:scheduling}.

\emph{Routing.} For each long-range link we consider up to $24$ shortest paths, with the generator
placed so that both halves travel about equally far, and assign links one at a time, longest
first, choosing the path that least increases the load on generators, relay nodes and lanes.
Routes are fixed once and reused every round.

\emph{Calendar.} The checks are split into six subcycles, three for each check type, such that
checks in the same subcycle share no data qubit. Subcycles start every \emph{initiation interval}
$\mathrm{II}$ and overlap: the singlets for subcycle $s+1$ are generated and transported while the
gates of subcycle $s$ run. Following software pipelining \cite{Lam_1988, Rau_1994}, every
operation reserves its resources on a periodic calendar, and a reservation is accepted only if
lane, generator, storage, exchange and readout capacities hold at every time step. Gate
operations are placed first; each singlet then receives a deadline, and its purification,
transport and generation are placed as late as possible before it, which keeps waiting short. We
search for the smallest $\mathrm{II}$ for which a complete calendar is found and validate the
result by reconstructing the occupancy of every resource. The round lasts
\begin{align}
    t_{\mathrm{cyc}} = 6\,\mathrm{II}.
\end{align}
The reported $\mathrm{II}$ is a validated feasible schedule, not a proof of optimality. The
calendar records every transit, park, wait and idle interval, and these durations set the
transport and dephasing errors of the circuit (Sec.~\ref{sec:stim}).

Table~\ref{tab:resources} collects the hardware and the per-round
workload of each configuration for the two codes we simulate. The upper
block is the physical device, which is the same for every gadget except
that the purification module is present only when purification is used.
The middle block counts what one syndrome round consumes. ``Singlets per
round'' counts every singlet that is generated and delivered in one round,
including the extra raw pairs consumed by purification; ``local lane
transits'' counts the single-lane moves of ancillas and GHZ arms between a
check centre and its neighbouring data sites, which need no singlet. The
lower block gives the round timing from the scheduler, together with the
lower bound on the initiation interval computed from resource workloads
alone (Sec.~\ref{sec:scheduling}). An achieved interval close to the
bound leaves little room for improvement under the same workload model.
A larger gap can reflect resource contention, operation dependencies or
limitations of the greedy placement procedure; the gap alone does not
identify which of these causes is responsible. The last rows give
the largest number of resources of each kind that the schedule ever uses
at once at any single node or lane, which confirms that the provisioned
inventory is sufficient.

\subsection{Stim circuits}
\label{sec:stim}

The schedule describes electrons and hardware resources, whereas Stim
simulates a quantum circuit. We connect the two descriptions by translating
scheduled gates and measurements into circuit instructions, and transport
and waiting into the corresponding error channels. Stim does not move an
electron through a geometric device model. Instead, the route determines
which transport errors are inserted, and the recorded waiting times
determine the storage and idle dephasing.

The circuit explicitly contains the data qubits and the ancillas needed
for the extraction operations represented at gate level. Two procedures are
instead replaced by an exact effective description: source generation and
verification (Appendix~\ref{app:source}) and link purification. For each, we
compute the joint distribution of the Bell-frame Pauli error on the delivered
pair and the flag outcomes. For purification this is done by propagating the
two input pairs through the noisy circuit of Fig.~\ref{fig:purification} as a
density matrix, including the noise of every gate and readout in that circuit.
Each link is then inserted into the Stim circuit as a set of mutually exclusive
error mechanisms over its Pauli error and flag detectors, so the internal
preparation and purification qubits need not appear as circuit qubits.
For a purified link, the transport, relay and waiting errors that each raw
pair accumulates before purification are included in the two input pairs of
this calculation, and the waiting of the purified pair until it is consumed is
added to its output. For an unpurified link, the transport and waiting errors
are combined in the same way with the source-test outcome. Everything after
delivery (fusion, bridges, CZ gates to the data, teleportation and readouts)
is simulated gate by gate. 

For each independent trial, Stim~\cite{Gidney_2021} samples the noisy
memory circuit and returns detection events and flips of the tracked
logical observables. A detection event means that a combination of
measurement outcomes disagrees with its expected fault-free parity.
For example, a change between successive measurements of the same check
can produce a detection event. The decoder sees these events, together
with the herald information used in our model, and predicts which logical
observables have flipped. A detector error model extracted from the
circuit describes how the modelled faults connect detection events to
logical flips.

For the BB codes, we use belief propagation followed by
ordered-statistics decoding (BP+OSD)~\cite{Panteleev_2021, Roffe_2020}.
Belief propagation estimates likely faults from the measured syndrome;
the ordered-statistics stage attempts to find a syndrome-consistent
solution when further processing is needed. This approach accommodates
the more general detector connections of these circuits without requiring
every fault to have the pairwise structure used by ordinary matching.
We compare the decoder's predicted logical flips with those sampled by
Stim. A trial is a logical block failure if any tracked logical observable
is predicted incorrectly. Repeating the experiment estimates the failure
probability of the complete memory protocol, rather than merely the
probability that an individual physical gate fails.

\section{Results}
\label{sec:Results}

We ask how best to measure bivariate-bicycle (BB) stabilizers on a
silicon node network. We compare direct transport of a mobile ancilla,
using CAbLECAR~\cite{chadwick2026cablecarefficientlyschedulingqldpc}, with
two ways of distributing entanglement before coupling to the data: the
chain GHZ gadget of Sec.~\ref{sec:chain} and the teleported-ancilla gadget
of Sec.~\ref{sec:teleport}. We call these the GHZ and teleport gadgets.
Distributed singlets can limit how transport faults propagate to the data,
but preparing and consuming them requires additional gates and readouts.
The comparison tests whether that protection compensates for those extra
opportunities for error.

The original CAbLECAR implementation and our gadgets also place the data qubits differently.
We therefore compare with CAbLECAR both on its own layout and on our node lattice, and then test
whether the order in which a mobile ancilla visits its data qubits matters separately from its
route length. Noise-removal tests identify which faults limit each scheme; the following
subsections test better gates, purification, source heralding and verification, and a surface-code reference
(Appendix~\ref{app:surface}) examines the same trade-off without long-range checks.

\subsection{Simulation setup}
\label{sec:results_setup}

All BB-code schemes use the hardware parameters in Table~\ref{tab:params}.
The gate, preparation, and measurement error probabilities are those of
Table~II of Ref.~\cite{chadwick2026cablecarefficientlyschedulingqldpc}; the
one-qubit gate error is $p_1=p_2/10=10^{-4}$, as in the CAbLECAR simulation code.
We use an echo coherence time $T_2^{\mathrm{echo}}=10$~ms and a shuttle
duration of $50$~ns per $10~\mu$m for every scheme. The CAbLECAR circuits are
generated with the authors' code, modified only to use this shuttle
duration, to include the optional shuttle bit-flip channel described below,
to use the same idle-noise model as our circuits, and, for the layout control of
Sec.~\ref{sec:results_layout}, to accept our data-qubit positions. In every circuit, idle noise
is applied to every waiting interval of every qubit, using the Pauli-twirled $T_1$,
$T_2^{\mathrm{echo}}$ channel of Table~\ref{tab:params}, whose dephasing probability is
$(1-e^{-t/T_2^{\mathrm{echo}}})/2$ for an interval $t$.

The two implementations share every operation that has a counterpart in both. Two-qubit gates
(CX in the CAbLECAR circuits, CZ in ours) take $0.1~\mu$s and are followed by two-qubit
depolarizing noise of strength $p_2$; Hadamard gates take $0.1~\mu$s with one-qubit
depolarizing noise $p_1$; preparation and reset take $0.5~\mu$s with a bit flip of probability
$p_\mathrm{prep}$; a PSB measurement takes $0.5~\mu$s with a bit flip of probability $p_\mathrm{meas}$;
shuttling takes $50$~ns per $10~\mu$m with the channel described below; and every waiting
interval receives the same idle channel. Four differences remain. First, the CAbLECAR circuits
use CX as the native two-qubit gate, whereas our gadgets use CZ. Second, the approximately
$1~\mu$m move of a CAbLECAR ancilla between a cell and its gate site is assigned $0.1~\mu$s and
a dephasing probability $p_\mathrm{sh}/10$, whereas a park-and-relaunch event at one of our
nodes is assigned $0.1~\mu$s and the dephasing of $1~\mu$m of transport. Third, the CAbLECAR
scheduler applies its own resource constraints and does not impose the per-node limits of
Sec.~\ref{sec:lattice} (two concurrent exchange operations, two readout channels). These
constraints can only lengthen our rounds, so this difference does not favour the singlet gadgets.
Fourth, every single-spin readout in our gadgets includes the CZ with a singlet-prepared PSB pair
described in Sec.~\ref{sec:lattice}, whereas the CAbLECAR circuits measure their ancillas directly,
as in the original simulations. We leave the CAbLECAR circuits unchanged: adding the same CZ to
them would only raise their error rates and lengthen their rounds, so this difference does not
favour the singlet gadgets either.

Each readout CZ carries the same depolarizing noise $p_2$ as every other CZ. Because the
read spin is discarded, this noise acts only through the outcome: a $Z$ on the paired spin always
flips it, and an $X$ or $Y$ flips it only when the pair should have stayed in $\ket{S}$, giving an
average flip probability $8p_2/15$. We therefore simulate each single-spin readout as a
$0.6~\mu$s measurement with flip probability $p_\mathrm{meas}\oplus8p_2/15\approx1.5\times10^{-3}$,
where $a\oplus b=a+b-2ab$ combines two independent flips, assuming that the double dot is
re-prepared in $\ket{S}$ at no additional cost (Appendix~\ref{app:readout}).

A common shuttle speed keeps the comparison about the circuits rather
than the hardware; because idle noise is small at these coherence times
(Sec.~\ref{sec:results_rounds}), the speed affects the round duration much more than the
logical error rate. All schemes, including CAbLECAR, are decoded with the same decoder,
described below; on CAbLECAR's own Gross-code circuits it reproduces the logical error rate of
the authors' decoder (BP+OSD-CS of order $10$) at about half the computational cost.

\begin{table}[!htbp]
  \centering
  \caption{Hardware parameters used for all architectures. Two-qubit gates are
  followed by two-qubit depolarizing noise, each of the $15$ non-identity two-qubit
  Pauli operators occurring with probability $p_2/15$; one-qubit gates are followed by
  one-qubit depolarizing noise of strength $p_1$. Preparation, reset and measurement
  errors are flips in the relevant basis. Readout (CZ+PSB) is the single-spin readout of our gadgets, a CZ with a
  PSB pair followed by a PSB measurement; $a\oplus b=a+b-2ab$ (Appendix~\ref{app:readout}).
  CAbLECAR circuits measure their ancillas directly. Shuttling applies independent $Z$ and $X$
  errors (Sec.~\ref{sec:results_setup}). An idle interval of duration $t$ receives the
  Pauli twirl of amplitude and phase damping, $p_X=p_Y=(1-e^{-t/T_1})/4$ and
  $p_Z=(1+e^{-t/T_1}-2e^{-t/T_2^{\mathrm{echo}}})/4$.}
  \label{tab:params}
  \begin{tabular}{lcc}
    \toprule
    Operation & Error probability & Duration \\
    \midrule
    One-qubit gate & $p_1=10^{-4}$ & $0.1~\mu$s \\
    Two-qubit gate (CZ) & $p_2=10^{-3}$ & $0.1~\mu$s \\
    Preparation / reset & $p_\mathrm{prep}=10^{-3}$ & $0.5~\mu$s \\
    Measurement (PSB) & $p_\mathrm{meas}=10^{-3}$ & $0.5~\mu$s \\
    Readout (CZ+PSB) & $p_\mathrm{meas}\oplus8p_2/15$ & $0.6~\mu$s \\
    Shuttle, per $10~\mu$m & $p_\mathrm{sh}$ ($Z$), $p_x$ ($X$) & $0.05~\mu$s \\
    Idle & $T_1 = 10$~s, $T_2^{\mathrm{echo}} = 10$~ms & -- \\
    \bottomrule
  \end{tabular}
\end{table}

The parameter varied throughout is the shuttle dephasing probability
$p_\mathrm{sh}$ per $10~\mu$m of transport, over the range
$10^{-4}\leq p_\mathrm{sh}\leq 10^{-2}$. Dephasing is an important error
mechanism reported for conveyor-mode shuttling \cite{Struck_2024, DeSmet_2025}, but spin
flips during transport cannot be excluded, and the two kinds of error propagate very
differently through a syndrome-extraction circuit. We therefore consider
three noise models: dephasing alone, and dephasing accompanied by a bit flip
per $10~\mu$m of transport with probability $p_x=p_\mathrm{sh}/10$ or
$p_x=p_\mathrm{sh}/3$. Comparing the three models shows how much of each
scheme's behaviour is caused by phase errors and how much by bit flips. The $X$ and $Z$
errors are independent, in our circuits and in the CAbLECAR circuits alike, so both occur,
giving a $Y$ error, with probability $p_\mathrm{sh}p_x$; over a route, the per-segment
probabilities are compounded, so $h$ segments with probability $p$ give
$\tfrac12[1-(1-2p)^{h}]$.

We simulate the $[[72,12,6]]$ and $[[144,12,12]]$ (Gross) codes using
Stim~\cite{Gidney_2021}. Each independent trial, or shot, is a memory
experiment of $d$ syndrome-extraction rounds: six rounds for the
$[[72,12,6]]$ code and twelve for the Gross code. Separate experiments
prepare and read out the logical information in the $X$ and $Z$ bases,
because a memory experiment in one basis is sensitive only to errors that
flip logical operators of that basis. The full syndrome history is decoded
by belief propagation with a serial schedule and $200$ iterations, followed
by ordered-statistics decoding (OSD-CS, order $4$). The same decoder
settings are used for every BB-code scheme. This holds the decoding
procedure fixed, although its effectiveness can still depend on the error
correlations produced by each circuit. For the singlet-based
gadgets, the source and purification flags recorded in each shot also
update that shot's error priors before decoding, in the same way that erasure~\cite{Wu_2022}
or flag~\cite{Chamberland_2020} information is used to reweight a decoder shot by shot. For each link, the effective description of Sec.~\ref{sec:stim} gives the
probability of every combination of delivered Pauli error and flag outcomes, computed exactly
over all fault combinations inside the source test and the purifier, including missed
detections, multiple faults, and faults whose effects on the flags cancel; no truncation is
applied. In the Stim circuit each link is a set of mutually exclusive error mechanisms, each
acting on the delivered pair and its flag detectors, and different links are independent. When
the circuit is converted to a decoding problem, each mechanism becomes a fault column with its
own prior. For each shot, every column that would raise a flag is first given prior zero. For
each link whose flags show a pattern $F$, the columns whose flag pattern is exactly $F$ are then
rescaled to their conditional distribution given $F$; if no single column produces $F$, as when
two independent faults each raise one flag, the columns whose patterns are contained in $F$ are
rescaled together. The flag detectors are then removed. Three approximations are involved: the
candidate errors of a flagged link are treated by belief propagation as independent rather than
mutually exclusive; a link's errors that raise no flag merge with other faults of the same
syndrome signature and keep their unconditioned prior; and patterns produced only by several
faults are handled by the containment rule above. To test
their combined effect, we decoded identical shots with the flag detectors kept in the decoding
problem, so that belief propagation sees the flag correlations directly; the logical error rate
was the same within statistical error, at about thirty times the cost. The decoder
is assumed to know the source model, as it is assumed to know every other error rate.

For each memory basis $B\in\{X,Z\}$, let $P_{\mathrm{shot}}^B$ be the
probability that at least one of the twelve logical qubits fails during the
complete $d$-round experiment. We convert this block-failure probability to
an effective per-round rate and add the two bases,
\begin{equation}
  P_L^B=1-\left(1-P_{\mathrm{shot}}^B\right)^{1/d},
  \qquad P_L=P_L^X+P_L^Z .
\end{equation}
Unless otherwise stated, the figures show this quantity: an effective rate per round for the
whole twelve-qubit block, used to compare experiments of different length. The sum over bases
estimates the rate of either kind of logical error; it is not a measured joint probability.

Each BB-code point uses between about $10^4$ and $2\times10^7$ shots, chosen to
observe roughly $100$ to $300$ failures where this is affordable. We compute $95\%$
Clopper--Pearson intervals~\cite{Clopper_1934} for each shot-failure probability, transform
their endpoints to per-round rates, and add the endpoints of the two bases
to obtain the displayed interval on $P_L$. When we quote a ratio of two rates, the range in
brackets is a conservative $95\%$ range obtained by dividing the lower endpoint of one interval
by the upper endpoint of the other, and vice versa. Open downward triangles are $95\%$ upper bounds where no failures were observed; open upward
triangles mark saturated points, where nearly every shot failed and which we do not use to
quantify differences.

\subsection{Round duration}
\label{sec:results_rounds}

The resource calendar of Sec.~\ref{sec:scheduling} gives the round durations
in Table~\ref{tab:rounds}. They differ substantially between schemes: the
unpurified teleport gadget completes a Gross-code round in $6.6~\mu$s,
whereas CAbLECAR needs $11.1~\mu$s on its own layout and $9.1~\mu$s on our
node lattice, where its routes are shorter. The readout CZ lengthens every single-spin readout of
our gadgets from $0.5$ to $0.6~\mu$s; this lengthens the teleport rounds, whose readouts lie on the
critical path, but not the GHZ rounds, which are limited by singlet generation.
Purification lengthens the teleport round to $18.0~\mu$s. The
reason is congestion rather than the doubled number of raw singlets: each
node has a single purification module, so two purified links that terminate
at the same node and are needed in the same subcycle must be purified one
after the other, and the check that uses them waits until both are ready.

The assumed coherence time is long compared with the round duration:
with $T_2^{\mathrm{echo}}=10$~ms, a $10~\mu$s idle interval contributes a
dephasing probability of about $5\times10^{-4}$. Removing idle noise
lowers the teleport logical error rate by at most about $15\%$, the CAbLECAR rate by
$10$ to $50\%$ and the GHZ rate by up to about $35\%$
(Sec.~\ref{sec:results_budget}). Waiting is therefore not the main
limitation in these simulations, although it can still have a noticeable
effect. We report logical error per round together with the durations in
Table~\ref{tab:rounds}. Neither quantity alone establishes which scheme
best preserves a memory for a fixed elapsed time, and a shorter round
should not automatically be interpreted as a lower logical error rate.

\begin{table}[!htbp]
  \centering
  \caption{Duration of one syndrome-extraction round, in $\mu$s.
  ``Purified'' applies phase purification to every link of two or more
  hops. The verified and unverified GHZ gadgets have the same round time.
  ``CAbLECAR, our layout'' is the CAbLECAR scheduler run on the node lattice
  of Sec.~\ref{sec:lattice} (Sec.~\ref{sec:results_layout}); ``pinned''
  additionally fixes each ancilla's data-visit order to the one used on
  CAbLECAR's own layout. Our gadgets include the CZ of every single-spin readout
  (Sec.~\ref{sec:results_setup}); the CAbLECAR circuits use direct readout.}
  \label{tab:rounds}
  \begin{tabular}{lcc}
    \toprule
    Scheme & $[[72,12,6]]$ & $[[144,12,12]]$ \\
    \midrule
    CAbLECAR, own layout & $9.4$ & $11.1$ \\
    CAbLECAR, our layout & $8.3$ & $9.1$ \\
    CAbLECAR, our layout, pinned & $12.5$ & $13.8$ \\
    Teleport, unpurified & $6.6$ & $6.6$ \\
    Teleport, purified & $18.0$ & $18.0$ \\
    GHZ, unpurified & -- & $9.0$ \\
    GHZ, purified & $13.95$ & $14.1$ \\
    \midrule
    Surface code, walk & \multicolumn{2}{c}{$2.35$ (all $d$)} \\
    Surface code, 4-GHZ & \multicolumn{2}{c}{$4.0$ (all $d$)} \\
    \bottomrule
  \end{tabular}
\end{table}

\subsection{Comparison with CAbLECAR on its own layout}
\label{sec:results_comparison}

\begin{figure*}[t]
  \centering
  \includegraphics[width=0.85\linewidth]{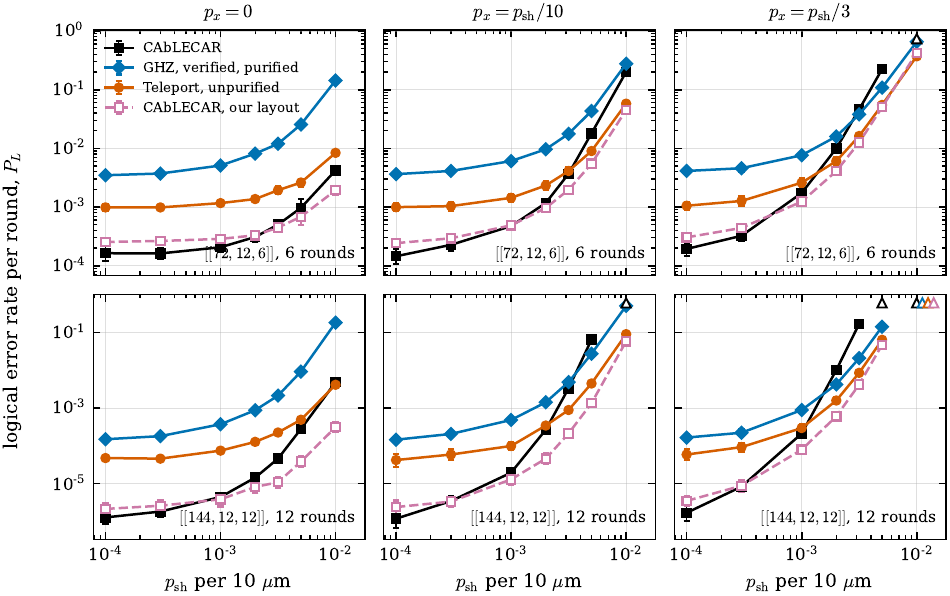}
  \caption{Logical error rate per round, $P_L=P_L^X+P_L^Z$, versus the
  shuttle dephasing probability $p_\mathrm{sh}$. Solid black squares:
  CAbLECAR on its own layout. Open purple squares with dashed lines: CAbLECAR
  on our node lattice, using the authors' scheduler and circuits with the
  data qubits placed at our data sites and one unit cell per $5~\mu$m lane.
  Blue diamonds: verified chain GHZ gadget with purification. Orange
  circles: teleport gadget without purification. Top row: $[[72,12,6]]$
  code, $6$ rounds; bottom row: Gross code, $12$ rounds. Columns: no shuttle
  bit flips, and bit flips with probability $p_\mathrm{sh}/10$ and
  $p_\mathrm{sh}/3$ per $10~\mu$m. Open upward triangles at the top edge
  mark saturated points, where more than $99\%$ of shots failed.}
  \label{fig:results_main}
\end{figure*}

\begin{figure}[t]
  \centering
  \includegraphics[width=1.0\linewidth]{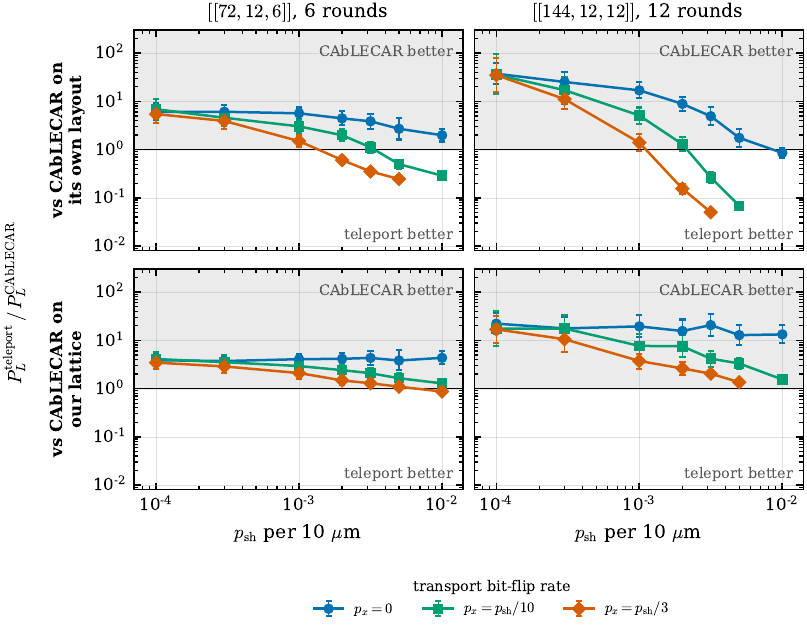}
  \caption{Ratio of the teleport-gadget and CAbLECAR logical error rates per
  round, $P_L=P_L^X+P_L^Z$. Top row: CAbLECAR on its own layout. Bottom row:
  CAbLECAR on our node lattice. Colours give the transport bit-flip rate.
  Values below one (white region) favour the teleport gadget. Points where
  neither architecture failed, or where either is saturated, are omitted.}
  \label{fig:results_ratio}
\end{figure}

Figure~\ref{fig:results_main} shows the logical error rate of each scheme as
a function of $p_\mathrm{sh}$, and Fig.~\ref{fig:results_ratio} shows the
ratio of the teleport rate to the CAbLECAR rate; a ratio below one favours
the teleport gadget. Unless stated otherwise, source heralding is enabled,
the teleport gadget is unpurified, and the GHZ gadget uses both phase
purification and central-fusion verification, which define the reference configurations used here. Purification helps
the GHZ gadget only above a noise-dependent crossover; its effect is
examined separately in Sec.~\ref{sec:results_purification}. In this subsection we
discuss the solid black curves, CAbLECAR on its own layout; the dashed purple
curves are the layout control of the next subsection.

With dephasing alone ($p_x=0$), CAbLECAR has the lower logical error rate at
almost every sampled point. For the Gross code at $p_\mathrm{sh}=10^{-4}$,
its rate is $1.3\times10^{-6}$ per round, compared with $4.7\times10^{-5}$
for the teleport gadget and $1.5\times10^{-4}$ for the GHZ gadget. At low
shuttle noise the two singlet-based gadgets reach a floor: reducing
$p_\mathrm{sh}$ further barely helps, because their logical failures are
caused by faults in other operations. The noise-removal tests of
Sec.~\ref{sec:results_budget} identify two-qubit gates as the main such
operation. Each bulk teleport check uses eight CZ gates and five readouts, each
readout with its own CZ, compared with six two-qubit gates and one direct
readout for a CAbLECAR ancilla, so the singlet gadgets start from a higher
baseline of gate and readout faults.

As $p_\mathrm{sh}$ grows, the gap narrows, and shuttle bit flips change the
picture. When a bit flip occurs on a CAbLECAR ancilla partway through its
route, the error can propagate through subsequent ancilla--data gates, producing
a correlated error on several data qubits that the code must then correct. In the singlet-based gadgets, most transport is done by singlet
halves before they interact with any data qubit, and the remaining moves
are short and local, so a bit flip during transport reaches fewer data
qubits. For the Gross code with $p_x=p_\mathrm{sh}/3$, the teleport gadget
consequently overtakes CAbLECAR near $p_\mathrm{sh}=1.1\times10^{-3}$. At
$p_\mathrm{sh}=2\times10^{-3}$ its rate is $1.6\times10^{-3}$ per round,
against $1.0\times10^{-2}$ for CAbLECAR, and at
$p_\mathrm{sh}=3.16\times10^{-3}$ the two differ by a factor of $20$ [$18$, $22$]. With
less frequent bit flips, $p_x=p_\mathrm{sh}/10$, the crossover moves to
near $p_\mathrm{sh}=2\times10^{-3}$. The
$[[72,12,6]]$ code behaves the same way, with a crossover near
$p_\mathrm{sh}=1.4\times10^{-3}$ for $p_x=p_\mathrm{sh}/3$.

Taken by itself, this comparison would suggest that the teleport gadget
wins whenever transport bit flips are frequent. That conclusion, however,
compares two different layouts as well as two different circuits, and the
next subsection shows that the layout accounts for most of the difference.

\begin{figure*}[t]
  \centering
  \includegraphics[width=0.85\linewidth]{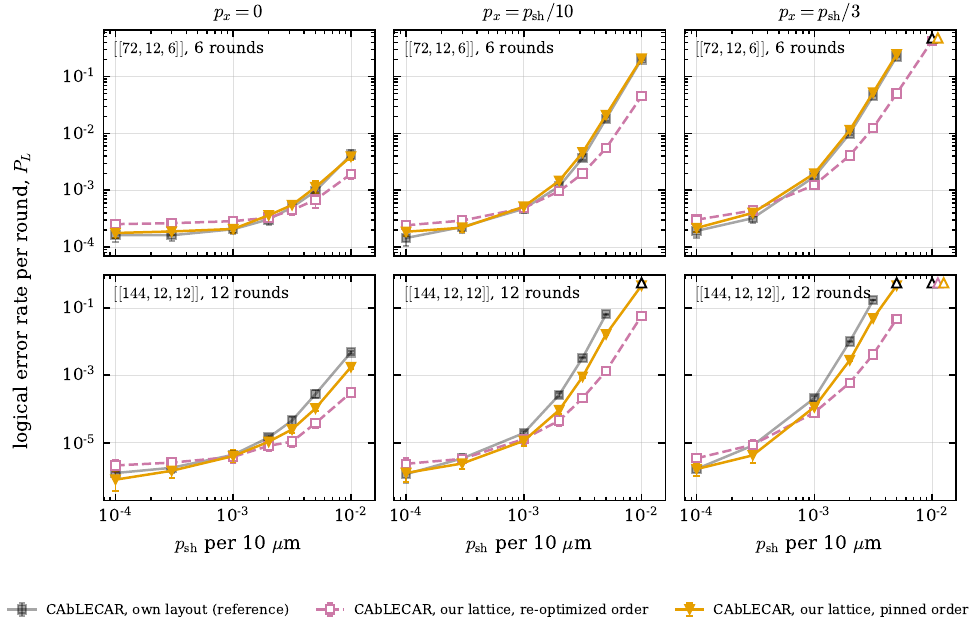}
  \caption{Visit-order test: logical error rate per round, $P_L=P_L^X+P_L^Z$,
  of CAbLECAR on our node lattice. Open purple squares with dashed lines: the
  scheduler chooses each ancilla's visit order to suit our lattice (the
  dashed curves of Fig.~\ref{fig:results_main}). Filled orange triangles with
  solid lines: each ancilla's visit order is fixed to the order the scheduler
  chooses on CAbLECAR's own layout. The faint grey curve is CAbLECAR on its
  own layout, shown for reference. Top row: $[[72,12,6]]$ code, $6$ rounds;
  bottom row: Gross code, $12$ rounds. Columns: no shuttle bit flips, and bit
  flips with probability $p_\mathrm{sh}/10$ and $p_\mathrm{sh}/3$. Open
  upward triangles at the top edge mark saturated points.}
  \label{fig:results_order}
\end{figure*}

\subsection{Layout control: CAbLECAR on our node lattice}
\label{sec:results_layout}

The architectures differ in the placement of their data qubits.
CAbLECAR uses a square grid of $10~\mu\mathrm{m}$ unit cells,
whereas our node lattice has $5~\mu\mathrm{m}$ lanes with data
sites and check centres arranged in a checkerboard pattern.
Moving between neighbouring data sites on our lattice therefore
requires two lanes, or $10~\mu\mathrm{m}$ of transport.
The shorter routes on our lattice arise from the placement of
each check's non-local data qubits, rather than from reducing
this local transport distance.

For each of the $144$ Gross-code checks, we calculate the shortest
path visiting its six data qubits along the allowed transport
directions. The mean length is $126~\mu\mathrm{m}$ on our lattice,
compared with $243~\mu\mathrm{m}$ on CAbLECAR's original grid.
These lengths characterize the geometry alone; the scheduled
routes can be longer because they must also accommodate the
other ancillas.

To compare extraction schemes on a common layout, we run the
CAbLECAR scheduler with the data-qubit positions of our node
lattice. We represent each $5~\mu\mathrm{m}$ lane as one grid step
and halve the transport time and shuttle-error probability
assigned to that step. This preserves the transport time and,
to first order in the error probabilities, the noise per unit
length. The approximately $1~\mu\mathrm{m}$ move between a cell
and its gate site retains its original noise contribution.
With this change of layout, the mean scheduled ancilla route
falls from $350~\mu\mathrm{m}$ to $153~\mu\mathrm{m}$ per round,
and the longest route falls from $570~\mu\mathrm{m}$ to
$300~\mu\mathrm{m}$.

The dashed purple curves in Fig.~\ref{fig:results_main} and the bottom row of
Fig.~\ref{fig:results_ratio} show the result. For the Gross code on our lattice, CAbLECAR has a lower
logical error rate than the teleport gadget at every simulated point and in
all three noise models. Without bit flips it is lower by a factor of $13$ to $22$ at every
point, and the $95\%$ interval of the ratio never extends below about $8$; at $p_\mathrm{sh}=10^{-2}$, where the teleport gadget had
been marginally ahead of CAbLECAR on its own layout, CAbLECAR on our lattice
is lower by a factor of $13$ [$8.8$, $21$]. With bit flips the gap narrows steadily as
$p_\mathrm{sh}$ grows, in agreement with the propagation argument of the
previous subsection, but it does not close before the rates saturate. For
$p_x=p_\mathrm{sh}/3$ the teleport rate is higher by a factor of $2.6$ [$1.9$, $3.6$] at
$p_\mathrm{sh}=2\times10^{-3}$, $2.0$ [$1.8$, $2.3$] at $3.16\times10^{-3}$ and $1.4$
[$1.3$, $1.4$] at $5\times10^{-3}$, where both schemes already fail several percent of rounds.
Relative to its own layout, CAbLECAR improves by a factor of $17$ [$15$, $19$] at
$p_\mathrm{sh}=2\times10^{-3}$ and $40$ [$39$, $42$] at $3.16\times10^{-3}$ for
$p_x=p_\mathrm{sh}/3$, and by up to $48$ [$41$, $58$] for $p_x=p_\mathrm{sh}/10$ (at
$p_\mathrm{sh}=5\times10^{-3}$). These gains accompany a reduction of more than a
factor of two in the mean scheduled transport distance.

The $[[72,12,6]]$ code (top row of Fig.~\ref{fig:results_main} and left
column of Fig.~\ref{fig:results_ratio}) tells the same story with a smaller margin.
Without bit flips, CAbLECAR on our lattice is lower than the teleport
gadget by a factor of $3.7$ to $4.4$ at every $p_\mathrm{sh}$. The margin is
smaller than for the Gross code because the $[[72,12,6]]$ code is smaller
and its checks are already compact on either lattice, so there is less
transport to save. With bit flips at $p_x=p_\mathrm{sh}/3$, the gap
narrows from a factor of $3.5$ [$2.6$, $4.7$] at $p_\mathrm{sh}=10^{-4}$ to $1.5$
[$1.2$, $1.8$] at $2\times10^{-3}$ and $1.3$ [$1.1$, $1.5$] at $3.16\times10^{-3}$. At
$p_\mathrm{sh}=5\times10^{-3}$ CAbLECAR is still marginally lower ($1.10$ [$1.01$, $1.19$]),
and at $10^{-2}$ the teleport gadget is lower by a factor of $1.15$ [$1.11$, $1.19$]. At these
points, however, both schemes fail between $5\%$ and $40\%$ of rounds, far above any useful
operating point, so this narrow reversal does not identify a practical regime for the
teleport gadget. It does confirm the trend seen for the Gross code: as
transport bit flips become more frequent, the singlet gadget's limited
propagation of those flips gradually offsets its extra gates and readouts.

\begin{figure*}[t]
  \centering
  \includegraphics[width=0.8\linewidth]{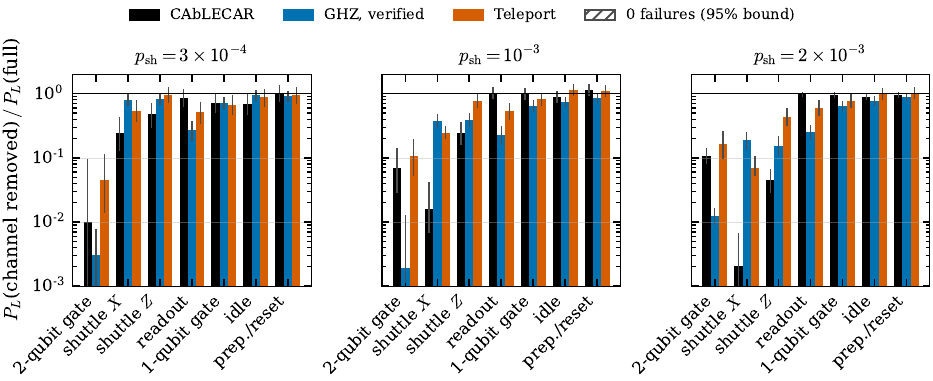}
  \caption{Logical error rate per round with one noise channel removed,
  relative to the full noise model, for the Gross code with
  $p_x = p_\mathrm{sh}/3$, $P_L = P_L^X + P_L^Z$. Black: CAbLECAR on its own
  layout; blue: GHZ gadget, verified and purified; orange: teleport gadget,
  unpurified. Whiskers are $95\%$ ranges.}
  \label{fig:results_budget}
\end{figure*}

At low shuttle noise, the ordering of the two CAbLECAR implementations
tends to reverse. For the $[[72,12,6]]$ code without shuttle bit flips, the original layout gives a
rate lower by a factor of $1.6$ [$1.0$, $2.5$] at $p_\mathrm{sh}=10^{-4}$ and $1.6$ [$1.1$,
$2.4$] at $3\times10^{-4}$. For the Gross code the central values are $1.7$ and $1.4$, with
ranges [$0.8$, $3.7$] and [$0.7$, $2.8$]. Only the $[[72,12,6]]$ point at $3\times10^{-4}$
excludes equality, but the direction is the same in both codes and at both noise levels. In this regime the rate on our lattice barely changes with
$p_\mathrm{sh}$, so it is set by gate, readout and idle noise rather than by
transport, and the shorter routes cannot help. Idle noise cannot account for
the difference either, because the round on our lattice is shorter.

A further difference between the two circuits is the order in which
each ancilla visits its data qubits. With shuttle errors switched off, the
two circuits contain the same $864$ error mechanisms, with total
probability agreeing to within $0.05\%$. However, CAbLECAR's scheduler is
free to choose each ancilla's visit order, and it chooses the order that
minimizes that ancilla's route on the lattice it is given. On our lattice it
chooses a different order for every ancilla in each of the two codes. The visit order matters
because of hook errors~\cite{Dennis_2002}. An error component that propagates through the ancilla--data gates can
reach several data qubits through the remaining interactions. A single
ancilla fault can therefore produce a correlated data error. The affected
qubits depend on both the Pauli component of the fault and when it occurs. Which data
qubits are affected, and therefore whether the correlated error lies close
to a logical operator, depends on the order of the interactions. Two
circuits with identical lists of fault locations can thus have different
numbers of low-weight failure paths.

\begin{figure*}[t]
  \centering
  \includegraphics[width=0.78\linewidth]{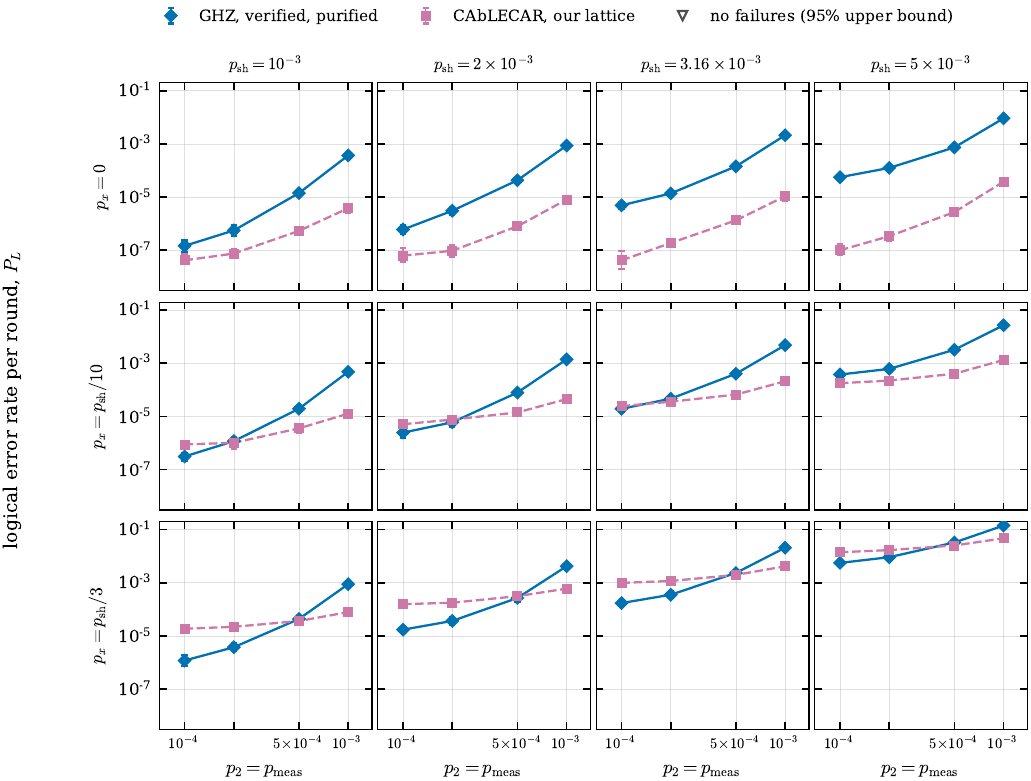}
  \caption{Logical error rate per round, $P_L=P_L^X+P_L^Z$, of the GHZ gadget (verified,
  purified; solid) and CAbLECAR on our node lattice (dashed) as the two-qubit gate and readout
  errors are lowered together, $p_2=p_\mathrm{meas}$. Gross code, $12$ rounds; all other
  parameters as in Table~\ref{tab:params}; $p_2=10^{-3}$ is the matched profile. Rows: transport
  bit-flip model; columns: shuttle noise $p_\mathrm{sh}$. Error bars are $95\%$ Clopper--Pearson
  intervals.}
  \label{fig:results_p2meas}
\end{figure*}

We test this explanation directly. CAbLECAR's scheduler can also be given a
fixed visit order for each ancilla, in which case it only routes the ancilla
between its data qubits in that order. We call this a \emph{pinned} order: the
scheduler may choose the route and timing, but cannot change which data
qubit is visited first, second, and so on. We extract the order the scheduler
chooses on CAbLECAR's own layout and impose it on our lattice. The hardware and noise parameters remain
fixed, but the routes, timings and resulting noise exposure can change. As a check that imposing an order does not alter
anything else, imposing the same order on CAbLECAR's own layout reproduces
the original circuit's error model: with shuttle noise switched off the
two are identical, and with it switched on their total error probabilities
agree to within $0.2\%$ for the Gross code and $2.5\%$ for the
$[[72,12,6]]$ code. Figure~\ref{fig:results_order} compares both versions on our lattice
with CAbLECAR on its own layout. With the order pinned, the low-noise
reversal disappears. For the $[[72,12,6]]$ code, where the reversal is clearest, the
pinned-order rate is $1.0$ to $1.3$ times the original-layout rate below
$p_\mathrm{sh}=10^{-3}$, with every range including one, while the re-optimized order remains
higher than the pinned one by a factor of $1.4$ [$1.01$, $2.0$] at $p_\mathrm{sh}=10^{-4}$
without bit flips. For the Gross code the pinned-order rate is $0.5$ to $1.1$ times the
original-layout rate below $10^{-3}$, again consistent with one. The
low-noise penalty of the re-optimized circuit therefore comes from its
visit order, as the hook-error argument predicts.

The pinned-order round is also longer than either of the other two
($13.8~\mu$s for the Gross code, Table~\ref{tab:rounds}), so its qubits idle
for longer; that it nonetheless reaches the lower floor is further evidence
that idle noise is not the cause of the reversal.

Pinning the order has a cost at high shuttle noise. The order chosen for
CAbLECAR's grid is not the shortest route on our lattice, so the pinned
routes are longer than the re-optimized ones: the sum of the probabilities of the modelled shuttle-error events per
round for the Gross code is $3.06$ with the pinned order,
compared with $2.28$ with the re-optimized order and $3.62$ on the original
layout. These sums can exceed one because they add probabilities over many
error locations; they are not probabilities that the round fails. The
pinned circuit therefore keeps part of the transport saving.
For the Gross code it is still lower than the original layout by a factor
of about $3$ to $4$ at high $p_\mathrm{sh}$ when bit flips are present (e.g.\ $3.4$ [$3.4$, $3.5$] at
$3.16\times10^{-3}$ with $p_x=p_\mathrm{sh}/3$), but it is up to $12$ [$10$, $15$] times worse
than the re-optimized circuit in the bit-flip regime. For the $[[72,12,6]]$
code the pinned order removes the transport saving entirely: its corresponding summed
shuttle-error probability ($1.43$) is similar to that of the original layout ($1.42$),
and so is its logical error rate. For this smaller code, the entire gain
from our lattice comes from the freedom to re-optimize the visit order.

The visit order thus controls a trade-off between two effects. An order
chosen for short routes reduces transport faults, which dominate at high
$p_\mathrm{sh}$. An order chosen for favourable hook errors reduces the
impact of gate and readout faults, which dominate at low $p_\mathrm{sh}$.
Neither CAbLECAR's scheduler nor our runs optimize for the second effect, and
the better of the two curves in Fig.~\ref{fig:results_order} is an
achievable rate at each $p_\mathrm{sh}$. A scheduler that considers both route length and
fault propagation might improve further; neither implementation is an optimized limit on
mobile-ancilla performance.

In summary, at the parameters studied a mobile ancilla on our node lattice gives the lower
logical error rate: the crossover of Sec.~\ref{sec:results_comparison} is mostly a consequence of
CAbLECAR's longer routes on its own layout. The singlet gadgets do limit how far transport bit
flips spread, which is why the gap narrows as bit flips become more frequent, but that protection
does not outweigh their additional gates and readouts.

\begin{table*}[t]
  \centering
  \caption{Logical error rate per round, $P_L=P_L^X+P_L^Z$, with two-qubit
  gate noise removed, including that of the readout CZs of our gadgets. Gross code, $12$ rounds, $p_x=p_\mathrm{sh}/3$; other
  parameters as in Table~\ref{tab:params}. Brackets give the summed $95\%$
  Clopper--Pearson endpoints; the number of observed failures (both bases)
  is given in parentheses.}
  \label{tab:p2off}
  \footnotesize
  \setlength{\tabcolsep}{3pt}
  \begin{tabular}{@{}lccc@{}}
    \toprule
    Scheme & $p_\mathrm{sh}=3\times10^{-4}$ & $10^{-3}$ & $2\times10^{-3}$ \\
    \midrule
    CAbLECAR, our layout & $6.1\times10^{-7}$ ($22$) & $1.5\times10^{-5}$ ($184$) & $1.4\times10^{-4}$ ($344$) \\
     & {\scriptsize$[3.2,\,11]\times10^{-7}$} & {\scriptsize$[1.2,\,1.9]\times10^{-5}$} & {\scriptsize$[1.2,\,1.7]\times10^{-4}$} \\
    CAbLECAR, own layout & $8.3\times10^{-8}$ ($1$) & $1.5\times10^{-5}$ ($15$) & $1.1\times10^{-3}$ ($130$) \\
     & {\scriptsize$[0.02,\,7.7]\times10^{-7}$} & {\scriptsize$[0.61,\,2.9]\times10^{-5}$} & {\scriptsize$[0.84,\,1.4]\times10^{-3}$} \\
    GHZ, verified, purified & $6.8\times10^{-7}$ ($9$) & $1.7\times10^{-6}$ ($20$) & $5.2\times10^{-5}$ ($100$) \\
     & {\scriptsize$[2.1,\,17]\times10^{-7}$} & {\scriptsize$[1.0,\,11]\times10^{-6}$} & {\scriptsize$[3.9,\,6.9]\times10^{-5}$} \\
    Teleport, unpurified & $4.2\times10^{-6}$ ($13$) & $3.1\times10^{-5}$ ($21$) & $2.6\times10^{-4}$ ($35$) \\
     & {\scriptsize$[1.3,\,10]\times10^{-6}$} & {\scriptsize$[1.6,\,5.6]\times10^{-5}$} & {\scriptsize$[1.5,\,4.1]\times10^{-4}$} \\
    \bottomrule
  \end{tabular}
\end{table*}

\subsection{Which noise channels limit performance?}
\label{sec:results_budget}

The layout comparison establishes which implementation performs best in
the sampled regime. To investigate what limits each extraction circuit,
we repeat the
Gross-code simulations with one noise channel removed at a time, for the
teleport gadget, the GHZ gadget and CAbLECAR on its own layout. All operation
durations and the resource schedule remain unchanged, so each test changes
the noise without also changing the waiting times.
Figure~\ref{fig:results_budget} gives the rate in each test divided by the
rate with the full noise model. A smaller ratio means that removing that
channel has a larger effect.

Two-qubit gate noise is the main limitation for all three schemes. In our gadgets this channel
includes the CZ of every single-spin readout, so removing it also lowers their effective readout
error to $p_\mathrm{meas}$; the readout test removes $p_\mathrm{meas}$ only. Removing
two-qubit gate noise lowers the teleport rate by factors of $6$ to $22$ over the tested points,
the GHZ rate by a factor of $80$ to $530$ (with wide ranges at the lowest $p_\mathrm{sh}$,
where few failures remain), and the CAbLECAR rate by $15$ [$6$, $41$] at
$p_\mathrm{sh}=10^{-3}$. At $p_\mathrm{sh}=3\times10^{-4}$, CAbLECAR with two-qubit gate noise
removed showed a single failure in $2\times10^{6}$ shots, a reduction of at least a factor of
about nine. Removing one-qubit gate noise, preparation noise or
idle noise individually lowers the rate by at most a factor of two; idle
noise is small because the $10$~ms echo coherence time is roughly three
orders of magnitude longer than a syndrome round.

\begin{figure*}[t]
  \centering
  \includegraphics[width=0.72\linewidth]{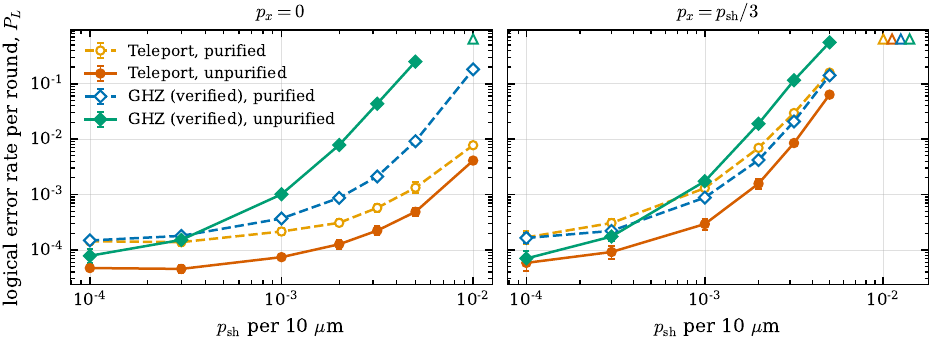}
  \caption{Effect of phase purification on the teleport gadget (orange) and
  the verified GHZ gadget (blue and green), Gross code, $12$ rounds,
  $P_L = P_L^X + P_L^Z$. Dashed lines with open markers: purification on
  every link of two or more hops. Solid lines with filled markers: source
  heralding only.}
  \label{fig:results_purification}
\end{figure*}

The schemes differ sharply in their sensitivity to shuttle bit flips. For
CAbLECAR, removing that channel lowers the rate by a factor of about $60$ at
$p_\mathrm{sh}=10^{-3}$ and about $500$ at $2\times10^{-3}$. Together with the gate-noise test, this indicates strong sensitivity to
both channels, consistent with failures involving combinations of shuttle
and gate faults. The tests do not identify the individual fault
combinations. These tests use the original CAbLECAR layout; they do not directly measure
the noise sensitivity of its node-lattice implementation. Shuttle dephasing matters for
CAbLECAR as well: at $p_\mathrm{sh}=2\times10^{-3}$, removing it lowers the
rate by a factor of $22$, because even with refocusing a phase error along a
long route flips the measured syndrome. The singlet-based gadgets are much
less sensitive to shuttle bit flips, though not immune at the largest
$p_\mathrm{sh}$, and more sensitive to readout errors, consistent with the
larger number of measurements in each check.

Because two-qubit gates are the main limitation, and because the GHZ gadget
uses the most of them, we ask whether better two-qubit gates could reverse the
conclusion of Sec.~\ref{sec:results_layout}. Table~\ref{tab:p2off} first compares
all schemes with two-qubit gate noise removed entirely, including CAbLECAR on our
lattice, for the Gross code with $p_x=p_\mathrm{sh}/3$. At
$p_\mathrm{sh}=3\times10^{-4}$, CAbLECAR on our lattice and the GHZ gadget are not
resolved (ratio $1.1$ [$0.2$, $5.3$]). At $2\times10^{-3}$ the GHZ gadget is lower, by a
factor of $2.8$ [$1.8$, $4.3$], based on $100$ observed failures.
Removing two-qubit gate noise improves the GHZ gadget by a factor of $80$ to $530$ but
CAbLECAR on our lattice by only a factor of $4$ to $14$ at these noise levels: its remaining
failures are driven by transport bit flips copied onto the data along its routes. The central
teleport rate remains above CAbLECAR on our lattice at every point even with perfect two-qubit
gates, although only at $p_\mathrm{sh}=3\times10^{-4}$ does the interval exclude equality.

Perfect gates are an idealisation, so we then lower the two-qubit gate and readout errors
together, $p_2=p_\mathrm{meas}=q$ with $q\in\{10^{-4},\,2\times10^{-4},\,5\times10^{-4}\}$, keeping
every other parameter of Table~\ref{tab:params}, for the GHZ gadget and CAbLECAR on our lattice
at four shuttle-noise levels and all three noise models (Fig.~\ref{fig:results_p2meas}). The
readout error is included because the GHZ gadget measures many more qubits per check; the readout
CZ of our gadgets follows the two-qubit gate error, so their single-spin readout error is
$q\oplus8q/15$. With frequent bit flips, $p_x=p_\mathrm{sh}/3$, the two curves cross near
$q=5\times10^{-4}$ at every $p_\mathrm{sh}$; interpolating places the crossover at
$q^\ast\approx4$--$5\times10^{-4}$, about half the matched value of $10^{-3}$, with
little dependence on $p_\mathrm{sh}$. Below it the GHZ gadget is clearly lower: at
$q=2\times10^{-4}$ by a factor of $4.8$ [$3.0$, $8.0$] at $p_\mathrm{sh}=2\times10^{-3}$ and
$3.3$ [$2.1$, $5.4$] at $3.16\times10^{-3}$, and at $q=10^{-4}$ by up to a factor of about $16$
($16$ [$7.7$, $33$] at $p_\mathrm{sh}=10^{-3}$).
With rarer bit flips, $p_x=p_\mathrm{sh}/10$, the two schemes become comparable only near
$q\approx10^{-4}$ to $2\times10^{-4}$: at $q=10^{-4}$ the GHZ gadget is lower by a factor of
$1.3$ to $2.8$ for $p_\mathrm{sh}\le3.16\times10^{-3}$, and the interval excludes equality only at
$p_\mathrm{sh}=10^{-3}$ ($2.8$ [$1.6$, $5.2$]).
Without bit flips, CAbLECAR remains lower at every resolved point, by a factor of $26$ to $270$
for $q\ge5\times10^{-4}$: a phase error on a travelling ancilla only flips its syndrome bit, so
there is little for distributed entanglement to protect against. Distributed entanglement
therefore wins on a common layout when transport bit flips are frequent and two-qubit gate and
readout errors are below about $4$--$5\times10^{-4}$.

\subsection{When does purification help?}
\label{sec:results_purification}

The preceding comparisons include several forms of resource protection.
We now examine them separately, beginning with phase purification.
Purification is intended to reduce phase errors on a delivered pair, but it
does so by adding operations that can themselves fail: a second raw
singlet, a CZ gate and a readout at each end of the link.
Figure~\ref{fig:results_purification} shows which effect wins for each
gadget on the Gross code. Source heralding is used in both cases;
purification is an additional step applied to every link of two or more
hops.

\begin{figure*}[t]
  \centering
  \includegraphics[width=0.72\linewidth]{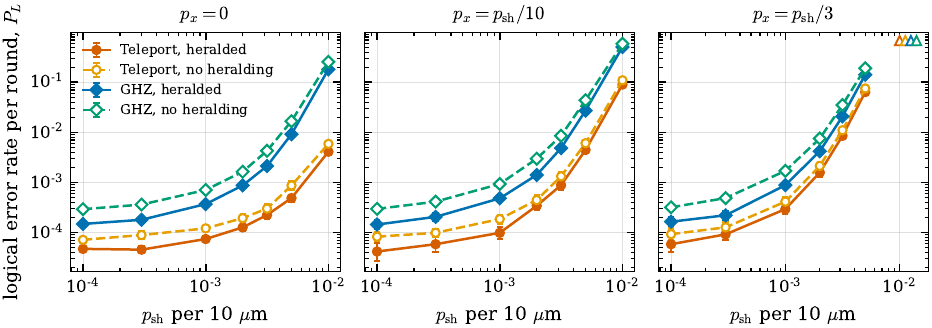}
  \caption{Effect of source heralding, Gross code, $12$ rounds,
  $P_L = P_L^X + P_L^Z$. Teleport gadget without purification and
  verified GHZ gadget with purification. Solid lines with filled
  markers: heralded sources (teleport red-orange, GHZ blue); dashed lines with
  open markers: heralding switched off (teleport yellow-orange, GHZ green).}
  \label{fig:results_heralding}
\end{figure*}

\begin{figure*}[t]
  \centering
  \includegraphics[width=0.72\linewidth]{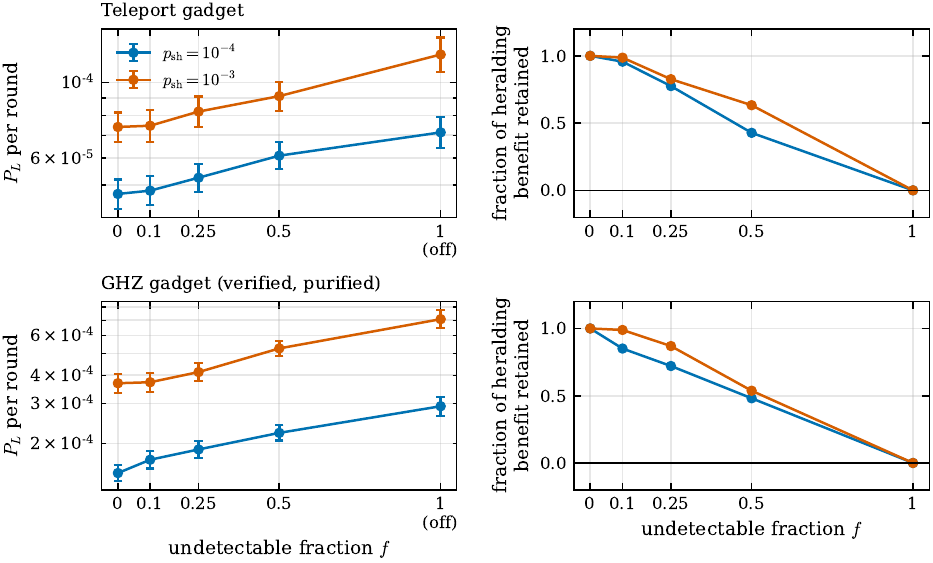}
  \caption{Dependence of heralding on the fraction $f$ of preparation failures that the
  source verification cannot detect. Top: teleport gadget (unpurified); bottom: GHZ gadget
  (verified, purified). Gross code, $12$ rounds, $p_x=0$,
  $\eta=0.999$, $P_L=P_L^X+P_L^Z$. Left: logical error rate per round, with $95\%$ intervals;
  $f=1$ is the heralding-off simulation. Right: fraction of the heralding benefit retained,
  $(P_L^{\mathrm{off}}-P_L(f))/(P_L^{\mathrm{off}}-P_L(0))$. Each point is based on $570$ to
  $1320$ failures.}
  \label{fig:results_srcmodel}
\end{figure*}

\begin{figure*}[t]
  \centering
  \includegraphics[width=0.72\linewidth]{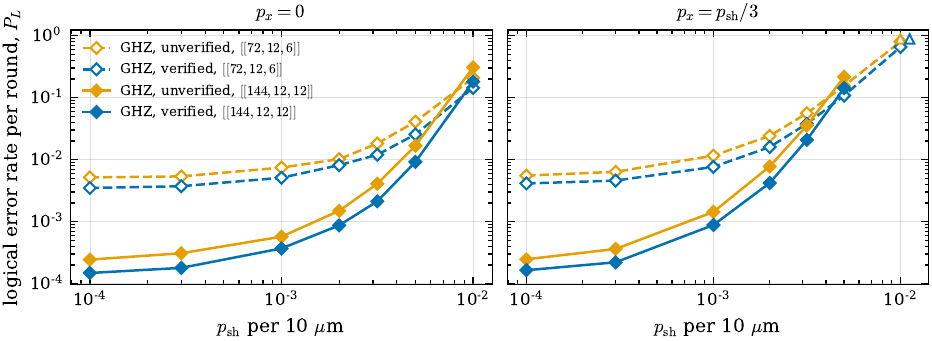}
  \caption{Chain GHZ gadget with (blue) and without (orange)
  central-fusion verification, both with purification,
  $P_L = P_L^X + P_L^Z$. Filled markers and solid lines: Gross code, $12$
  rounds; open markers and dashed lines: $[[72,12,6]]$ code, $6$ rounds.}
  \label{fig:results_verification}
\end{figure*}

The two gadgets respond in opposite ways. For the teleport gadget,
purification increases the logical error rate by a factor of $1.9$ to $4.4$ at
every sampled point below saturation. For the verified GHZ gadget,
purification helps above a crossover between $p_\mathrm{sh}=3\times10^{-4}$ and $10^{-3}$,
reaching a factor-of-$21$ reduction at $p_\mathrm{sh}=3.16\times10^{-3}$, and
increases the rate by up to a factor of about two below that point.

The circuit-level propagation of link errors helps explain this difference.
In the teleport gadget, a phase error on a delivered singlet mainly corrupts
the teleported ancilla and therefore flips the measured syndrome bit. Repeated syndrome rounds give the decoder information with which to
distinguish a measurement fault from a persistent data error. In the
sampled regime, reducing these link errors does not compensate for the
additional faults introduced by purification. In the chain GHZ gadget, the purified long-range links are the outer bridges: in the bulk the
central bridge spans a single lane and is never purified, and of the $324$ purified links of the
Gross layout, $288$ are outer bridges, $34$ are inner pair links and $2$ are central
bridges. A phase error on an outer bridge flips its fusion record, which leaves an incorrect
frame on one arm and hence an error on the data qubit coupled to it (Sec.~\ref{sec:fusion}), and
these bridges carry the longest transport and hence most of the phase noise. To confirm that
the benefit comes from them, we repeated the Gross simulations at $p_x=0$ and
$p_x=p_\mathrm{sh}/3$, purifying either only the outer bridges or only the inner links (pair links and central bridges; Fig.~\ref{fig:results_purif_split}, Appendix~\ref{app:figures}). Purifying the outer
bridges alone recovers the benefit: its rate is within about $20\%$ of full purification at
every sampled point, and slightly lower when bit flips are frequent, since its round is
$13.2~\mu$s rather than $14.1~\mu$s. Purifying only the inner links leaves the rate within about
$25\%$ of the unpurified value. Suppressing these errors can justify the extra operations once shuttle
noise is sufficiently high. The main comparison therefore uses an
unpurified teleport gadget and a purified GHZ gadget as fixed reference
configurations, while the present comparison shows where the latter
choice helps or hurts. The general lesson is that improving the
fidelity of an individual link does not by itself guarantee a better logical
memory; the protection must target errors that would otherwise become
harmful after propagating through the check.

\subsection{Benefit of source heralding}
\label{sec:results_heralding}

Source heralding tests each pair before it leaves the generator, whereas
purification acts on a link after transport. To isolate its effect,
Fig.~\ref{fig:results_heralding} compares Gross-code simulations with
source heralding enabled and disabled. In the disabled case, generators
deliver every pair without the local preparation check, and the decoder
receives no source flags. The teleport gadget remains unpurified and the
GHZ gadget remains purified.

Heralding lowers the central logical error rate of both gadgets at every point below
saturation and for all three noise models, significantly at every GHZ point and at most teleport
points. Without shuttle bit flips at $p_\mathrm{sh}=10^{-4}$, it improves the teleport rate by a
factor of $1.5$ [$1.2$, $1.9$] and the GHZ rate by a factor of $2.0$ [$1.65$, $2.35$]. At
$p_\mathrm{sh}=10^{-2}$ the improvement falls to factors of $1.2$ to $1.4$. This trend follows from
where the check acts: heralding detects faulty preparations, which form a
significant part of the gate-limited floor at low shuttle noise, but it
cannot detect errors that occur later during transport, which dominate at
high shuttle noise. The comparison changes both the source procedure and
the information available to the decoder, so it does not separate the value
of the decoder flags from that of rejecting faulty pairs.

These improvements use the most favourable source model of Sec.~\ref{sec:singlets}, in which
every faulty pair is caught by the spin check ($f=0$), and the heralding-off curve
corresponds to the opposite limit $f=1$. Figure~\ref{fig:results_srcmodel} shows both
gadgets for intermediate fractions of undetectable failures. The logical error rate rises
smoothly with $f$ rather than jumping to the heralding-off value. For the teleport gadget at
$p_\mathrm{sh}=10^{-4}$, heralding retains about $95\%$ of its benefit at $f=0.1$, $80\%$ at
$f=0.25$ and $45\%$ at $f=0.5$, and somewhat more at $10^{-3}$. The GHZ gadget behaves similarly,
retaining $85$ to $99\%$ at $f=0.1$, $70$ to $85\%$ at $f=0.25$ and about $50\%$ at
$f=0.5$ at the two noise levels. Heralding therefore remains useful unless undetectable failures are a large
fraction of all faulty pairs, because even at $f=0.1$ an accepted pair is faulty with
probability $10^{-4}$, still small compared with the other faults of the gadget. Varying the detection efficiency between $\eta=0.9$ and $0.999$ at $f=0$ changed
the rate by less than about $25\%$ in exploratory runs, so the undetectable fraction, not the
sensor fidelity, is the parameter that matters. A less favourable source can only raise the logical error rates of the singlet gadgets. To
test whether this changes the comparison with CAbLECAR at lower gate and readout error
(Fig.~\ref{fig:results_p2meas}), we repeated the GHZ simulations with $f=0.1$ and $0.25$ at
$p_x=p_\mathrm{sh}/3$ for every $p_\mathrm{sh}$ and $q\in\{10^{-4},\,2\times10^{-4},\,5\times10^{-4}\}$
(Fig.~\ref{fig:results_srcrobust}, Appendix~\ref{app:figures}). At $f=0.25$ the GHZ rate rises by
a factor of $1.1$ to $1.7$, most at low $p_\mathrm{sh}$, where source faults are a larger share of
the total, and the advantage over CAbLECAR survives: for $q\le2\times10^{-4}$ the GHZ gadget stays
lower at every point, at $f=0.25$ by a factor of $1.55$ [$1.4$, $1.8$] to $9.3$ [$5.4$, $16$], with
every interval excluding equality. At $q=5\times10^{-4}$, near the crossover, the two are
comparable at $f=0$ except at $p_\mathrm{sh}=5\times10^{-3}$, where CAbLECAR is lower by a factor
of $1.3$ [$1.2$, $1.4$]; at $f=0.25$ CAbLECAR is lower at all four $p_\mathrm{sh}$, significantly at three, by up
to a factor of $1.7$.

\subsection{Benefit of central-fusion verification}
\label{sec:results_verification}

Source heralding checks the initial resource, but cannot detect faults
introduced later during fusion. To examine protection at that later stage,
Fig.~\ref{fig:results_verification} compares the purified chain GHZ gadget
with and without the central-fusion verification circuit of
Sec.~\ref{sec:chain} (Fig.~\ref{fig:verification}, Appendix~\ref{app:figures}). The circuit adds a detector that checks the
consistency of the central fusion record before the GHZ arms couple to the
data. It targets one specific pathway, in which an incorrect fusion record
leaves a correlated error on three data qubits.

Verification lowers the Gross-code logical error rate by a factor of $1.5$ to $1.9$ for
$p_\mathrm{sh}\geq2\times10^{-3}$, in all three noise models; for example, at
$p_\mathrm{sh}=3.16\times10^{-3}$ with $p_x=0$ the rate changes from
$4.1\times10^{-3}$ to $2.1\times10^{-3}$, a factor of $1.9$ [$1.5$, $2.5$]. At lower
$p_\mathrm{sh}$ the central values are similar ($1.5$ to $1.7$), and every point
excludes equality. For the $[[72,12,6]]$ code the improvement is a factor of $1.2$ to
$1.6$, significant at every sampled point. The benefit disappears only as the
failure rate saturates. The verification operations fit within the existing
schedule; they require the resident verifier dot, two CZ gates and one
readout. The improvement is limited because the detector covers only part of
the GHZ preparation faults: it misses an $X$ error on arm $a_0$ before
fusion, does not check the outer bridges, and can be applied only where the
required arms meet at a check centre, which is the case for $110$ of the
$144$ Gross-code checks.

\subsection{Rotated surface-code reference}
\label{sec:results_surface}

We also simulated the rotated surface code ($d=4,6,8$) on the same fabric and with the same
hardware parameters, as a reference in which every check is local; the gadgets, the results and
the sampling details are given in Appendix~\ref{app:surface}. For
$10^{-3}\le p_\mathrm{sh}\le10^{-2}$ the walking-ancilla gadget is lower than the 4-GHZ gadget by a
factor of $1.3$ to $4.8$, growing with distance: when all checks are local, there is little transport for distributed
entanglement to protect against.

\section{Conclusion}
\label{sec:conclusion}

We developed a singlet fabric for measuring bivariate-bicycle stabilizers on a silicon spin-qubit
network, built a chain GHZ gadget and a teleported-ancilla gadget on it, scheduled both under
explicit hardware capacities, and compared their memory performance with direct ancilla
transport using the CAbLECAR scheduler, under identical gate, readout and transport parameters.
Every single-spin readout in our gadgets includes the CZ with a singlet-prepared PSB pair that such
a measurement requires, which the CAbLECAR circuits do not, so the comparison is conservative for
the singlet gadgets in this respect.

Against CAbLECAR on its own layout, the teleport gadget wins when transport frequently flips
spins: an error on a travelling ancilla is copied onto every data qubit it visits afterwards,
whereas the singlet gadgets transport their resources before coupling them to the data. The
layout control shows that this advantage comes mainly from where the data qubits are placed.
Our node lattice keeps neighbouring data qubits $10~\mu$m apart but brings each check's distant
data qubits much closer; running CAbLECAR on it roughly halves its routes, lowers its logical
error rate by up to a factor of about $50$, and makes direct transport better than both singlet
gadgets at every sampled Gross-code point. The same control showed that the order in which an
ancilla visits its data qubits matters separately from route length, by changing the hook errors
its faults produce, so the shortest route is not necessarily the best schedule.

Two-qubit gate faults limit every scheme, and the protections built into the singlet gadgets
must be judged on the complete memory circuit. Source heralding improves both gadgets by a
factor of $1.5$ to $2$ at low shuttle noise and keeps most of that benefit unless faults the verification cannot see
dominate the faulty pairs; central-fusion verification lowers the GHZ
error rate by a factor of $1.5$ to $1.9$ at every sampled shuttle-noise level, at no cost in round time; and phase purification helps the
GHZ gadget only at high shuttle noise while making the teleport gadget worse throughout. Better
gates and readout change the comparison. When transport bit flips are frequent, the GHZ gadget overtakes direct
transport on a common layout once two-qubit gate and readout errors fall below about
$4$--$5\times10^{-4}$, at every sampled shuttle-noise level; with rarer bit flips the two become
comparable only near $10^{-4}$--$2\times10^{-4}$, and without bit flips direct transport always wins. Distributed
entanglement could therefore pay off on a common layout, but only with roughly twofold better
gates and readout and with transport that frequently flips spins.

\section*{Code availability}
The circuit generators, scheduler and analysis scripts used in this work
are available on GitHub at
\url{https://github.com/KNFLWNF/singlet-fabric-qldpc}.

\begin{acknowledgments}
This research was undertaken thanks in part to funding from the Canada First Research
Excellence Fund (Transformative Quantum Technologies) and Canada's Natural Sciences and
Engineering Research Council (NSERC).
\end{acknowledgments}

\bibliographystyle{apsrev4-2}
\bibliography{references}

\clearpage
\appendix

\section{Scheduling details}
\label{app:scheduling}

Each distributed-entanglement link connects two endpoints determined by the
stabilizer-measurement gadget. A singlet is generated on a conveyor lane
between these endpoints, and its two electrons are transported in opposite
directions towards their destinations.

For each link, we construct up to 24 candidate shortest paths through the
physical array. The number 24 limits how many alternative routes we
consider; it is not a path length. Several shortest paths may connect the
same endpoints because horizontal and vertical moves can be arranged in
different orders. Each path is a sequence of nearest-neighbour conveyor lanes.
The candidate generator positions are restricted so that the two singlet halves travel approximately equal
distances. This limits the difference in transport time and accumulated
shuttling error between the two halves.

Choosing a shortest path for each link independently can concentrate traffic
on a small number of generators, nodes, or lanes. We therefore assign routes
using a greedy load-balancing procedure. Here, ``greedy'' means that one
demand is assigned at a time, using the loads created by the assignments
already made.

Demands are processed in nonincreasing order of shortest-path length, with
randomized ordering among demands of equal length. Longer routes are
considered first because they occupy more transport and relay resources.
For each demand, we evaluate every candidate path and central generator
position using the loads that would result from adding that assignment.

The routing cost is
\begin{align}
    C =
    \left(
        G_{\max},
        \sum_e G_e^2,
        R_{\max},
        \sum_v R_v^2,
        L_{\max},
        \sum_e L_e^2
    \right),
\end{align}
where $G_e$ counts routed demands assigned to the generator on lane $e$,
$R_v$ counts routed demands passing through node $v$ as an intermediate
relay, and $L_e$ counts routed demands using conveyor lane $e$.
The quantities $G_{\max}$, $R_{\max}$, and $L_{\max}$ are the largest
loads in their respective resource classes. For example, $G_{\max}=5$
means that the busiest generator has been assigned five routed demands.
These are counts of assigned demands, not five jobs that must occur at
the same instant; the timetable determines when the jobs occur.

Candidate assignments are compared lexicographically: the first cost
component is compared first, and a later component is considered only if
all earlier components are equal. Thus, the procedure first reduces the
largest generator load. Among assignments with the same largest generator
load, it favours a more even distribution of generator usage through the
sum of squared loads. Relay-node loads and conveyor-lane loads are then
compared in the same way. This is a priority rule rather than a weighted
sum: a candidate with a smaller maximum generator load is preferred even
if it has a larger lane load. The squared-load terms discourage placing
additional demands on resources that are already heavily used.

These counters describe the routing demands, rather than the complete
physical workload. In particular, extra singlet copies required for
purification and the resources used by local gadget operations are
accounted for during temporal scheduling.

The route-assignment procedure is repeated for 24 seeded trials, with
different randomized tie-breaking among equal-length demands. The
assignment with the smallest final routing cost is retained. The selected
routes and generator locations are then reused in every syndrome round;
they do not change in response to physical errors or decoder outcomes.

For the Gross-code geometry considered here, the maximum selected route
length is 34 conveyor lanes for the chain GHZ gadget and 30 lanes for the
teleported-ancilla gadget. The two constructions require different
distributed-entanglement endpoints and therefore produce different routes.
The effect of those routes on contention is assessed through their resource
loads and the resulting temporal schedule.

The geometrically local parts of a stabilizer measurement do not require
a global route search, but they still occupy hardware and must be included
in the schedule.

For the chain GHZ gadget, the local ancilla resource is assembled using
short, single-lane singlets generated near the check centre. Long-range
singlets provide bridges that are consumed during fusion. The GHZ arms
that subsequently interact with the data are generated locally or undergo
only short-range transport.

For the teleported-ancilla gadget, the syndrome ancilla is conveyed between
neighbouring check and data nodes to perform the local interactions.
These local steps do not require distributed singlets. Teleportation is
used when the ancilla must reach a non-local connection, including
connections made non-local by the physical boundaries.

Both constructions therefore combine long-range communication with local
generation, transport, gates, and measurements. All of these operations
contribute to the resource calendar. Local operations can be an important
bottleneck because several steps may repeatedly use the same lanes or
processing resources near an active check centre.

The stabilizer checks are divided into six groups, measured in six
subcycles. A subcycle is therefore one part of the complete round, not a
separate round in which every stabilizer is measured. We use
$\mathrm{II}$, the \emph{initiation interval}, to denote the time between
the starts of consecutive subcycles. This term, and the modulo reservation
table we use to enforce resource capacities across the period, are borrowed
from software pipelining in compiler design \cite{Lam_1988, Rau_1994}, where
the same problem arises of overlapping successive iterations of a loop body
subject to fixed functional-unit capacities.

A subcycle does not require all communication for the previous subcycle
to finish before new communication begins. Instead, operations are
pipelined: for example, a singlet needed by subcycle $s+1$ may be generated
and transported while the gadget operations of subcycle $s$ are still
running. Whether such overlap is possible depends on the resources
required by the two operations.

The scheduler represents resource usage on a periodic calendar with
discrete time steps. Each reservation identifies a resource, the interval
during which it is occupied, and the amount of capacity required.
A conveyor lane can accommodate one scheduled transit at a time, and
a singlet-generator bank can accommodate one scheduled generation job
at a time. Nodes have the finite storage, exchange-gate, and readout
capacities specified in Sec.~\ref{sec:lattice}.

A proposed operation is accepted only if its reservations fit within
these capacities. For a transported singlet, this requires checking the
generation interval, each subsequent lane transit and port occupancy,
and the storage intervals at relay and destination nodes. The resources
along a route need to be available during their respective occupancy
intervals; they need not all be free simultaneously at the launch time.

Scheduling proceeds backwards from the times at which entangled resources
are consumed. The stabilizer gadget operations are placed first relative
to the subcycle boundaries. These operations include fusion, coupling to
data qubits, Bell measurements, and ancilla readout. Operations sharing
a node are serialized where required by its exchange or readout capacity.

Once these operations have been placed, they define delivery deadlines
for the distributed resources. Purification operations are placed next,
as late as possible before the purified pairs are needed. A purification
can occupy an interval only if the purification module and a readout
channel are free at both endpoints, and the raw halves waiting for it must
fit in the endpoint dot pools.

Generation and transport are then scheduled to deliver the required
singlets before their deadlines, with longer and more highly contended
links considered first. For each link, the scheduler searches backwards
for the latest launch time whose associated reservations fit the
calendar. If a shared resource is occupied, the link must be moved to
an earlier available interval.

This as-late-as-possible placement aims to reduce storage between delivery
and consumption. It finds the latest placement allowed by reservations
already committed; it does not establish a global minimum of total
storage time. Likewise, if an operation cannot be placed, that scheduling
attempt fails, but this does not prove that no feasible schedule exists
at the chosen initiation interval.

A smaller initiation interval permits faster syndrome extraction but
leaves less room for operations that compete for the same resources.
We first calculate a lower bound on $\mathrm{II}$ from the modelled
resource workloads and gadget constraints. For example, a generator
assigned many jobs must have enough time to complete them, while a heavily
used node must have sufficient capacity for its relay, purification,
gate, and readout operations.

Starting from this lower bound, we find a larger interval for which the
scheduler produces a complete calendar. We then narrow the search bracket
by bisection. At each trial interval, up to three seeded operation orderings
are attempted. If an attempt succeeds, its calendar is retained and the
search continues towards a smaller interval. If all attempts fail, the
search proceeds towards a larger interval.

The search stops when the bracket width is approximately $2\%$ of the
initial lower bound, subject to the discrete time resolution of the
calendar. This is a search stopping criterion, not a guarantee that the
result lies within $2\%$ of the optimal schedule.

Because placement uses a greedy heuristic, its success is not guaranteed
to vary monotonically with the trial interval. The reported initiation
interval is therefore a validated feasible upper bound on the minimum
achievable interval under the modelled constraints, rather than a proof
of global optimality.

For each completed schedule, resource-occupancy tables are independently
reconstructed from the emitted reservations. These tables are checked
against the declared lane, generator, storage, exchange, and readout
capacities. Aggregate lane, generator, and relay usage is also checked
against the corresponding demand accounting.

The calendar is periodic, so these checks include operations whose
reservations extend across the boundary between successive syndrome
rounds. A resource occupied by the end of one round cannot also be
assigned beyond its capacity at the beginning of the next.

A reported calendar is therefore verified to satisfy the resource
constraints represented in the scheduling model. The hardware capacities
used for these checks are those specified in Sec.~\ref{sec:lattice};
the validation does not establish that this inventory is the smallest
possible one.

Since a complete syndrome round contains six subcycles, its steady-state
duration is
\begin{align}
    t_{\mathrm{cyc}} = 6\,\mathrm{II}.
\end{align}
The full resource pattern repeats after this period. Individual
communication operations may begin before the subcycle that consumes
their outputs, so the syndrome-round period should be distinguished
from the latency of an individual link and from the initial time needed
to fill the pipeline.

The accepted calendar supplies the durations used for timing-dependent
noise. In addition to intrinsic gate, readout, and shuttling errors,
storage and inactivity contribute dephasing according to the assumed
echo-protected coherence time $T_2^{\mathrm{echo}}$.

The accounting distinguishes several intervals: dwell at intermediate
relay nodes, storage of a singlet half after arrival, storage of a
purified output before consumption, and idle periods of gadget ancillas
and data qubits. Keeping these contributions separate allows the noise
model to reflect the actual timing of each scheduled operation.

A park-and-relaunch event at an intermediate node contributes a fixed
$0.1~\mu\mathrm{s}$ dwell, together with the dephasing assigned to
$1~\mu\mathrm{m}$ of equivalent conveyor transport. The equivalent
transport length is a noise-model contribution, rather than an
additional physical route segment.

For each data qubit, idle time is calculated over the complete syndrome
round, subtracting the durations of its active gate operations.
Transport, storage, and idle durations are recorded separately and
passed to the circuit-level simulation. Consequently, any additional
waiting introduced by resource contention contributes to the modelled
error exposure as well as to the achievable syndrome-extraction rate.

\section{Source and purification model}
\label{app:source}

Each generator first loads a second electron into a singly occupied dot by energy-selective loading,
$(0,1)\to(0,2)$, and a charge sensor confirms the $(0,2)$ charge state \cite{Petta_2005, Veldhorst_2018}. A
failed load leaves $(0,1)$ and produces no pair. It is not a Pauli error: the link is supplied by the other
generator of the bank, or, if both fail to load, the load is repeated. With load-failure probability of
order $10^{-3}$ the second case occurs with probability of order $10^{-6}$ per link, negligible next to the
gate and readout faults, and we do not simulate it separately.

A loaded pair can still be in the wrong spin state, for instance a triplet loaded thermally or through an
imperfect energy window, or one that has not relaxed. These are detected by a spin-blockade step: the pair
is separated into $(1,1)$ and returned towards $(0,2)$, which a singlet can re-enter and a triplet cannot,
and the sensor reads the outcome. Errors after this step, chiefly in the final separation of the verified
pair, cannot be detected by it.

The probabilities below describe the pair at the end of the verification sequence, when it is accepted
or rejected. Errors that occur afterwards, during storage at the generator, transport to the endpoints,
parking and consumption, are applied separately by the transport, idle and gate channels of
Table~\ref{tab:params}; source verification cannot detect them, and they are not included in
$P(\mathrm{bad}\mid\mathrm{accept})$.

We describe this verification by three parameters. With probability $p_{\mathrm{src}}$ a loaded pair is
faulty, so the \emph{raw preparation fidelity} is $F_{\mathrm{raw}}=1-p_{\mathrm{src}}$; a fraction $f$ of
faulty pairs are undetectable by the blockade step and the rest are detectable,
\begin{align}
  p_s = f\,p_{\mathrm{src}}, \qquad p_c=(1-f)\,p_{\mathrm{src}} .
\end{align}
The \emph{detection efficiency} $\eta$ is the probability that a detectable fault is flagged. A pair that
behaves as a singlet, either a good one or an undetectable fault, is accepted with probability $\eta'$ and
wrongly rejected otherwise. In the circuit, a faulty pair of either kind is modelled as a uniformly random
Pauli error $X$, $Y$ or $Z$ on one half of the pair, since $(P\otimes P)\ket{\Psi^-}=\pm\ket{\Psi^-}$ makes
an error on either half equivalent to one on the other. The probability that one generator's pair is
rejected is
\begin{align}
  p_{\mathrm{fail}} = p_c\,\eta + (1-p_c)(1-\eta'),
\end{align}
and Bayes' rule gives the probability that a pair is faulty given the test
outcome,
\begin{align}
  P(\mathrm{bad}\mid\mathrm{accept}) &= \frac{p_c(1-\eta)+p_s\,\eta'}{1-p_{\mathrm{fail}}},
  \nonumber\\
  P(\mathrm{bad}\mid\mathrm{reject}) &= \frac{p_c\,\eta + p_s(1-\eta')}{p_{\mathrm{fail}}} .
\end{align}
The residual error of an accepted pair therefore has two parts: detectable faults that slipped past the
check, of order $p_c(1-\eta)$, and every undetectable fault, of order $p_s$. Unless undetectable faults are
very rare, the second part dominates, and the check cannot reduce it. The blockade step makes triplet
preparations detectable, so $f$ is set mainly by errors in the final separation and by the blockade
readout itself.

In the simulations we take $p_{\mathrm{src}}=p_{\mathrm{prep}}$ and
$\eta=\eta'=1-p_{\mathrm{meas}}$ (Table~\ref{tab:params}), so
$p_{\mathrm{src}}=10^{-3}$ and $\eta=0.999$. The main results use $f=0$: every faulty pair is
detectable, an accepted pair is faulty with
probability $\approx1.0\times10^{-6}$, and a rejected pair with probability
$\approx0.5$, since about half of all rejections are false alarms. This is the
most favourable case for heralding. In the opposite limit $f=1$ the check
detects nothing, every accepted pair carries the raw error $p_{\mathrm{src}}$,
and the flags carry essentially no information; this is equivalent, to
within the false-alarm rate, to switching source heralding off. Already
$f=0.1$ raises $P(\mathrm{bad}\mid\mathrm{accept})$ to $1.0\times10^{-4}$, a
hundred times the $f=0$ value. When source heralding is switched off
(Sec.~\ref{sec:results_heralding}), there is one generator per lane, no check
and no flag, and every delivered pair carries the raw error probability
$p_{\mathrm{src}}=10^{-3}$. Section~\ref{sec:results_heralding} uses the two
limits to bound the benefit of heralding.

\emph{Purification.} For an idealized channel containing only independent phase flips with
probability $p$, the probability of accepting the parity check is
\begin{align}
    P_{\mathrm{acc}}
    &=
    (1-p)^2+p^2 ,
\end{align}
Acceptance means that either neither pair has a phase error or both
pairs do. A single erroneous pair gives unequal parities and is rejected.
Among the accepted cases, the retained pair is wrong only when both raw
pairs were wrong. Its conditional phase-error probability is therefore
\begin{align}
    p_{Z|\mathrm{acc}}
    &=
    \frac{p^2}{(1-p)^2+p^2}.
\end{align}
Thus, for $p\ll1$, the accepted branch suppresses the targeted phase error
from first to second order, $p_{Z|\mathrm{acc}}\sim p^2$.

\section{Single-spin readout model}
\label{app:readout}

\begin{figure}[t]
\centering
\begin{quantikz}[row sep=0.3cm, column sep=0.45cm]
\lstick{$q$} & \ctrl{1} & \qw & \qw \rstick{discarded} \\
\lstick[2]{$\ket{S}_{cd}$} & \control{} & \gate[2]{\mathrm{PSB}} & \cw \rstick{$S$ / $T$} \\
 & \qw & \qw &
\end{quantikz}
\caption{Single-spin readout~\cite{Buonacorsi_2019}. The spin $q$ stays in a pool dot; the double
dot $(c,d)$ of a readout channel is prepared in $\ket{S}$, a CZ acts between $q$ and $c$, and PSB
reads the pair. The outcome is $S$ (unblocked) for $q=\ket{\uparrow}$ and $T$ (blocked) for
$q=\ket{\downarrow}$.}
\label{fig:readout}
\end{figure}
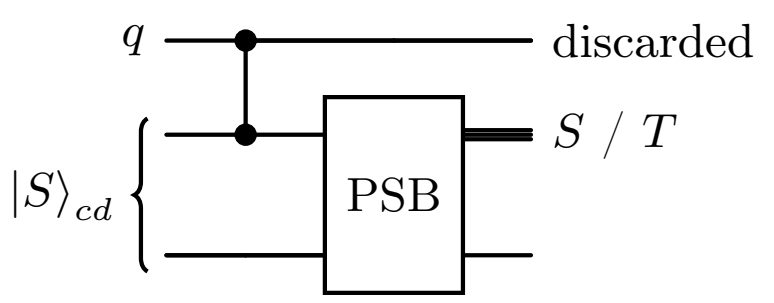

Pauli spin blockade distinguishes the singlet of a pair from its three triplets, but it cannot
read a single spin directly. Pairing an unknown spin with a reference $\ket{\uparrow}$ does not
work: $\ket{\uparrow\uparrow}=\ket{T_+}$ is blocked, but $\ket{\downarrow\uparrow}=
(\ket{T_0}-\ket{S})/\sqrt2$ is blocked only half of the time, so a blocked outcome does not
determine the spin. We therefore use the sequence of Ref.~\cite{Buonacorsi_2019}
(Fig.~\ref{fig:readout}). The double dot $(c,d)$ is prepared in
$\ket{S}=(\ket{\uparrow\downarrow}-\ket{\downarrow\uparrow})/\sqrt2$, and a CZ acts between the
spin $q$ to be read and $c$. It applies $Z_c$ only if $q=\ket{\downarrow}$, and
$Z_c\ket{S}=\ket{T_0}$, so the pair is left in $\ket{S}$ if $q=\ket{\uparrow}$ and in $\ket{T_0}$ if
$q=\ket{\downarrow}$. PSB on the pair then gives the spin. An $X$-basis readout is preceded by a
Hadamard on $q$. The sequence takes $t_2+t_\mathrm{meas}=0.6~\mu$s. The source test at the
generator bank already measures a pair and needs no CZ.

\emph{Noise of the readout CZ.} The CZ is followed by two-qubit depolarizing noise: each of the
$15$ non-identity Paulis $P_q\otimes P_c$ occurs with probability $p_2/15$. The spin $q$ is not
used again before it is reset, so only the factor $P_c$ matters. Sorting the Paulis by this factor,
\begin{center}
\begin{tabular}{lll}
\toprule
$P_c$ & Paulis & Effect on the outcome \\
\midrule
$I$ & $XI,\ YI,\ ZI$ & none \\
$Z$ & $IZ,\ XZ,\ YZ,\ ZZ$ & always flipped \\
$X,\,Y$ & $IX,\ XX,\ YX,\ ZX,$ & flipped if $q=\ket{\uparrow}$ \\
 & $IY,\ XY,\ YY,\ ZY$ & \\
\bottomrule
\end{tabular}
\end{center}
A $Z$ on $c$ exchanges $\ket{S}$ and $\ket{T_0}$ and therefore always reverses the outcome. An $X$
or $Y$ on $c$ maps both $\ket{S}$ and $\ket{T_0}$ into the span of $\ket{T_+}$ and $\ket{T_-}$; for
example $X_c\ket{S}=(\ket{\downarrow\downarrow}-\ket{\uparrow\uparrow})/\sqrt2$. These states are
blocked and read as $T$, which is wrong if $q=\ket{\uparrow}$ (the pair should be $\ket{S}$) and
correct if $q=\ket{\downarrow}$ (it should be $\ket{T_0}$). The CZ therefore flips the outcome with
probability
\begin{align}
  P(\mathrm{flip}\mid\uparrow)=\tfrac{12}{15}p_2, \qquad
  P(\mathrm{flip}\mid\downarrow)=\tfrac{4}{15}p_2 ,
\end{align}
whose average is $8p_2/15$. We simulate this as a symmetric flip of probability $8p_2/15$. Every
single-spin readout in our gadgets, of a bridge, a Bell measurement, an arm, an ancilla or a
sacrificial half, has a uniformly random outcome, so the symmetric flip reproduces the error rate
of each readout and neglects only the dependence of the flip on the outcome value.

\emph{Combined readout error.} The PSB measurement itself errs independently with probability
$p_\mathrm{meas}$. The readout is wrong when exactly one of the two flips occurs, since two flips
cancel, so its flip probability is
\begin{align}
  p_\mathrm{meas}\oplus\tfrac{8}{15}p_2
  &\equiv p_\mathrm{meas}+\tfrac{8}{15}p_2-2\,p_\mathrm{meas}\tfrac{8}{15}p_2 \nonumber\\
  &=1.53\times10^{-3}
\end{align}
for the parameters of Table~\ref{tab:params}. In the sweep of Fig.~\ref{fig:results_p2meas}, with
$p_2=p_\mathrm{meas}=q$, it is $q\oplus8q/15$.

\emph{Pair preparation.} We assume that each readout pair is re-prepared in $\ket{S}$ within the
readout cycle of its channel at no additional time or error. A pair prepared in a triplet instead
would flip the outcome with probability at most $p_\mathrm{prep}$ per readout; we neglect this
contribution.

\section{Rotated surface-code reference}
\label{app:surface}

\begin{figure*}[t]
  \centering
  \includegraphics[width=0.94\linewidth]{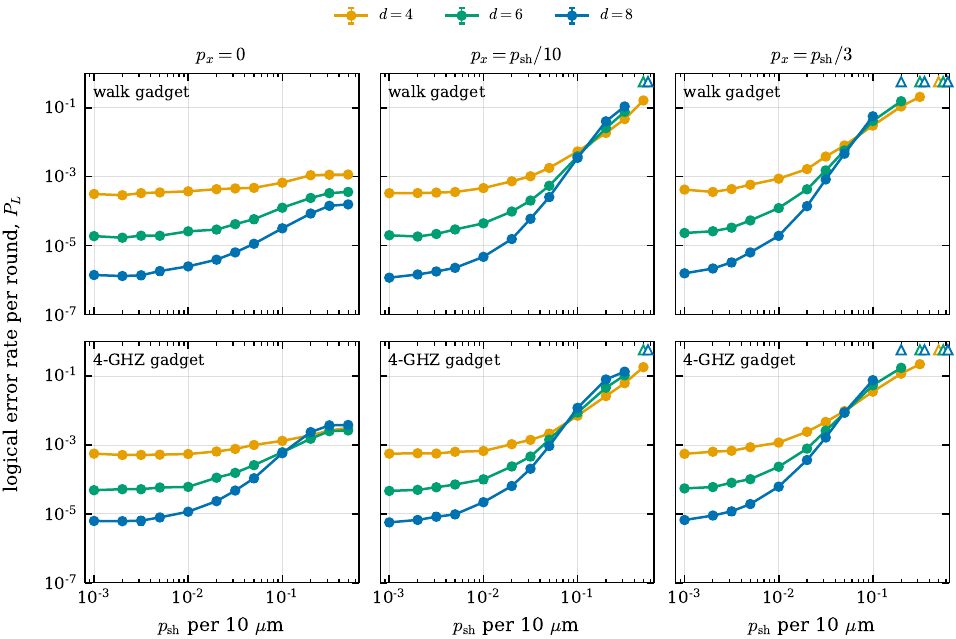}
  \caption{Rotated surface code on the singlet fabric, logical error rate per
  round for $d$ rounds, $P_L = P_L^X + P_L^Z$, PyMatching, for
  $10^{-3} \le p_\mathrm{sh} \le 0.5$. Top row: walking-ancilla gadget;
  bottom row: 4-GHZ gadget. Colours give the code distance. Open downward
  triangles are $95\%$ upper bounds where no failures were observed; open
  upward triangles mark saturated points (Appendix~\ref{app:surface_sampling}).}
  \label{fig:results_surface}
\end{figure*}

\subsection{Weight-4 gadgets}
\label{sec:weight4}

For the rotated surface-code reference, the GHZ
construction reduces to four arms supplied by
two singlets and joined by one bridge-mediated
fusion. Each arm couples to one data qubit.
The sequential construction needs no teleportation:
one ancilla walks between the four neighbouring
data qubits and is measured at the end.

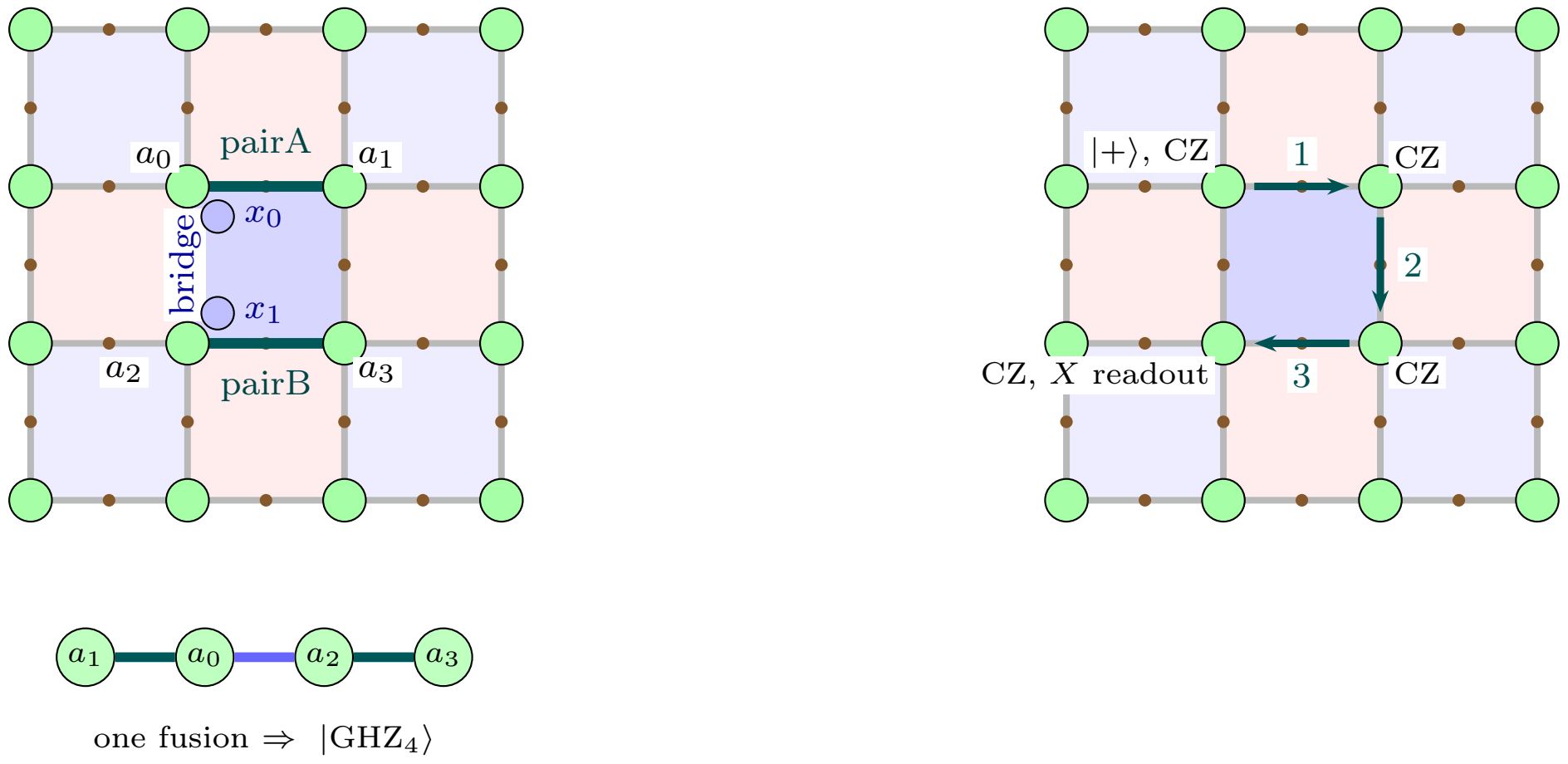
\begin{figure*}[t]
\centering
\begin{tikzpicture}[every node/.style={font=\footnotesize}]
  \def\s{1.25}
  \newcommand{\patch}[1]{%
    \coordinate (NW) at (1*\s,2*\s); \coordinate (NE) at (2*\s,2*\s);
    \coordinate (SW) at (1*\s,1*\s); \coordinate (SE) at (2*\s,1*\s);
    \foreach \fx in {0,1,2}{\foreach \fy in {0,1,2}{
      \pgfmathtruncatemacro{\pty}{mod(\fx+\fy,2)}
      \ifnum\pty=0 \fill[blue!7] (\fx*\s,\fy*\s) rectangle ({(\fx+1)*\s},{(\fy+1)*\s});
      \else \fill[red!7] (\fx*\s,\fy*\s) rectangle ({(\fx+1)*\s},{(\fy+1)*\s}); \fi}}
    \fill[blue!16] (1*\s,1*\s) rectangle (2*\s,2*\s);
    \foreach \x in {0,...,3}{\foreach \y in {0,...,3}{
      \ifnum\x<3 \draw[gray!55, line width=1.5pt] (\x*\s,\y*\s) -- ({(\x+1)*\s},\y*\s);
        \fill[brown!70!black] ({(\x+0.5)*\s},\y*\s) circle (0.05); \fi
      \ifnum\y<3 \draw[gray!55, line width=1.5pt] (\x*\s,\y*\s) -- (\x*\s,{(\y+1)*\s});
        \fill[brown!70!black] (\x*\s,{(\y+0.5)*\s}) circle (0.05); \fi}}
    #1
    \foreach \x in {0,...,3}{\foreach \y in {0,...,3}{
      \node[draw, circle, fill=green!35, minimum size=0.34cm, inner sep=0] at (\x*\s,\y*\s) {};}}
  }
\begin{scope}
  \patch{
    \draw[teal!70!black, line width=2.4pt] (NW) -- (NE);
    \draw[teal!70!black, line width=2.4pt] (SW) -- (SE);
    \draw[blue!60, line width=2.4pt] (NW) -- (SW);
    \node[teal!55!black, anchor=south] at ($(NW)!0.5!(NE)+(0,0.10)$) {pairA};
    \node[teal!55!black, anchor=north] at ($(SW)!0.5!(SE)+(0,-0.10)$) {pairB};
    \node[blue!60!black, fill=white, inner sep=1.2pt, rotate=90] at ($(NW)!0.5!(SW)+(-0.02,0)$) {bridge};
    \node[draw, circle, fill=blue!25, minimum size=0.26cm, inner sep=0] at ($(NW)+(0.24,-0.24)$) {};
    \node[draw, circle, fill=blue!25, minimum size=0.26cm, inner sep=0] at ($(SW)+(0.24,0.24)$) {};
    \node[blue!55!black, anchor=west] at ($(NW)+(0.34,-0.24)$) {$x_0$};
    \node[blue!55!black, anchor=west] at ($(SW)+(0.34,0.24)$) {$x_1$};
  }
  \node[anchor=south east, fill=white, inner sep=1.2pt] at ($(NW)+(-0.06,0.10)$) {$a_0$};
  \node[anchor=south west, fill=white, inner sep=1.2pt] at ($(NE)+(0.06,0.10)$) {$a_1$};
  \node[anchor=north east, fill=white, inner sep=1.2pt] at ($(SW)+(-0.30,-0.10)$) {$a_2$};
  \node[anchor=north west, fill=white, inner sep=1.2pt] at ($(SE)+(0.06,-0.10)$) {$a_3$};
  \begin{scope}[shift={(0.35*\s,-1.25)}]
    \foreach \i/\n in {0/a_1, 1/a_0, 2/a_2, 3/a_3}{
      \node[draw, circle, fill=green!25, minimum size=0.46cm, inner sep=0] (c\i) at (\i*0.95,0) {\scriptsize $\n$};}
    \draw[teal!70!black, line width=2.2pt] (c0) -- (c1);
    \draw[teal!70!black, line width=2.2pt] (c2) -- (c3);
    \draw[blue!60, line width=2.2pt] (c1) -- (c2);
    \node[anchor=north, align=center] at (1.425,-0.42) {\scriptsize one fusion $\Rightarrow\ \ket{\mathrm{GHZ}_4}$};
  \end{scope}
  \node[anchor=north, text width=5.6cm, align=justify] at (1.5*\s,-2.55)
    {(a) Chain gadget. Two singlets on the north and south lanes of the plaquette supply
     all four arms, and a single bridge on the west lane fuses them into
     $a_1\!-\!a_0\!-\!a_2\!-\!a_3$. Three singlets total (two arm pairs and one bridge), six readouts
     (two bridge, four arm), six CZs.};
\end{scope}

\begin{scope}[shift={(6.6*\s,0)}]
  \patch{
    \draw[-{Stealth[length=5pt]}, teal!65!black, line width=1.7pt, shorten >=7pt, shorten <=7pt] (NW) -- (NE);
    \draw[-{Stealth[length=5pt]}, teal!65!black, line width=1.7pt, shorten >=7pt, shorten <=7pt] (NE) -- (SE);
    \draw[-{Stealth[length=5pt]}, teal!65!black, line width=1.7pt, shorten >=7pt, shorten <=7pt] (SE) -- (SW);
    \node[teal!65!black, fill=white, inner sep=1.2pt] at ($(NW)!0.5!(NE)+(0,0.26)$) {1};
    \node[teal!65!black, fill=white, inner sep=1.2pt] at ($(NE)!0.5!(SE)+(0.26,0)$) {2};
    \node[teal!65!black, fill=white, inner sep=1.2pt] at ($(SW)!0.5!(SE)+(0,-0.26)$) {3};
  }
  \node[anchor=south east, fill=white, inner sep=1.2pt] at ($(NW)+(-0.06,0.10)$) {\scriptsize $\ket{+}$, CZ};
  \node[anchor=south west, fill=white, inner sep=1.2pt] at ($(NE)+(0.06,0.10)$) {\scriptsize CZ};
  \node[anchor=north east, fill=white, inner sep=1.2pt] at ($(SW)+(-0.06,-0.10)$) {\scriptsize CZ, $X$ readout};
  \node[anchor=north west, fill=white, inner sep=1.2pt] at ($(SE)+(0.06,-0.10)$) {\scriptsize CZ};
  \node[anchor=north, text width=5.6cm, align=justify] at (1.5*\s,-2.55)
    {(b) Walk gadget. One ancilla is born $\ket{+}$ at a corner and is conveyed once around
     the plaquette, applying a CZ at each of the four data sites and being read out at the
     last. No singlet and no teleportation is used. One readout, four CZs, three one-lane
     transits.};
\end{scope}
\end{tikzpicture}
\caption{The weight-4 reduction of both gadgets, on the all-data-node lattice used for the
surface-code baseline. Every site carries a data qubit and there are no check nodes; the
plaquettes (shaded, $X$ and $Z$ alternating in a checkerboard) are faces of the lane graph
rather than nodes of it, so each gadget lives entirely on the four lanes bounding one face.
The chain construction of Sec.~\ref{sec:fusion} retains one fusion and has no
routed bridges, and the teleported-ancilla gadget of Sec.~\ref{sec:gadgets} loses its
teleportation steps altogether, reducing to a bare walk.}
\label{fig:surfacegadgets}
\end{figure*}

The surface-code layout uses data nodes without
the separate BB check-centre nodes. This comparison
tests the cost of GHZ-based extraction when
long-range communication is absent
(Fig.~\ref{fig:surfacegadgets}).

\subsection{Results}
\label{app:surface_results}

The BB-code comparisons involve long-range checks. To examine the same
trade-off when those connections are absent,
Fig.~\ref{fig:results_surface} shows the rotated surface code on the same
fabric and with the same hardware parameters, as a reference in which every
check is local. We simulate distances $d=4,6,8$, each for $d$ rounds, using
the walking-ancilla and 4-GHZ gadgets of Sec.~\ref{sec:weight4}. These
circuits are decoded with minimum-weight perfect matching
(PyMatching~\cite{Higgott_2025}) rather than BP+OSD. This decoder does not use the herald flags, but it is
fast enough to sample up to $2\times10^7$ shots per point.

Every check involves only neighbouring sites, so each ancilla moves only a
few lanes per round, and we extend the sweep to $p_\mathrm{sh}=0.5$. Below approximately $p_\mathrm{sh}=10^{-2}$, gate and readout
faults set the logical error rate, which barely changes with
$p_\mathrm{sh}$. In this regime, each increase of the distance by two lowers
the rate by roughly an order of magnitude: for the walk gadget, from
$2.9\times10^{-4}$ at $d=4$ to $1.7\times10^{-5}$ at $d=6$ and
$1.3\times10^{-6}$ at $d=8$ per round.

The two kinds of shuttle error act very differently here. A phase error on a
walking ancilla commutes with its CZ gates, so it cannot spread to the data;
it only flips the measured syndrome bit, which the decoder treats as a
measurement error. Dephasing alone is therefore only weakly damaging: without
bit flips, the walk-gadget curves remain ordered by distance all the way to
$p_\mathrm{sh}=0.5$, and the 4-GHZ curves cross only near
$p_\mathrm{sh}=0.1$. A bit flip on an ancilla, by contrast, does spread to
the data qubits it visits next, and adding bit flips moves the crossings to
lower noise: near $p_\mathrm{sh}=0.07$ for the 4-GHZ gadget and $0.15$ for
the walk with $p_x=p_\mathrm{sh}/10$, and around $0.05$ to $0.07$ for both
with $p_x=p_\mathrm{sh}/3$. These finite-distance crossings indicate where
increasing the code distance stops improving the rate; they are not a
precise determination of an asymptotic threshold. Below the crossings, the
walk gadget is lower than the 4-GHZ gadget by a factor that grows with distance: for
$10^{-3}\le p_\mathrm{sh}\le10^{-2}$ it is $1.3$ to $1.8$ at $d=4$, $1.9$ to $3.0$ at $d=6$ and $3.0$ to $4.8$
at $d=8$, consistent with its smaller number of readouts:
when all checks are local, there is little transport for distributed
entanglement to protect against.

\subsection{Sampling details}
\label{app:surface_sampling}

Points with a shot-failure fraction above $45\%$ are marked as saturated,
since a single logical observable fails half the time when the outcome is
random. Each point uses up to $2\times10^7$ shots, stopping at
$300$ failures if that occurs first. Each surface-code patch encodes a
single logical qubit, so these rates refer to a different memory than the
twelve-qubit BB-block rates. For perspective, the Gross code encodes twelve
logical qubits at distance twelve in $144$ data qubits, the number of data
qubits that a single distance-twelve rotated surface code needs for one.

\section{Additional figures}
\label{app:figures}

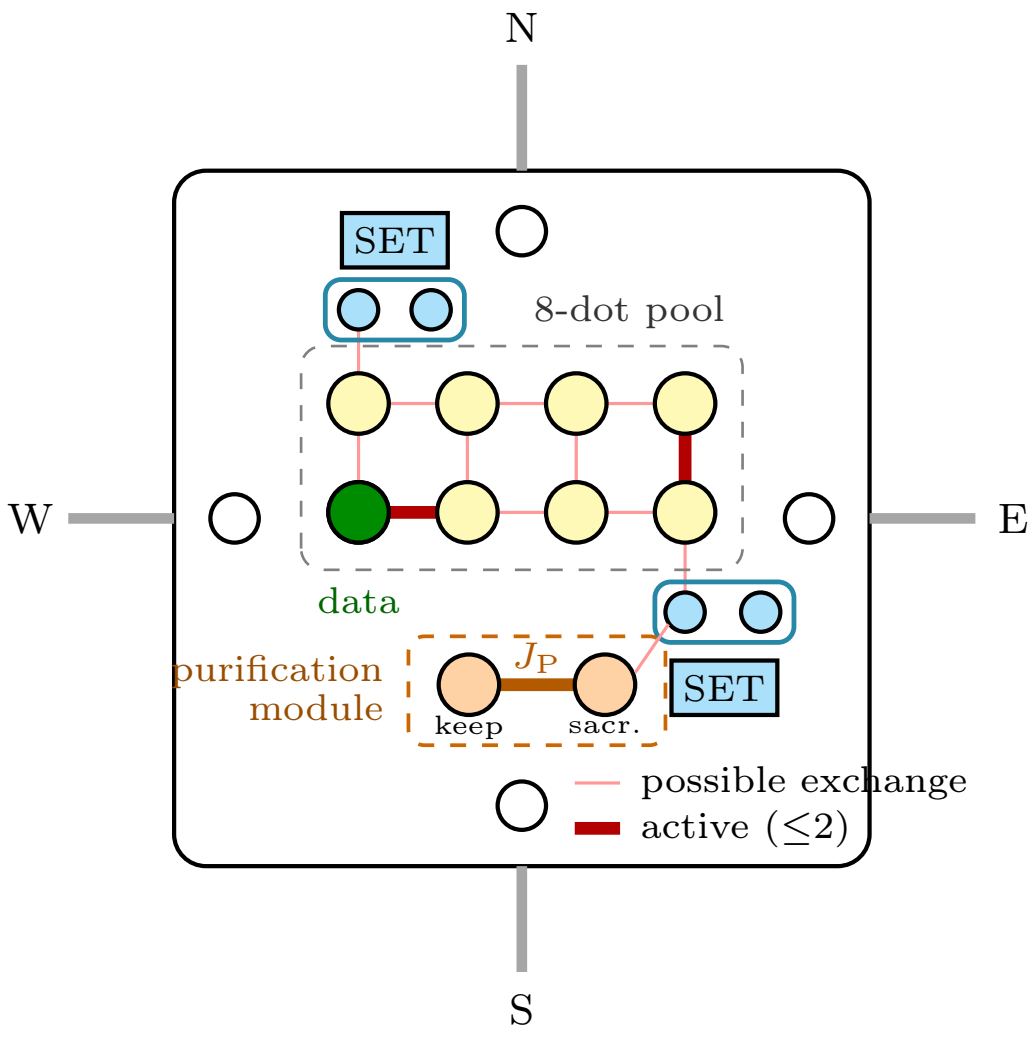
\begin{figure*}[t]
\centering
\begin{tikzpicture}[every node/.style={font=\footnotesize}]
\begin{scope}
  \def\d{0.72}\def\y{0.4}
  \draw[rounded corners=6pt, line width=0.8pt] (-2.3,-2.3) rectangle (2.3,2.3);
  \foreach \a/\l in {90/N, 270/S, 0/E, 180/W}{
    \draw[gray!70, line width=2pt] (\a:2.3) -- (\a:3.0);
    \node at (\a:3.25) {\l};
    \draw[fill=white, line width=0.8pt] (\a:1.9) circle (0.16);
  }
  \foreach \i in {0,...,2}{\foreach \j in {0,1}{
    \draw[red!40, line width=0.6pt] ({(\i-1.5)*\d+0.2},{(\j-0.5)*\d+\y}) -- ({(\i-0.5)*\d-0.2},{(\j-0.5)*\d+\y});}}
  \foreach \i in {0,...,3}{
    \draw[red!40, line width=0.6pt] ({(\i-1.5)*\d},{-0.5*\d+0.2+\y}) -- ({(\i-1.5)*\d},{0.5*\d-0.2+\y});}
  \draw[red!70!black, line width=2.4pt] ({-1.5*\d+0.2},{-0.5*\d+\y}) -- ({-0.5*\d-0.2},{-0.5*\d+\y});
  \draw[red!70!black, line width=2.4pt] ({1.5*\d},{-0.5*\d+0.2+\y}) -- ({1.5*\d},{0.5*\d-0.2+\y});
  \draw[red!40, line width=0.6pt] ({-1.5*\d},{0.5*\d+0.2+\y}) -- ({-1.5*\d},{1.25});
  \draw[cyan!60!black, line width=0.9pt, rounded corners=3pt] ({-1.5*\d-0.22},{1.18}) rectangle (-0.38,{1.58});
  \draw[fill=cyan!30, line width=0.8pt] ({-1.5*\d},{1.38}) circle (0.13);
  \draw[fill=cyan!30, line width=0.8pt] (-0.6,{1.38}) circle (0.13);
  \draw[red!40, line width=0.6pt] ({1.5*\d},{-0.5*\d-0.2+\y}) -- ({1.5*\d},{-0.49});
  \draw[cyan!60!black, line width=0.9pt, rounded corners=3pt] (0.88,-0.82) rectangle (1.80,-0.42);
  \draw[fill=cyan!30, line width=0.8pt] ({1.5*\d},-0.62) circle (0.13);
  \draw[fill=cyan!30, line width=0.8pt] (1.58,-0.62) circle (0.13);
  \foreach \i in {0,...,3}{\foreach \j in {0,1}{
    \draw[fill=yellow!35, line width=0.8pt] ({(\i-1.5)*\d},{(\j-0.5)*\d+\y}) circle (0.2);}}
  \draw[fill=green!55!black, line width=0.8pt] ({-1.5*\d},{-0.5*\d+\y}) circle (0.2);
  \node[anchor=north, green!40!black] at ({-1.5*\d},{-0.5*\d-0.38+\y}) {\scriptsize data};
  \draw[dashed, gray, rounded corners=4pt, line width=0.5pt] ({-1.5*\d-0.38},{-0.5*\d-0.38+\y}) rectangle ({1.5*\d+0.38},{0.5*\d+0.38+\y});
  \node[anchor=south east, gray!40!black] at ({1.5*\d+0.38},{0.5*\d+0.38+\y}) {\scriptsize 8-dot pool};
  \draw[fill=cyan!30, line width=0.8pt] (-1.19,1.66) rectangle (-0.49,2.02); \node at (-0.84,1.84) {\scriptsize SET};
  \draw[fill=cyan!30, line width=0.8pt] (0.99,-1.30) rectangle (1.69,-0.94); \node at (1.34,-1.12) {\scriptsize SET};
  \draw[dashed, orange!80!black, line width=0.7pt, rounded corners=3pt] (-0.75,-1.5) rectangle (0.95,-0.78);
  \draw[orange!70!black, line width=2.4pt] (-0.15,-1.1) -- (0.35,-1.1);
  \draw[fill=orange!35, line width=0.8pt] (-0.35,-1.1) circle (0.2);
  \draw[fill=orange!35, line width=0.8pt] (0.55,-1.1) circle (0.2);
  \node[orange!70!black] at (0.1,-0.92) {\scriptsize $J_\mathrm{P}$};
  \node at (-0.35,-1.38) {\tiny keep};
  \node at (0.55,-1.38) {\tiny sacr.};
  \draw[red!40, line width=0.6pt] (0.75,-1.02) -- (0.97,-0.70);
  \node[orange!60!black, anchor=east, align=right] at (-0.8,-1.12) {\scriptsize purification\\[-2pt]\scriptsize module};
  \draw[red!40, line width=0.6pt] (0.35,-1.75) -- (0.65,-1.75); \node[anchor=west] at (0.67,-1.75) {\scriptsize possible exchange};
  \draw[red!70!black, line width=2.4pt] (0.35,-2.05) -- (0.65,-2.05); \node[anchor=west] at (0.67,-2.05) {\scriptsize active (${\le}2$)};
\end{scope}

\end{tikzpicture}
\caption{Bulk data node as in Fig.~\ref{fig:node}(b), with link purification enabled. The purification module (orange) is a pair of work dots with a dedicated exchange coupling $J_\mathrm{P}$, separate from the pool's exchange couplings (red; thin: possible, thick: active). The retained and sacrificial halves of a purified link are loaded into it, it executes this endpoint's half of the purification circuit of Fig.~\ref{fig:purification}, and the sacrificial half is read out through one of the node's two Pauli-spin-blockade channels: a CZ with one spin of the channel's singlet-prepared double dot (cyan), which is then read by its sensor (SET); the same channel otherwise reads the adjacent pool dot. The module is occupied for $3t_1+2t_2+t_\mathrm{meas}=1.0~\mu$s per purified link; raw halves waiting for it are held in the dot pool.}
\label{fig:node_purifier}
\end{figure*}

\begin{figure*}[t]
\centering
\begin{quantikz}[row sep={0.72cm,between origins}, column sep=0.55cm]
\lstick{$a_0$}          & \ctrl{1}\gategroup[4,steps=2,style={dashed,rounded corners,inner xsep=2pt}]{central fusion}   & \qw        & \ctrl{5}\gategroup[6,steps=2,style={dashed,rounded corners,fill=blue!8,inner xsep=2pt},background,label style={label position=below,anchor=north,yshift=-0.2cm}]{verification}   & \qw        & \qw      & \qw      & \rstick{to $d_0$} \\
\lstick{$x_0$}          & \control{} & \qw        & \qw        & \qw        & \gate{H} & \meter{} & \setwiretype{c} \rstick{$m_{x_0}$} \\
\lstick{$x_1$}          & \qw        & \ctrl{1}   & \qw        & \qw        & \gate{H} & \meter{} & \setwiretype{c} \rstick{$m_{x_1}$} \\
\lstick{$a_2$}          & \qw        & \control{} & \qw        & \qw        & \qw      & \qw      & \rstick{to $d_2$} \\
\lstick{$a_3$}          & \qw        & \qw        & \qw        & \ctrl{1}   & \qw      & \qw      & \rstick{to $d_3$} \\
\lstick{$v\ \ket{+}$}   & \qw        & \qw        & \control{} & \control{} & \gate{H} & \meter{} & \setwiretype{c} \rstick{$m_v$}
\end{quantikz}
\caption{Verification of the central fusion at a bulk check centre. Wires $a_0$ and $a_2$ are arms from the singlets $(a_0,a_1)$ and $(a_2,a_3)$; the bridge singlet $(x_0,x_1)$ mediates the central fusion, whose readouts $m_{x_0}\oplus m_{x_1}$ record the parity $Z_{a_0}Z_{a_2}$. The verifier $v$, held in the resident verifier dot, is prepared in $\ket{+}$ and coupled to $a_0$ and $a_3$ by two CZs; its $X$ readout gives $m_v = Z_{a_0}Z_{a_3}$ without measuring either arm. Because $Z_{a_2}Z_{a_3}$ is fixed by the singlet $(a_2,a_3)$, the detector $D_v = m_v\oplus m_{x_0}\oplus m_{x_1}$ is deterministic and equals $0$ without faults. A readout error on $x_0$ or $x_1$, which would otherwise leave the silent data error $Z_{d_2}Z_{d_3}Z_{d_5}$, flips $D_v$. The verification CZs commute with the fusion CZs, so they may be placed in any CZ layer before $a_0$ and $a_3$ leave the centre; the remaining chain (outer fusions, data CZs and arm readouts) is as in Fig.~\ref{fig:chaincircuit}.}
\label{fig:verification}
\end{figure*}
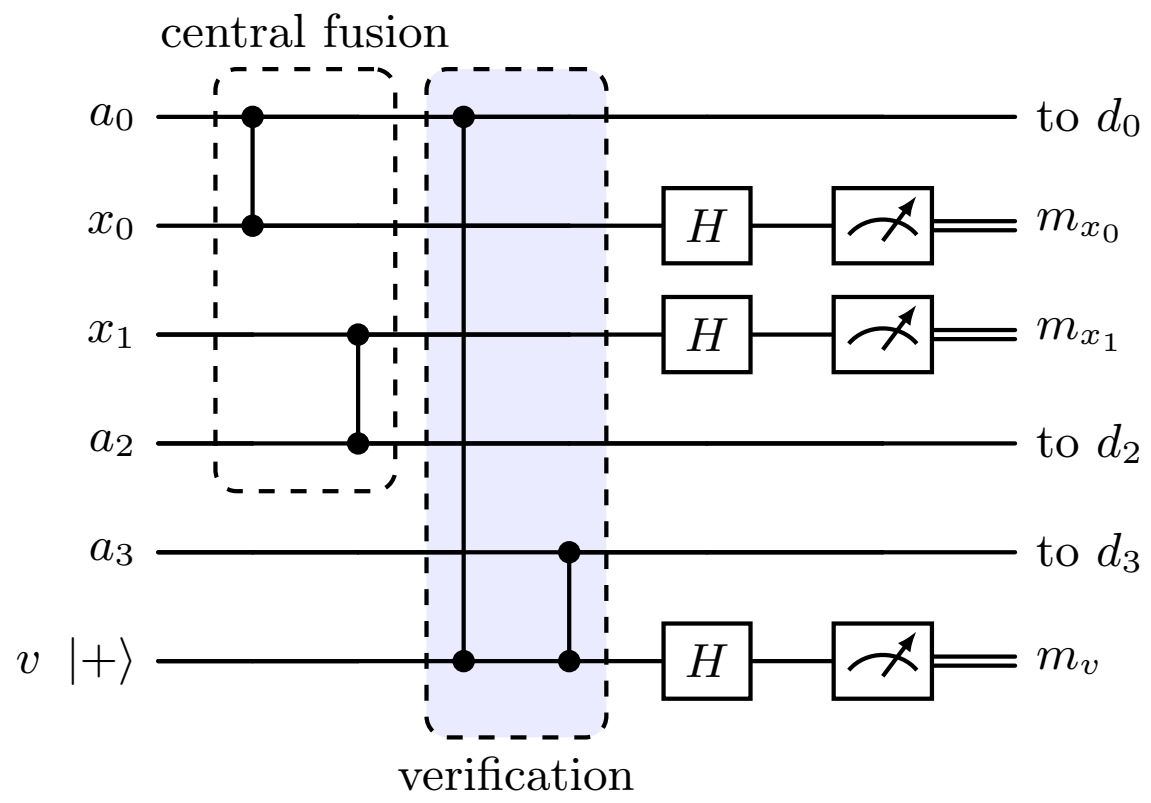

\begin{figure*}[t]
  \centering
  \includegraphics[width=0.72\linewidth]{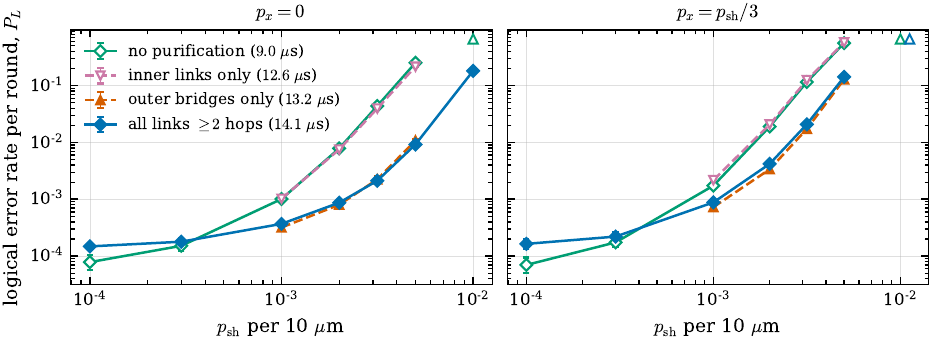}
  \caption{Which purified links matter for the verified GHZ gadget, Gross code, $12$ rounds,
  $P_L=P_L^X+P_L^Z$. Solid lines: no purification (green, open) and purification of every link of
  two or more hops (blue, filled), as in Fig.~\ref{fig:results_purification}. Dashed lines:
  purification of the outer bridges only (orange, filled) or of the inner links only, i.e.\ the
  pair links and central bridges (pink, open). Round durations are given in the legend.}
  \label{fig:results_purif_split}
\end{figure*}

\begin{figure*}[t]
  \centering
  \includegraphics[width=0.85\linewidth]{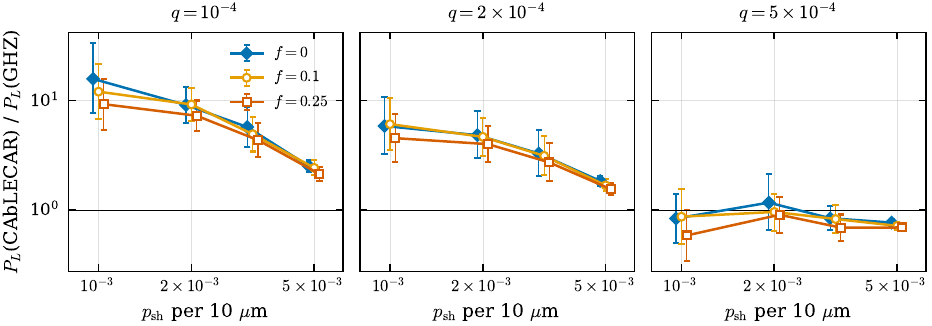}
  \caption{Ratio of the logical error rates of CAbLECAR on our node lattice and the verified,
  purified GHZ gadget, $P_L=P_L^X+P_L^Z$, for sources whose undetectable fraction of faulty pairs
  is $f=0$ (filled), $0.1$ and $0.25$ (open). Gross code, $12$ rounds, $p_x=p_\mathrm{sh}/3$;
  panels: two-qubit gate and readout error $q=p_2=p_\mathrm{meas}$; all other parameters as in
  Table~\ref{tab:params}. Values above one favour the GHZ gadget. Error bars are conservative
  $95\%$ ranges; points are offset slightly in $p_\mathrm{sh}$ for clarity.}
  \label{fig:results_srcrobust}
\end{figure*}

Figure~\ref{fig:node_purifier} shows a node with the purification module,
Fig.~\ref{fig:verification} the central-fusion verification circuit
(Sec.~\ref{sec:results_verification}), and Fig.~\ref{fig:results_purif_split} the test of which
purified links give the GHZ gadget its benefit (Sec.~\ref{sec:results_purification}), and
Fig.~\ref{fig:results_srcrobust} the comparison with CAbLECAR at lower gate and readout error for
a less favourable source (Sec.~\ref{sec:results_heralding}).

\end{document}